\documentclass[prl,reprint,groupedaddress,superscriptaddress]{revtex4-1}
\usepackage{amsmath,bm}
\usepackage{physics}
\usepackage{graphicx}
\usepackage{hyperref}
\usepackage{times}
\usepackage{xcolor}
\usepackage{siunitx}
\usepackage{layouts}
\usepackage{gensymb}
\usepackage{mathtools}
\usepackage{tabularx}

\usepackage{etoolbox}
\makeatletter
\patchcmd{\frontmatter@abstract@produce}
  {\vskip200\p@\@plus1fil
   \penalty-200\relax
   \vskip-200\p@\@plus-1fil}
  {}
  {}
  {}
\makeatother

\newcommand{\subref}[2]{\hyperref[#1]{\ref*{#1}(#2)}}
\definecolor{refblue}{HTML}{2E2E91}
\hypersetup{colorlinks=true, citecolor=refblue, urlcolor=refblue, linkcolor=refblue}

\newcommand{\wit}[1]{\textrm{#1}}

\begin{document}

\title{Cavity-QED determination of the natural linewidth of the $^{87}$Sr millihertz clock transition with 30~$\mu$Hz resolution}

\author{Juan A. Muniz}
\email{jmunizq2@gmail.com}
\affiliation{JILA, NIST, and Dept. of Physics, University of Colorado, 440 UCB, Boulder, CO 80309, USA}
\affiliation{Instituto de F\'isica, Facultad de Ingener\'ia, Universidad de la Rep\'ublica, J.H. y Reissig 565, 11300 Montevideo, Uruguay}
\author{Dylan J. Young}
\affiliation{JILA, NIST, and Dept. of Physics, University of Colorado, 440 UCB, Boulder, CO 80309, USA}
\author{Julia R. K. Cline}
\affiliation{JILA, NIST, and Dept. of Physics, University of Colorado, 440 UCB, Boulder, CO 80309, USA}
\author{James K. Thompson}
\affiliation{JILA, NIST, and Dept. of Physics, University of Colorado, 440 UCB, Boulder, CO 80309, USA}

\date{\today}

\begin{abstract}

We present a new method for determining the intrinsic natural linewidth or lifetime of exceptionally long-lived optical excited states. Such transitions are key to the performance of state-of-the-art atomic clocks, have potential applications in searches for fundamental physics and gravitational wave detectors, as well as novel quantum many-body phenomena. With longer lifetime optical transitions, sensitivity is increased, but so far it has proved challenging to determine the natural lifetime of many of these long lived optical excited states because standard population decay detection techniques become experimentally difficult. Here, we determine the ratio of a poorly known ultranarrow linewidth transition ($^3$P$_0$ to $^1$S$_0$ in $^{87}$Sr) to that of another narrow well known transition ($^3$P$_1$ to $^1$S$_0$) by coupling the two transitions to a single optical cavity and performing interleaved nondestructive measurements of the interaction strengths of the atoms with cavity modes near each transition frequency. We use this approach to determine the natural linewidth of the clock transition $^3$P$_0$ to $^1$S$_0$ in $^{87}$Sr to be  $\gamma_0/(2\pi) = 1.35(3)$~mHz or $\tau= 118(3)$~s. The 30~$\mu$Hz resolution implies that we could detect states with lifetimes just below 2 hours, and with straightforward future improvements, we could detect states with lifetimes up to 15 hours, using measurement trials that last only a few hundred milliseconds, eliminating the need for long storage times in optical potentials.

\end{abstract}


\maketitle 
Ultranarrow linewidth optical transitions have become the new standard for precision optical metrology, providing fast phase evolution, long coherence times, and intrinsic insensitivity to key environmental perturbations that have allowed remarkable fractional accuracy at the $10^{-18}$ level \cite{Ludlow_2015,Oelker_2019,Campbell_2017,Ushijima_2015,Takano_2016,Grotti_2018,Schioppo_2016,Brewer_2019}, and potential applications for fundamental physics, such as gravitational wave detection using matter-wave interferometry or dark-matter searches \cite{Safronova_2016,Tino2019,Hu_2017,Aguila_2018, Arvanitaki_2018,Wcislo_2018}, quantum many-body physics \cite{Norcia_SS_2018,Muniz2020,Kolkowitz_SOC_2016,Bromley_2018,Goban_2018,Senaratne_2018}, and novel cavity QED applications for superradiant lasing \cite{Norcia_SR_2016,Norcia_SRFreq_2018} and spin squeezing on an optical clock transition \cite{Pedrozo_2020}. Precise knowledge of the intrinsic linewidth of these transitions is important to understand the ultimate limits on quantum coherence offered by various atomic transitions and species.

Although the expected lifetime of these states could surpass 100~s \cite{Yasuda_2004,Jensen_2011,Rosenband_2007,Barton_2000,Staanum_2004,Kreuter_2004,Becker_2001,Walhout_1994}, various mechanisms preclude the observation of the natural excited state lifetime, such as black-body radiation-induced decay \cite{Walhout_1995,Yasuda_2004} or scattering due to optical lattice light used to trap the atoms \cite{Dorscher_2018,Hutson_2019}, preventing the application of standard population decay techniques to determine their lifetimes. For example, state-of-the-art optical lattice clocks have only demonstrated coherence up to $\sim$10~s \cite{Campbell_2017,Hutson_2019} which is limited mostly by Raman scattering of the lattice light off an excited state \cite{Hutson_2019}. In fact, most of the systems where these long excited state lifetimes have been measured consist of atoms trapped without optical potentials, such as magnetic or ion traps \cite{Yasuda_2004,Jensen_2011,Rosenband_2007,Barton_2000,Staanum_2004,Kreuter_2004,Becker_2001,Walhout_1994}. 

To date, the two reported values for the $^3$P$_0$ excited state lifetime in $^{87}$Sr are $\tau = 330(140)$~s from Ref.~\cite{Dorscher_2018} obtained from population decay measurements from excited metastable states, and $\tau = 145(40)$~s from Ref.~\cite{Boyd_2007} obtained from effective atomic models and measurements of differential Land\'e g-factors between ground and excited clock states, while ab-initio calculations estimate a lifetime between 110-130~s \cite{Porsev_2004,Santra_2004}. With the implementation of new potential landscapes for operating with reduced lattice-induced scattering \cite{Hutson_2019,Norcia_2019,Madjarov_2019} that can suppress these effects, and with reference optical cavities whose coherence times start to approach the minute time scale \cite{Matei_2017,Zhang_2017,Robinson_2019}, the full enhancement of these ultranarrow optical transitions can be achieved. 

In this letter, we present a series of measurements that allow us to directly determine the natural lifetime of the excited clock state $^3$P$_0$ ($\ket{e_0}$) in $^{87}$Sr. Our technique consists of precisely and simultaneously measuring the \emph{ratio} of single photon Rabi frequencies along two optical transitions---the millihertz transition ($^1$S$_0=\ket{g}\to ^3$P$_0$ = $\ket{e_0}$) and the 7.5~kHz transition ($^1$S$_0 \to ^3$P$_1$  = $\ket{e_1}$)---using a common atomic ensemble inside an optical resonator (see Fig.~\subref{fig:fig1}{a}). These Rabi frequencies, denoted by $2g_{0,1}$ for light-matter coupling strengths $g_{0,1}$ along the millihertz and 7.5~kHz transitions respectively, depend on the electric dipole moment of the atoms ($d$) along with well-known and independently characterized geometric factors \cite{TanjiSuzuki_2011}, such as the cavity's mode waist ($\wit{w}$) and length ($\wit{L}$). The natural linewidth $\gamma_0$ can then be linked to the known natural linewidth $\gamma_1$ from the measured coupling strength ratio as:
\begin{equation} \label{eq:gammaR}
    \frac{\gamma_0}{\gamma_1} = \left(\frac{\wit{L}_0}{\wit{L}_1}\right)\left(\frac{\wit{w}_0}{\wit{w}_1}\right)^2\left(\frac{\omega_{A0}}{\omega_{A1}}\right)^2\left(\frac{g_0}{g_1}\right)^2,
\end{equation}
where $\omega_A$ is the (well-known) atomic transition frequency. Note that, for this manuscript, we generically use subscripts $0$ and $1$ to denote quantities for the clock transition (with wavelength $\lambda_0 = 698.44$~nm) and the 7.5~kHz transition (with wavelength $\lambda_1 = 689.45$~nm) respectively, as shown in Fig.~\subref{fig:fig1}{a}. Calculating this ratio as opposed to just the millihertz transition Rabi frequency allows for cancellation of many common noise and systematic errors such as atom number fluctuations, inhomogeneous atom-cavity coupling, cavity and laser frequency noise, and finite ensemble size effects.

To accomplish the above, we perform consecutive measurements of the dispersive cavity resonance frequency shift $\Delta\omega_1$ (or equivalently the multi-pass phase shift $\Delta\varphi_1$) on the 7.5~kHz transition, as well as the  dispersive phase shift $\Delta\varphi_0$ on the millihertz transition. These phase shifts depend directly on the light-matter coupling strength, scaling as $\Delta\varphi_0/\Delta\varphi_1\propto (g_0/g_1)^2$.
\cite{Jaynes_1963,Kimble_1998}. 

Ultimately, this involves measuring the phase shift experienced by far off-resonant light passing through an atomic medium [Fig.~\subref{fig:fig1}{b}], which arises from the interference between the incident and the scattered fields. The optical cavity magnifies this phase shift due to the multiple round trips \cite{TanjiSuzuki_2011} and defines the spatial modes that interact with the atoms. However, an optical resonator introduces different systematic effects that need to be taken into account to precisely determine $(g_0/g_1)$. Remarkably, due to the ultranarrow linewidth of the clock transition, our phase shift measurements $\Delta\varphi_0$ are dispersive, i.e., the probe is sufficiently detuned from the atomic transition, while also being in the resolved motional sideband limit, as the axial vibration frequency is much larger than the probe's detuning. This scenario does not have precedent in the atomic quantum non-demolition (QND) measurements community, and impacts in our measurement as an effective reduction of the single photon Rabi frequencies that constitutes the largest systematic correction (by $6\%$) in our analysis.

\begin{figure}[!tb]
\includegraphics[width=3.375in]{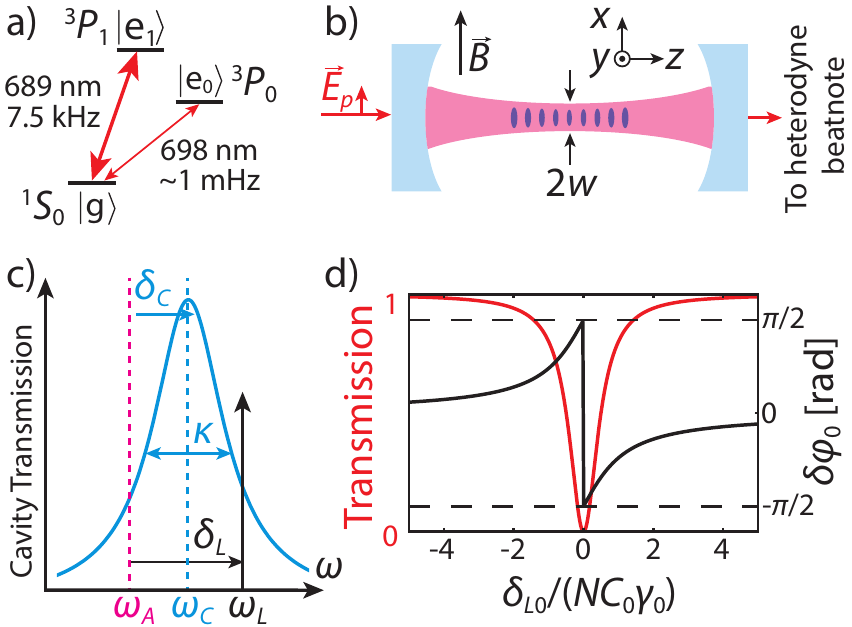}
\caption{{\bf{System description.}} (a) Relevant energy levels in $^{87}$Sr. (b) Atoms are trapped inside an intracavity optical lattice. The probe electric field $\vec{E}_p$ is polarized along the magnetic field $\vec{B}$ direction ($\hat{x}$) and has mode waist radius $\wit{w}$. Phase detection is done via heterodyne measurements. (c) Experimental frequencies showing the cavity center frequency $\omega_C$, laser frequency $\omega_L$ and their respective detunings, $\delta_C$ and $\delta_{L}$ from the atomic transition frequency $\omega_A$. Light leaks out of the cavity at a total rate $\kappa$. (d) For the clock transition (subindex 0), simulated cavity transmission and phase shift $\delta\varphi_{0}$ against probe detuning $\delta_{L0}$ for $\delta_{C0} = 0$ in units of $NC_0\gamma_0$. At a given detuning the phase shift produced by the atomic ensemble is proportional to $\gamma_0$.}
\label{fig:fig1}
\end{figure}

Our system, also described in \cite{Norcia_SR_2016,Norcia_SRFreq_2018}, consists of an ensemble of up to $10^5$ $^{87}$Sr atoms confined within a high finesse optical cavity by a $\lambda_\text{trap}=813$~nm, near-magic wavelength intracavity optical lattice, as sketched in Fig.~\subref{fig:fig1}{b} \cite{Ye_2008,Campbell_2017}. We optically pump the atoms into a 50/50 spin mixture of the $^1$S$_0$ nuclear Zeeman levels $m_F=\pm 9/2$, with less than 5\% of atoms remaining in the other 8 $m_F$ states. The 813~nm trap is 185~$\mu$K deep, with measured axial trap frequency $\omega_z/(2\pi) = 230(1)$~kHz. The atoms have an axial temperature $T_z = 14(1)$~$\mu$K, and their mean vibrational quantum number is $\bar{n}_z = 0.9(1)$, obtained using sideband spectroscopy \cite{Blatt_2009}. 

Both transitions fall into the so-called bad cavity regime, where the $\sim 150~$kHz cavity linewidth ($\kappa$) is larger than the excited state linewidth ($\gamma$). To understand the dispersive measurements we perform, we consider $N$ equally coupled two level atomic dipoles with transition frequency $\omega_A$ that interact with one cavity mode at frequency $\omega_C$, detuned by $\delta_C = \omega_C - \omega_A$ from the atomic transition, as shown in Fig.~\subref{fig:fig1}{c}. 

For the 7.5~kHz transition our system typically satisfies $NC_1\gamma_1\gg \kappa_1$, with $C_1 = (2g_1)^2/(\gamma_1\kappa_1)$ denoting the single atom cooperativity parameter that characterizes the cavity-enhanced interactions on this transition \cite{Kimble_1998,Chen_2014}. This gives a resolved collective vacuum Rabi splitting for a resonant cavity mode ($\delta_{C1} = 0$). For this experiment, we instead operate in the dispersive limit ($\delta_{C1} \gg \sqrt{N}g_1$), where the cavity resonance frequency experiences a shift $\delta\omega_1 = N g_1^2/\delta_{C1}$ \cite{Chen_2014} due to the presence of $N$ atoms in the ground state. This frequency shift corresponds to an equivalent multi-pass phase shift $\delta\varphi_1 = \delta\omega_1 /(\kappa_1/2) = N C_1 \gamma_1 / 2\delta_{C1}$. This regime has been explored in many different QND platforms \cite{Braginsky_1996,Appel_2009,Chen_2010,Schleier-Smith_2010,Bohnet2014,Hosten2016,Cox2016}.

The millihertz optical transition falls into a less common regime for ensemble-cavity experiments, where the collective vacuum Rabi splitting is unresolved and $NC_0\gamma_0\ll \kappa_0$, with $C_0 = (2g_0)^2/(\gamma_0\kappa_0)$ the single atom coopertivity parameter. Dispersive measurements are realized by observing the multi-pass phase shift, $\delta\varphi_0$, of the transmitted probe light detuned $\delta_{L0}$ from the atomic transition $\omega_{A0}$. In this case $\delta\varphi_{0} = -N C_0 \gamma_0 /(2\delta_{L0}) - (\delta_{C0}-\delta_{L0})/(\kappa_0/2)$, for small angles ($\kappa_0\gg (\delta_{C0},\delta_{L0}) \gg N C_0 \gamma_0$). The first term that contributes to $\delta\varphi_0$ is the phase shift induced on the probe light by the atoms, the quantity that we wish to measure. The second term is a phase shift that arises when the probe light is not on resonance with an empty cavity, representing a background that must be subtracted.

The normalized power transmission (red line) and phase shift $\delta\varphi_0$ (black line) for a weak probe in the presence of atoms for $\delta_{C0} = 0$ are shown in Fig.~\subref{fig:fig1}{d}. In the weak excitation limit, the transmission drops as the incident and scattered electric fields destructively interfere over a characteristic frequency width of order $NC_0\gamma_0$, while $\delta\varphi_0$ shows a narrow feature around $\delta_{L0} = 0$ of order $\gamma_0$. As $|\delta_{L0}|$ increases, the cavity-like phase shift starts to be significant compared to the atomic-like phase shift: for our chosen probe detuning $|\delta_{L0}|/(2\pi) = 1$~kHz, the former is $\sim 15$~mrad while the latter is $\sim 40$~mrad.

To measure $\delta\varphi_0$ we select a TEM$_{00}$ cavity mode and adjust the cavity length to be on resonance with the clock transition, i.e. $\delta_{C0} =0$. At the clock transition wavelength $\lambda_0$, the cavity's linewidth is $\kappa_0 = 2\pi\times 140.9(3)$~kHz, while the free spectral range (FSR) is $\Delta_{\textrm{FSR},0}=2\pi\times 3.71461(3)$~GHz. The different probe tones used to determine the phase shifts $\delta\varphi_{0,1}$ are created using an in-fiber electro-optical modulator before being coupled to the cavity, and are polarized along the quantization direction $\hat{x}$, established by a static magnetic field $\vec{B} = B_0\hat{x}$ with $B_0 \sim 100$~mG [Fig.~\subref{fig:fig1}{a}].

We investigate cavity transmission characteristics in the presence of atoms for the ultranarrow transition in Fig.~\ref{fig:fig2}. The atomic clock transition is addressed with light from a stabilized state-of-the-art sub-10~mHz linewidth laser \cite{Matei_2017,Robinson_2019,Oelker_2019}. The power transmission of a near-resonant probe, detuned by $\delta_{L0}$ from the atomic transition, exhibits two distinct peaks [Fig.~\subref{fig:fig2}{a}], associated with the $m_F = \pm 9/2$ ground states in the presence of a magnetic field \cite{Boyd_2007}. We attribute the absence of full absorption in this example data to an overly large probe power causing atoms to transition to the excited state $^3$P$_0$. The imbalance in the depth of the absorption features is attributed to imbalance on the relative $m_F = \pm 9/2$ populations.

To gain partial immunity to systematic uncertainty in the atomic transition frequency (i.e. uncertainty in $\delta_{L0}$) as well as laser frequency noise, we probe the cavity with two symmetrically detuned tones at $\delta_{L0}\pm\delta_{p0}$ and measure their phase shifts, i.e. $\delta\varphi_0(\delta_{L0}\pm\delta_{p0})$, by creating a heterodyne beat note. Typically $\delta_{p0}/(2\pi) = 1$~kHz. The \emph{difference} between these two phases encodes the atomic contribution that we would like to measure. To extract this atomic contribution and further reduce sensitivity to various sources of frequency noise, we simultaneously measure the phase shift of an identical pair of tones that probe a consecutive TEM$_{00}$ cavity mode $(\delta\varphi_0(\Delta_{\textrm{FSR},0}+\delta_{L0}\pm\delta_{p0}))$. Finally we compute the \emph{pair-wise difference} $\Delta\varphi_0 = \left(\delta\varphi_0(\delta_{L0}+\delta_{p0})-\delta\varphi_0(\delta_{L0}-\delta_{p0})\right)-\left(\delta\varphi_0(\Delta_{\textrm{FSR},0}+\delta_{L0}+\delta_{p0})-\delta\varphi_0(\Delta_{\textrm{FSR},0}+\delta_{L0}-\delta_{p0})\right)$. For the rest of the paper $\Delta\varphi_0$ refers to this quantity. Note that on resonance ($\delta_{L0}=0$) the phase shift is $\Delta\varphi_0 = -NC_0\gamma_0/\delta_{p0}$.

In Fig.~\subref{fig:fig2}{b} we measure $\Delta\varphi_{0}$ against $\delta_{L0}$ for atoms initially in $\ket{g}$ when the two symmetric tones are applied (blue markers). The sharp resonances near $\delta_{L0} = \pm \delta_{p0}$ occur when one of the tones is near resonant with the atoms, while for this magnetic field ($\sim 100$~mG) and probe power, the Zeeman level resonances are not resolved. Importantly, by using two tones, the measured phase shift is now only quadratically sensitive to the detuning $\delta_{L0}$ when $|\delta_{L0}|\ll\delta_p$. Furthermore, we measure $\Delta\varphi_{0}$ after having adiabatically transferred the atoms to $\ket{e_0}$ \cite{Norcia_SR_2016,Norcia_SRFreq_2018}, and remove the remaining atoms in $\ket{g}$ using the strong $^1$S$_0$-$^1$P$_1$ transition at 461~nm (red markers). We clearly observe $\Delta\varphi_{0}$ switching sign along with the atomic inversion ($N\to -N$), as well as a reduction of the signal, in agreement with the measured adiabatic transfer efficiency.

\begin{figure}[!t]
\includegraphics[width=3.375in]{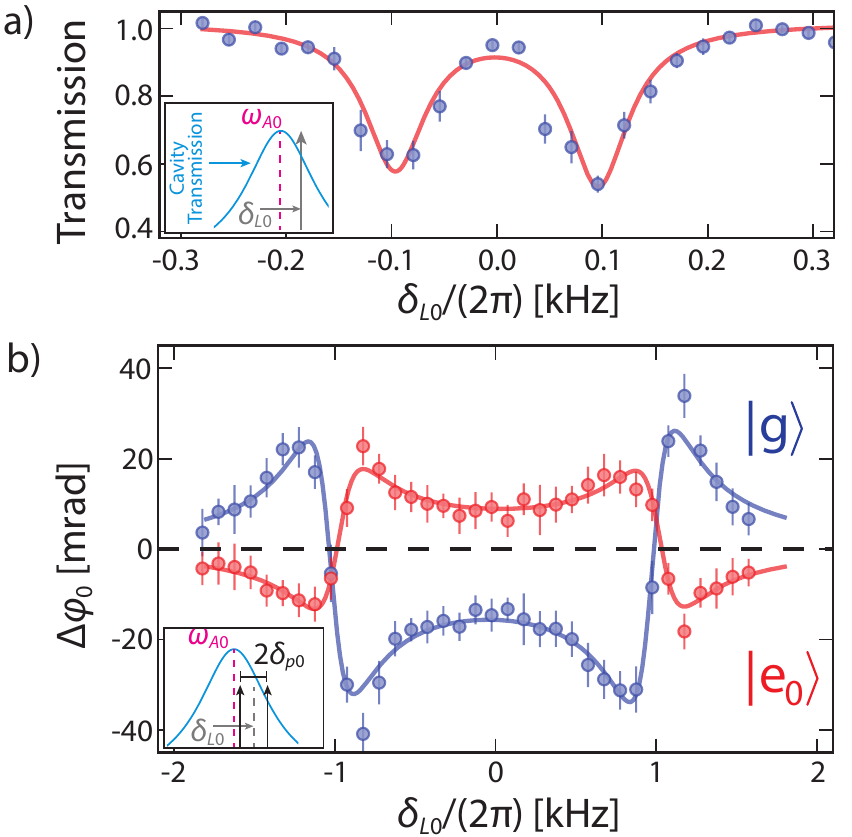}
\caption{{\bf{Probing the  clock transition.}} (a) Transmitted power for a single tone versus its detuning $\delta_{L0}$ from $\omega_{A0}$ (see inset). The two dips correspond to atoms in the ground $m_F=\pm 9/2$ states in the presence of a 200~mG magnetic field that creates a 200~Hz nuclear Zeeman splitting in the optical transition frequency. (b) Atomic induced phase shift $\Delta\varphi_{0}$ as $\delta_{L0}$ is scanned (see inset) using two sidebands at $\delta_{p0}/(2\pi) = 1$~kHz. Blue (Red) markers are for atoms initially in $\ket{g}$ ($\ket{e_0}$). Solid lines are empirical fits that take into account finite excitation fraction.}
\label{fig:fig2}
\end{figure}

We now center our attention on the measurement of $\gamma_0$. Using phase shifts induced on probe light by ultranarrow transitions has previously been proposed for laser frequency stabilization \cite{Martin_2011,Christensen_2015} but in the saturated and resonant configuration, which
is intrinsically destructive. Here, we aim to determine the ratio of light-matter coupling rates $(g_0/g_1)^2$ from dispersive phase shift measurements according to 
\begin{equation}
    \label{eq:DeltaR}
    \left(\frac{g_0}{g_1}\right)^2 = -\left(\frac{\Delta\varphi_{0}}{\delta\varphi_1}\right)\left(\frac{\kappa_0}{\kappa_1}\right)\left(\frac{\delta_{p0}}{\delta_{C1}}\right),
\end{equation}
allowing us to calculate $\gamma_0$ from the known $\gamma_1$ via Eq.~\ref{eq:gammaR}. Our scheme relies on non-destructive interleaved measurements of $\Delta\varphi_0$ and $\delta\varphi_1$ during a single shot such that the inhomogeneity of the atom-cavity coupling
\cite{Schleier-Smith_2010,Chen_2010} and fluctuations in atom number $N$ are common to both measurements and cancel in the final computed ratio. As the cavity length is already stabilized to be on resonance with the clock transition ($\delta_{C0} = 0$), the closest cavity mode will be detuned by $\delta_{C1}$ from the 689~nm transition $\ket{g} \to \ket{e_1}$ [Fig.~\subref{fig:fig3}{a}(ii)]. The cavity linewidth at 689~nm is $\kappa_1/(2\pi) = 153.0(4)$~kHz.

We set the cavity length such that the closest mode to the excited $^3$P$_1$ $F^\prime = 9/2$ state is detuned by $\delta_{C1}/(2\pi) = 277.5(8)$~MHz. The cavity phase shift $\delta\varphi_1$ is computed by measuring the cavity frequency shift $\delta\omega_1$ of the TEM$_{00}$ mode detuned by $\delta_{C1}$ from $\omega_{A1}$. To probe $\delta\omega_1$, we scan the frequency of a weak $\pi$-polarized probe across the cavity resonance. As before, in order to gain further insensitivity with respect to cavity and laser frequency noise, we simultaneously probe a consecutive longitudinal TEM$_{00}$ mode of the cavity at frequency $\delta_{
C1}-\Delta_{\textrm{FSR},1}$, and compute the \textit{difference} $\Delta\varphi_1=\delta\varphi_1(\delta_{C1})-\delta\varphi_1(\delta_{C1}-\Delta_{\textrm{FSR},1})$ (see SI \cite{SI} for details). From now on we will refer to $\Delta\varphi_1$ as this measured quantity.

\begin{figure}[!htb]
\includegraphics[width=3.375in]{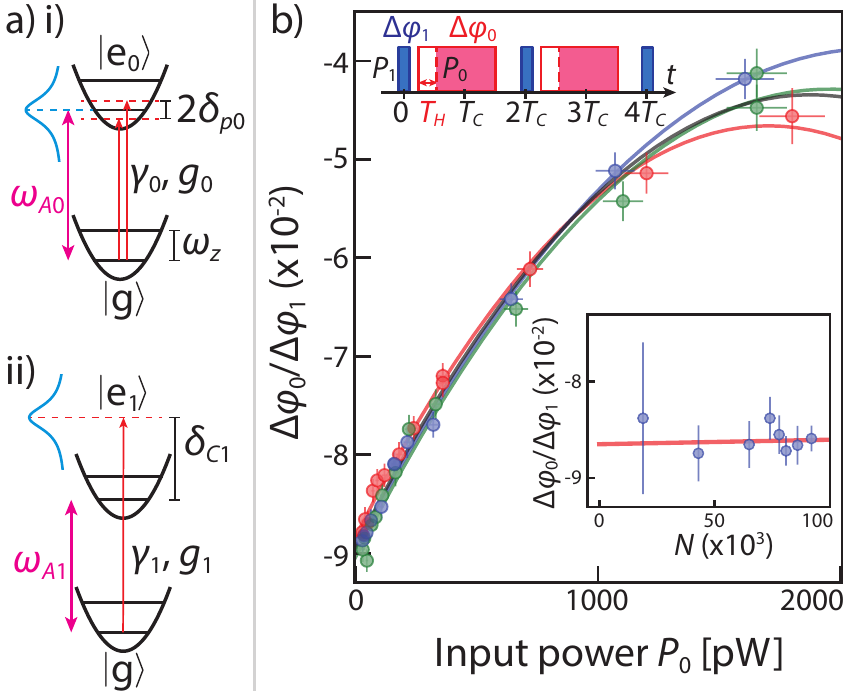}
\caption{{\bf{Measuring $(g_0/g_1)$}}. (a) i) For the phase shift measurement $\Delta\varphi_0$, the two probe tones are detuned by only 1~kHz from the atomic transition, to be compared to the spacing between the axial motional levels $\omega_z/(2\pi) = 230~$kHz. In this resolved-sideband regime, the probe tones experience a differential phase shift primarily from the carrier transition that does not change the motional quantum number. Due to finite axial confinement, the carrier transition strength is reduced by 6\%, the largest correction factor that must be applied to our measurement. ii) In contrast, for the phase shift measurement $\Delta\varphi_1$ the probe tones are far detuned and the probe experiences a phase shift due to interacting with all motional sideband transitions. (b) Ratio $\Delta\varphi_{0}/\Delta\varphi_1$ measurement, from the interleaved pulses sequence (top inset) as $\textit{P}_0$ is changed. Different colors correspond to different measurement sets (markers), over different days, and their color matching solid lines are quadratic polynomial fits on $\textit{P}_0$. Statistical errors ($1\sigma$) are indicated by the errorbars. Top inset shows the measurement sequence, which alternates three 2~ms $\Delta\varphi_1$ measurements between two 25~ms $\Delta\varphi_0$ measurements (first $T_H =5$~ms are removed for the extraction of $\Delta\phi_0$). The solid black line is a global fit to the three measurements. The bottom inset shows $\Delta\varphi_{0}/\Delta\varphi_1$ (markers) and its weighted linear fit (solid line), for a fixed set $(\textit{P}_0,\textit{P}_1)$, as atom number $N$ is changed.}
\label{fig:fig3}
\end{figure}

The ratio of $(g_0/g_1)^2$ is directly encoded in the $\Delta\varphi_{0}/\Delta\varphi_1$ measurement, absent atomic excitations, as described in Eq.~\ref{eq:DeltaR}, taking into account the details of the measurement and atomic structure. To approach the zero-power limit where no atomic excitations are created we measure $\Delta\varphi_{0}/\Delta\varphi_1$ as the probe optical powers are reduced. In order to gain insensitivity to atom loss from the lattice (lifetime about 500~ms), we alternate three short ($\sim 2~$ms) $\Delta\varphi_1$ measurements with two longer ($\sim 25~$ms) $\Delta\varphi_0$ measurements, as indicated in Fig.~\subref{fig:fig3}{b} top inset. From these five measurements we build a suitable estimator for the ratio $\Delta\varphi_{0}/\Delta\varphi_1$ and extract the ratio $(g_0/g_1)^2$.

Experimental results are shown in Fig.~\subref{fig:fig3}{b}. Probe optical powers $\textit{P}_0$ and $\textit{P}_1$ for the 698~nm and 689~nm probes, respectively, are reduced to interpolate to the zero-power value for $\Delta\varphi_0/\Delta\varphi_1$. Three different measurement sets (markers) are shown for consistency and repeatability in Fig.~\subref{fig:fig3}{b}, each fitted with a quadratic polynomial on $\textit{P}_0$ (solid lines) with reduced $\chi_\nu^2$ near 1 for all sets. For these data sets, we have verified that the $\textit{P}_1$ was already sufficiently low to avoid creating excitations in $\ket{e_1}$ (see \cite{SI}). Each of these sets were taken on different days and with independent cavity alignments to the clock transition. A simultaneous fit to the three sets is shown as a solid black line in Fig.~\subref{fig:fig3}{c}. Using different estimators and fit methods, we consistently measure a zero-power crossing ratio $\left(\Delta\varphi_{0}/\Delta\varphi_1\right)_{exp} = -8.95(9)\times10^{-2}$.

\begin{table}[tb!]
\caption{\label{tab:corrections_main} Largest identified corrections and uncertainties.}
\begin{ruledtabular}
\begin{tabular}{lcc}
\textrm{Effect}&
\textrm{Affects}&
\textrm{Correction}\\
\colrule
Finite axial confinement & $\Delta\varphi_{0}/\Delta\varphi_{1}$ &  1.062(4) \\
Cavity birefringence & $\Delta\varphi_{0}/\Delta\varphi_{1}$ & 1.012(5)\\
Atomic resonance uncertainty  & $\Delta\varphi_{0}$ & 0.994(6) \\
Cavity resonance offset & $\Delta\varphi_{0}$ & 1.008(6) \\
\end{tabular}
\end{ruledtabular}
\end{table}

We note that the spread of the zero-power values for different sets is consistent with the effect of the estimated uncertainty on our ability to tune $\delta_{C0}$ to zero for each data set. The bottom inset in Fig~\subref{fig:fig3}{b} shows $\Delta\varphi_0/\Delta\varphi_1$ for different atom number $N$ and a fixed powers $P_0$ and $P_1$, with the red line indicating a linear weighted fit. The variation of the measured values suggests that we can constrain any unknown first order variation with $N$ or $N^{-1}$ to the 2\% level, within our final error, limited by signal to noise. Since there is not an underlying model for why these scaling would exist (beyond offsets in $\Delta\varphi_{0}$ and $\Delta\varphi_1$ accounted for separately), no adjustment to the quoted uncertainty is applied.

To precisely determine the excited clock state linewidth from the measured $\left(\Delta\varphi_{0}/\Delta\varphi_1\right)_{exp}$, several systematic effects need to be accounted for. A detailed description is given in \cite{SI}, but here we focus on a few corrections [Table~\ref{tab:corrections_main}]. The largest systematic correction that must be applied arises from the fact that the phase shift measurements $\Delta\varphi_0$ are made in a resolved sideband regime in which the probe detunings $\delta_{p0}/(2\pi) =\pm1$~kHz are much less than the axial trapping frequency $\omega_z/(2\pi) = 230(1)~$kHz, as shown in Fig.~\subref{fig:fig3}{a}. To be in a dispersive regime requires $\delta_{p0}\gg NC_0\gamma_0$; for most fully allowed optical transitions, this typically implies $\delta_{p0}\gg\omega_z$ when $NC_0\gg1$. However, here $\gamma_0$ being so small allows us to operate in the dispersive regime, probing the carrier transition, even when $\delta_{p0}\ll\omega_z$. For our atomic sample, the correction to the measured $\Delta\varphi_0/\Delta\varphi_1$ is $1.062(4)$, where we also take into account the inhomogeneous coupling between probes and atoms across the optical lattice.

The cavity also possesses intrinsic birefringence which modifies both phase shifts and thus changes $\Delta\varphi_0/\Delta\varphi_1$. Rather than a single polarization-independent cavity resonance, birefringence creates two normal modes split by frequencies $\delta_{b0}$ and $\delta_{b1}$ at $\lambda_0$ and $\lambda_1$ respectively. If $\theta_b$ is the opening angle between the probe beam polarization ($\hat{x}$) and the birefringent eigenmode axis on the Poincar\'e sphere, the correction on each phase shift scales as $(\delta_{bi}/(\kappa_i/2) \sin\theta_b)^2$ for $i\in\{0,1\}$. Including relevant measurement details, we determine a correction factor on the phase shift ratio of $1.012(5)$.

Furthermore, the atomic phase shift measurement $\Delta\varphi_{0}$ is quadratically sensitive to uncertainty in the detuning $\delta_{L0}/(2\pi) = 0\pm100$~Hz [Fig.~\subref{fig:fig2}{b}]. Corrections on the measured value of $\Delta\varphi_{0}$ from this effect scale as $(1-\left(\delta_{L0}/\delta_{p0}\right)^2)$. Similarly, $\Delta\varphi_0$ depends quadratically on the cavity resonance condition, i.e., how close $\delta_{C0}$ is to 0. This correction scales as $(1+\left(\delta_{C0}/(\kappa_0/2)\right)^2)$, where typically $\left|\delta_{C0}\right|/(2\pi)\lesssim 10$~kHz.

Considering all the other systematic effects studied in \cite{SI}, the correction factor to $\left(\Delta\varphi_{0}/\Delta\varphi_1\right)_{exp}$ is $F_C = 1.074(16)$. The corrected ratio is then $\Delta\varphi_{0}/\Delta\varphi_1 = -9.61(17)\times 10^{-2}$. Using Eq.~\ref{eq:DeltaR} and taking into account additional factors to create a more realistic description of the $\Delta\varphi_{0,1}$ measurements, such as the hyperfine structure in $\ket{e_1}$, we extract $(g_0/g_1)^2 = 1.83(3)\times10^{-7}$. Finally, using Eq.~\ref{eq:gammaR} we determine $\gamma_0/\gamma_1 = 1.81(3)\times10^{-7}$, where the waist ($\wit{w}$) and length ($\textrm{L}$) for each mode are independently characterized \cite{SI,Yariv1991}. Using the value of $\gamma_1/(2\pi) = 7.48(1)$~kHz measured in Ref.~\cite{Nicholson_2015}, we finally find $\gamma_0/(2\pi) = 1.35(3)$~mHz for the clock excited state natural linewidth. This value implies an excited state lifetime of $\tau_0 = 118(3)$~s, in agreement with Ref.~\cite{Boyd_2007} but in disagreement with Ref.~\cite{Dorscher_2018}. Ab-initio atomic structure calculations place the atomic linewidth at $1.4(5)$~mHz in Ref.~\cite{Santra_2004}, and $1.2$~mHz in Ref.~\cite{Porsev_2004}. We also note that the value used for $\gamma_1$ reported in Ref.~\cite{Nicholson_2015} is consistent with previous less precise determinations, with relative errors at the 2\% level, from decay and photoassociation measurements \cite{Drozdowski_1997,PhysRevLett.96.203201}. 

In conclusion, we show that cavity enhanced dispersive measurements can be used to realize spectroscopic measurements on ultranarrow optical transitions that, if used in conjunction with measurements on another transition, also exhibit high insensitivity to key systematic effects. We report state dependent phase shifts in agreement with our model. With further improvements on our detection setup, this scheme could be used as an atom counting tool directly on the clock transition, in contrast to other systems \cite{Lodewyck_2009,Beguin_2014,Norcia_2016,Vallet_2017}.

Finally, we report a resolution of 30~$\mu$Hz, which implies we could measure excited state lifetimes of up to 90 minutes in comparable integration time $\sim 100$~ms. With reasonable improvements in our setup we could expect to measure up to 15 hour lifetimes, i.e. by reducing $\delta_{p0}$ to increase the signal, if no systematic effects are taken into account. For instance, it could be used to directly measure the magnetic field dependent linewidth of the Sr bosonic isotopes \cite{Taichenachev_2006,Origlia_2018}, determine the Sr $^3$P$_2$ excited state lifetime or even longer lived states such as the predicted $\sim 10~\mu$Hz nuclear transition on $^{229}$Th being pursued as a next generation clock \cite{Peik_2003,vonderWense_2016,Seiferle_2019}.

\begin{acknowledgments}
We acknowledge helpful discussions with Graham Greve, Matthew Norcia, Christian Sanner and Jun Ye's group for providing narrow linewidth light from their clock laser. This work is supported by the NSF JILA-PFC PHY-1734006 grants, DARPA Extreme Sensing, and NIST. J.R.K.C. acknowledges financial support from NSF GRFP.
\end{acknowledgments}

\bibliography{bibliography.bib}

\begin{thebibliography}{77}%
\makeatletter
\providecommand \@ifxundefined [1]{%
 \@ifx{#1\undefined}
}%
\providecommand \@ifnum [1]{%
 \ifnum #1\expandafter \@firstoftwo
 \else \expandafter \@secondoftwo
 \fi
}%
\providecommand \@ifx [1]{%
 \ifx #1\expandafter \@firstoftwo
 \else \expandafter \@secondoftwo
 \fi
}%
\providecommand \natexlab [1]{#1}%
\providecommand \enquote  [1]{``#1''}%
\providecommand \bibnamefont  [1]{#1}%
\providecommand \bibfnamefont [1]{#1}%
\providecommand \citenamefont [1]{#1}%
\providecommand \href@noop [0]{\@secondoftwo}%
\providecommand \href [0]{\begingroup \@sanitize@url \@href}%
\providecommand \@href[1]{\@@startlink{#1}\@@href}%
\providecommand \@@href[1]{\endgroup#1\@@endlink}%
\providecommand \@sanitize@url [0]{\catcode `\\12\catcode `\$12\catcode
  `\&12\catcode `\#12\catcode `\^12\catcode `\_12\catcode `\%12\relax}%
\providecommand \@@startlink[1]{}%
\providecommand \@@endlink[0]{}%
\providecommand \url  [0]{\begingroup\@sanitize@url \@url }%
\providecommand \@url [1]{\endgroup\@href {#1}{\urlprefix }}%
\providecommand \urlprefix  [0]{URL }%
\providecommand \Eprint [0]{\href }%
\providecommand \doibase [0]{http://dx.doi.org/}%
\providecommand \selectlanguage [0]{\@gobble}%
\providecommand \bibinfo  [0]{\@secondoftwo}%
\providecommand \bibfield  [0]{\@secondoftwo}%
\providecommand \translation [1]{[#1]}%
\providecommand \BibitemOpen [0]{}%
\providecommand \bibitemStop [0]{}%
\providecommand \bibitemNoStop [0]{.\EOS\space}%
\providecommand \EOS [0]{\spacefactor3000\relax}%
\providecommand \BibitemShut  [1]{\csname bibitem#1\endcsname}%
\let\auto@bib@innerbib\@empty
\bibitem [{\citenamefont {Ludlow}\ \emph {et~al.}(2015)\citenamefont {Ludlow},
  \citenamefont {Boyd}, \citenamefont {Ye}, \citenamefont {Peik},\ and\
  \citenamefont {Schmidt}}]{Ludlow_2015}%
  \BibitemOpen
  \bibfield  {author} {\bibinfo {author} {\bibfnamefont {A.~D.}\ \bibnamefont
  {Ludlow}}, \bibinfo {author} {\bibfnamefont {M.~M.}\ \bibnamefont {Boyd}},
  \bibinfo {author} {\bibfnamefont {J.}~\bibnamefont {Ye}}, \bibinfo {author}
  {\bibfnamefont {E.}~\bibnamefont {Peik}}, \ and\ \bibinfo {author}
  {\bibfnamefont {P.~O.}\ \bibnamefont {Schmidt}},\ }\href {\doibase
  10.1103/RevModPhys.87.637} {\bibfield  {journal} {\bibinfo  {journal} {Rev.
  Mod. Phys.}\ }\textbf {\bibinfo {volume} {87}},\ \bibinfo {pages} {637}
  (\bibinfo {year} {2015})}\BibitemShut {NoStop}%
\bibitem [{\citenamefont {Oelker}\ \emph {et~al.}(2019)\citenamefont {Oelker},
  \citenamefont {Hutson}, \citenamefont {Kennedy}, \citenamefont {Sonderhouse},
  \citenamefont {Bothwell}, \citenamefont {Goban}, \citenamefont {Kedar},
  \citenamefont {Sanner}, \citenamefont {Robinson}, \citenamefont {Marti},
  \citenamefont {Matei}, \citenamefont {Legero}, \citenamefont {Giunta},
  \citenamefont {Holzwarth}, \citenamefont {Riehle}, \citenamefont {Sterr},\
  and\ \citenamefont {Ye}}]{Oelker_2019}%
  \BibitemOpen
  \bibfield  {author} {\bibinfo {author} {\bibfnamefont {E.}~\bibnamefont
  {Oelker}}, \bibinfo {author} {\bibfnamefont {R.~B.}\ \bibnamefont {Hutson}},
  \bibinfo {author} {\bibfnamefont {C.~J.}\ \bibnamefont {Kennedy}}, \bibinfo
  {author} {\bibfnamefont {L.}~\bibnamefont {Sonderhouse}}, \bibinfo {author}
  {\bibfnamefont {T.}~\bibnamefont {Bothwell}}, \bibinfo {author}
  {\bibfnamefont {A.}~\bibnamefont {Goban}}, \bibinfo {author} {\bibfnamefont
  {D.}~\bibnamefont {Kedar}}, \bibinfo {author} {\bibfnamefont
  {C.}~\bibnamefont {Sanner}}, \bibinfo {author} {\bibfnamefont {J.~M.}\
  \bibnamefont {Robinson}}, \bibinfo {author} {\bibfnamefont {G.~E.}\
  \bibnamefont {Marti}}, \bibinfo {author} {\bibfnamefont {D.~G.}\ \bibnamefont
  {Matei}}, \bibinfo {author} {\bibfnamefont {T.}~\bibnamefont {Legero}},
  \bibinfo {author} {\bibfnamefont {M.}~\bibnamefont {Giunta}}, \bibinfo
  {author} {\bibfnamefont {R.}~\bibnamefont {Holzwarth}}, \bibinfo {author}
  {\bibfnamefont {F.}~\bibnamefont {Riehle}}, \bibinfo {author} {\bibfnamefont
  {U.}~\bibnamefont {Sterr}}, \ and\ \bibinfo {author} {\bibfnamefont
  {J.}~\bibnamefont {Ye}},\ }\href {\doibase 10.1038/s41566-019-0493-4}
  {\bibfield  {journal} {\bibinfo  {journal} {Nature Photonics}\ }\textbf
  {\bibinfo {volume} {13}},\ \bibinfo {pages} {714} (\bibinfo {year}
  {2019})}\BibitemShut {NoStop}%
\bibitem [{\citenamefont {Campbell}\ \emph {et~al.}(2017)\citenamefont
  {Campbell}, \citenamefont {Hutson}, \citenamefont {Marti}, \citenamefont
  {Goban}, \citenamefont {Darkwah~Oppong}, \citenamefont {McNally},
  \citenamefont {Sonderhouse}, \citenamefont {Robinson}, \citenamefont {Zhang},
  \citenamefont {Bloom},\ and\ \citenamefont {Ye}}]{Campbell_2017}%
  \BibitemOpen
  \bibfield  {author} {\bibinfo {author} {\bibfnamefont {S.~L.}\ \bibnamefont
  {Campbell}}, \bibinfo {author} {\bibfnamefont {R.~B.}\ \bibnamefont
  {Hutson}}, \bibinfo {author} {\bibfnamefont {G.~E.}\ \bibnamefont {Marti}},
  \bibinfo {author} {\bibfnamefont {A.}~\bibnamefont {Goban}}, \bibinfo
  {author} {\bibfnamefont {N.}~\bibnamefont {Darkwah~Oppong}}, \bibinfo
  {author} {\bibfnamefont {R.~L.}\ \bibnamefont {McNally}}, \bibinfo {author}
  {\bibfnamefont {L.}~\bibnamefont {Sonderhouse}}, \bibinfo {author}
  {\bibfnamefont {J.~M.}\ \bibnamefont {Robinson}}, \bibinfo {author}
  {\bibfnamefont {W.}~\bibnamefont {Zhang}}, \bibinfo {author} {\bibfnamefont
  {B.~J.}\ \bibnamefont {Bloom}}, \ and\ \bibinfo {author} {\bibfnamefont
  {J.}~\bibnamefont {Ye}},\ }\href {\doibase 10.1126/science.aam5538}
  {\bibfield  {journal} {\bibinfo  {journal} {Science}\ }\textbf {\bibinfo
  {volume} {358}},\ \bibinfo {pages} {90} (\bibinfo {year} {2017})}\BibitemShut
  {NoStop}%
\bibitem [{\citenamefont {Ushijima}\ \emph {et~al.}(2015)\citenamefont
  {Ushijima}, \citenamefont {Takamoto}, \citenamefont {Das}, \citenamefont
  {Ohkubo},\ and\ \citenamefont {Katori}}]{Ushijima_2015}%
  \BibitemOpen
  \bibfield  {author} {\bibinfo {author} {\bibfnamefont {I.}~\bibnamefont
  {Ushijima}}, \bibinfo {author} {\bibfnamefont {M.}~\bibnamefont {Takamoto}},
  \bibinfo {author} {\bibfnamefont {M.}~\bibnamefont {Das}}, \bibinfo {author}
  {\bibfnamefont {T.}~\bibnamefont {Ohkubo}}, \ and\ \bibinfo {author}
  {\bibfnamefont {H.}~\bibnamefont {Katori}},\ }\href
  {https://doi.org/10.1038/nphoton.2015.5} {\bibfield  {journal} {\bibinfo
  {journal} {Nature Photonics}\ }\textbf {\bibinfo {volume} {9}},\ \bibinfo
  {pages} {185 EP } (\bibinfo {year} {2015})}\BibitemShut {NoStop}%
\bibitem [{\citenamefont {Takano}\ \emph {et~al.}(2016)\citenamefont {Takano},
  \citenamefont {Takamoto}, \citenamefont {Ushijima}, \citenamefont {Ohmae},
  \citenamefont {Akatsuka}, \citenamefont {Yamaguchi}, \citenamefont
  {Kuroishi}, \citenamefont {Munekane}, \citenamefont {Miyahara},\ and\
  \citenamefont {Katori}}]{Takano_2016}%
  \BibitemOpen
  \bibfield  {author} {\bibinfo {author} {\bibfnamefont {T.}~\bibnamefont
  {Takano}}, \bibinfo {author} {\bibfnamefont {M.}~\bibnamefont {Takamoto}},
  \bibinfo {author} {\bibfnamefont {I.}~\bibnamefont {Ushijima}}, \bibinfo
  {author} {\bibfnamefont {N.}~\bibnamefont {Ohmae}}, \bibinfo {author}
  {\bibfnamefont {T.}~\bibnamefont {Akatsuka}}, \bibinfo {author}
  {\bibfnamefont {A.}~\bibnamefont {Yamaguchi}}, \bibinfo {author}
  {\bibfnamefont {Y.}~\bibnamefont {Kuroishi}}, \bibinfo {author}
  {\bibfnamefont {H.}~\bibnamefont {Munekane}}, \bibinfo {author}
  {\bibfnamefont {B.}~\bibnamefont {Miyahara}}, \ and\ \bibinfo {author}
  {\bibfnamefont {H.}~\bibnamefont {Katori}},\ }\href
  {https://doi.org/10.1038/nphoton.2016.159} {\bibfield  {journal} {\bibinfo
  {journal} {Nature Photonics}\ }\textbf {\bibinfo {volume} {10}},\ \bibinfo
  {pages} {662 EP } (\bibinfo {year} {2016})}\BibitemShut {NoStop}%
\bibitem [{\citenamefont {Grotti}\ \emph {et~al.}(2018)\citenamefont {Grotti},
  \citenamefont {Koller}, \citenamefont {Vogt}, \citenamefont {H{\"a}fner},
  \citenamefont {Sterr}, \citenamefont {Lisdat}, \citenamefont {Denker},
  \citenamefont {Voigt}, \citenamefont {Timmen}, \citenamefont {Rolland},
  \citenamefont {Baynes}, \citenamefont {Margolis}, \citenamefont {Zampaolo},
  \citenamefont {Thoumany}, \citenamefont {Pizzocaro}, \citenamefont {Rauf},
  \citenamefont {Bregolin}, \citenamefont {Tampellini}, \citenamefont
  {Barbieri}, \citenamefont {Zucco}, \citenamefont {Costanzo}, \citenamefont
  {Clivati}, \citenamefont {Levi},\ and\ \citenamefont
  {Calonico}}]{Grotti_2018}%
  \BibitemOpen
  \bibfield  {author} {\bibinfo {author} {\bibfnamefont {J.}~\bibnamefont
  {Grotti}}, \bibinfo {author} {\bibfnamefont {S.}~\bibnamefont {Koller}},
  \bibinfo {author} {\bibfnamefont {S.}~\bibnamefont {Vogt}}, \bibinfo {author}
  {\bibfnamefont {S.}~\bibnamefont {H{\"a}fner}}, \bibinfo {author}
  {\bibfnamefont {U.}~\bibnamefont {Sterr}}, \bibinfo {author} {\bibfnamefont
  {C.}~\bibnamefont {Lisdat}}, \bibinfo {author} {\bibfnamefont
  {H.}~\bibnamefont {Denker}}, \bibinfo {author} {\bibfnamefont
  {C.}~\bibnamefont {Voigt}}, \bibinfo {author} {\bibfnamefont
  {L.}~\bibnamefont {Timmen}}, \bibinfo {author} {\bibfnamefont
  {A.}~\bibnamefont {Rolland}}, \bibinfo {author} {\bibfnamefont {F.~N.}\
  \bibnamefont {Baynes}}, \bibinfo {author} {\bibfnamefont {H.~S.}\
  \bibnamefont {Margolis}}, \bibinfo {author} {\bibfnamefont {M.}~\bibnamefont
  {Zampaolo}}, \bibinfo {author} {\bibfnamefont {P.}~\bibnamefont {Thoumany}},
  \bibinfo {author} {\bibfnamefont {M.}~\bibnamefont {Pizzocaro}}, \bibinfo
  {author} {\bibfnamefont {B.}~\bibnamefont {Rauf}}, \bibinfo {author}
  {\bibfnamefont {F.}~\bibnamefont {Bregolin}}, \bibinfo {author}
  {\bibfnamefont {A.}~\bibnamefont {Tampellini}}, \bibinfo {author}
  {\bibfnamefont {P.}~\bibnamefont {Barbieri}}, \bibinfo {author}
  {\bibfnamefont {M.}~\bibnamefont {Zucco}}, \bibinfo {author} {\bibfnamefont
  {G.~A.}\ \bibnamefont {Costanzo}}, \bibinfo {author} {\bibfnamefont
  {C.}~\bibnamefont {Clivati}}, \bibinfo {author} {\bibfnamefont
  {F.}~\bibnamefont {Levi}}, \ and\ \bibinfo {author} {\bibfnamefont
  {D.}~\bibnamefont {Calonico}},\ }\href {\doibase 10.1038/s41567-017-0042-3}
  {\bibfield  {journal} {\bibinfo  {journal} {Nature Physics}\ }\textbf
  {\bibinfo {volume} {14}},\ \bibinfo {pages} {437} (\bibinfo {year}
  {2018})}\BibitemShut {NoStop}%
\bibitem [{\citenamefont {Schioppo}\ \emph {et~al.}(2016)\citenamefont
  {Schioppo}, \citenamefont {Brown}, \citenamefont {McGrew}, \citenamefont
  {Hinkley}, \citenamefont {Fasano}, \citenamefont {Beloy}, \citenamefont
  {Yoon}, \citenamefont {Milani}, \citenamefont {Nicolodi}, \citenamefont
  {Sherman}, \citenamefont {Phillips}, \citenamefont {Oates},\ and\
  \citenamefont {Ludlow}}]{Schioppo_2016}%
  \BibitemOpen
  \bibfield  {author} {\bibinfo {author} {\bibfnamefont {M.}~\bibnamefont
  {Schioppo}}, \bibinfo {author} {\bibfnamefont {R.~C.}\ \bibnamefont {Brown}},
  \bibinfo {author} {\bibfnamefont {W.~F.}\ \bibnamefont {McGrew}}, \bibinfo
  {author} {\bibfnamefont {N.}~\bibnamefont {Hinkley}}, \bibinfo {author}
  {\bibfnamefont {R.~J.}\ \bibnamefont {Fasano}}, \bibinfo {author}
  {\bibfnamefont {K.}~\bibnamefont {Beloy}}, \bibinfo {author} {\bibfnamefont
  {T.~H.}\ \bibnamefont {Yoon}}, \bibinfo {author} {\bibfnamefont
  {G.}~\bibnamefont {Milani}}, \bibinfo {author} {\bibfnamefont
  {D.}~\bibnamefont {Nicolodi}}, \bibinfo {author} {\bibfnamefont {J.~A.}\
  \bibnamefont {Sherman}}, \bibinfo {author} {\bibfnamefont {N.~B.}\
  \bibnamefont {Phillips}}, \bibinfo {author} {\bibfnamefont {C.~W.}\
  \bibnamefont {Oates}}, \ and\ \bibinfo {author} {\bibfnamefont {A.~D.}\
  \bibnamefont {Ludlow}},\ }\href {https://doi.org/10.1038/nphoton.2016.231}
  {\bibfield  {journal} {\bibinfo  {journal} {Nature Photonics}\ }\textbf
  {\bibinfo {volume} {11}},\ \bibinfo {pages} {48 EP} (\bibinfo {year}
  {2016})}\BibitemShut {NoStop}%
\bibitem [{\citenamefont {Brewer}\ \emph {et~al.}(2019)\citenamefont {Brewer},
  \citenamefont {Chen}, \citenamefont {Hankin}, \citenamefont {Clements},
  \citenamefont {Chou}, \citenamefont {Wineland}, \citenamefont {Hume},\ and\
  \citenamefont {Leibrandt}}]{Brewer_2019}%
  \BibitemOpen
  \bibfield  {author} {\bibinfo {author} {\bibfnamefont {S.~M.}\ \bibnamefont
  {Brewer}}, \bibinfo {author} {\bibfnamefont {J.-S.}\ \bibnamefont {Chen}},
  \bibinfo {author} {\bibfnamefont {A.~M.}\ \bibnamefont {Hankin}}, \bibinfo
  {author} {\bibfnamefont {E.~R.}\ \bibnamefont {Clements}}, \bibinfo {author}
  {\bibfnamefont {C.~W.}\ \bibnamefont {Chou}}, \bibinfo {author}
  {\bibfnamefont {D.~J.}\ \bibnamefont {Wineland}}, \bibinfo {author}
  {\bibfnamefont {D.~B.}\ \bibnamefont {Hume}}, \ and\ \bibinfo {author}
  {\bibfnamefont {D.~R.}\ \bibnamefont {Leibrandt}},\ }\href {\doibase
  10.1103/PhysRevLett.123.033201} {\bibfield  {journal} {\bibinfo  {journal}
  {Phys. Rev. Lett.}\ }\textbf {\bibinfo {volume} {123}},\ \bibinfo {pages}
  {033201} (\bibinfo {year} {2019})}\BibitemShut {NoStop}%
\bibitem [{\citenamefont {Safronova}\ \emph {et~al.}(2018)\citenamefont
  {Safronova}, \citenamefont {Budker}, \citenamefont {DeMille}, \citenamefont
  {Kimball}, \citenamefont {Derevianko},\ and\ \citenamefont
  {Clark}}]{Safronova_2016}%
  \BibitemOpen
  \bibfield  {author} {\bibinfo {author} {\bibfnamefont {M.~S.}\ \bibnamefont
  {Safronova}}, \bibinfo {author} {\bibfnamefont {D.}~\bibnamefont {Budker}},
  \bibinfo {author} {\bibfnamefont {D.}~\bibnamefont {DeMille}}, \bibinfo
  {author} {\bibfnamefont {D.~F.~J.}\ \bibnamefont {Kimball}}, \bibinfo
  {author} {\bibfnamefont {A.}~\bibnamefont {Derevianko}}, \ and\ \bibinfo
  {author} {\bibfnamefont {C.~W.}\ \bibnamefont {Clark}},\ }\href {\doibase
  10.1103/RevModPhys.90.025008} {\bibfield  {journal} {\bibinfo  {journal}
  {Rev. Mod. Phys.}\ }\textbf {\bibinfo {volume} {90}},\ \bibinfo {pages}
  {025008} (\bibinfo {year} {2018})}\BibitemShut {NoStop}%
\bibitem [{\citenamefont {Tino}\ \emph {et~al.}(2019)\citenamefont {Tino},
  \citenamefont {Bassi}, \citenamefont {Bianco}, \citenamefont {Bongs},
  \citenamefont {Bouyer}, \citenamefont {Cacciapuoti}, \citenamefont
  {Capozziello}, \citenamefont {Chen}, \citenamefont {Chiofalo}, \citenamefont
  {Derevianko}, \citenamefont {Ertmer}, \citenamefont {Gaaloul}, \citenamefont
  {Gill}, \citenamefont {Graham}, \citenamefont {Hogan}, \citenamefont {Iess},
  \citenamefont {Kasevich}, \citenamefont {Katori}, \citenamefont {Klempt},
  \citenamefont {Lu}, \citenamefont {Ma}, \citenamefont {M{\"u}ller},
  \citenamefont {Newbury}, \citenamefont {Oates}, \citenamefont {Peters},
  \citenamefont {Poli}, \citenamefont {Rasel}, \citenamefont {Rosi},
  \citenamefont {Roura}, \citenamefont {Salomon}, \citenamefont {Schiller},
  \citenamefont {Schleich}, \citenamefont {Schlippert}, \citenamefont
  {Schreck}, \citenamefont {Schubert}, \citenamefont {Sorrentino},
  \citenamefont {Sterr}, \citenamefont {Thomsen}, \citenamefont {Vallone},
  \citenamefont {Vetrano}, \citenamefont {Villoresi}, \citenamefont {von
  Klitzing}, \citenamefont {Wilkowski}, \citenamefont {Wolf}, \citenamefont
  {Ye}, \citenamefont {Yu},\ and\ \citenamefont {Zhan}}]{Tino2019}%
  \BibitemOpen
  \bibfield  {author} {\bibinfo {author} {\bibfnamefont {G.~M.}\ \bibnamefont
  {Tino}}, \bibinfo {author} {\bibfnamefont {A.}~\bibnamefont {Bassi}},
  \bibinfo {author} {\bibfnamefont {G.}~\bibnamefont {Bianco}}, \bibinfo
  {author} {\bibfnamefont {K.}~\bibnamefont {Bongs}}, \bibinfo {author}
  {\bibfnamefont {P.}~\bibnamefont {Bouyer}}, \bibinfo {author} {\bibfnamefont
  {L.}~\bibnamefont {Cacciapuoti}}, \bibinfo {author} {\bibfnamefont
  {S.}~\bibnamefont {Capozziello}}, \bibinfo {author} {\bibfnamefont
  {X.}~\bibnamefont {Chen}}, \bibinfo {author} {\bibfnamefont {M.~L.}\
  \bibnamefont {Chiofalo}}, \bibinfo {author} {\bibfnamefont {A.}~\bibnamefont
  {Derevianko}}, \bibinfo {author} {\bibfnamefont {W.}~\bibnamefont {Ertmer}},
  \bibinfo {author} {\bibfnamefont {N.}~\bibnamefont {Gaaloul}}, \bibinfo
  {author} {\bibfnamefont {P.}~\bibnamefont {Gill}}, \bibinfo {author}
  {\bibfnamefont {P.~W.}\ \bibnamefont {Graham}}, \bibinfo {author}
  {\bibfnamefont {J.~M.}\ \bibnamefont {Hogan}}, \bibinfo {author}
  {\bibfnamefont {L.}~\bibnamefont {Iess}}, \bibinfo {author} {\bibfnamefont
  {M.~A.}\ \bibnamefont {Kasevich}}, \bibinfo {author} {\bibfnamefont
  {H.}~\bibnamefont {Katori}}, \bibinfo {author} {\bibfnamefont
  {C.}~\bibnamefont {Klempt}}, \bibinfo {author} {\bibfnamefont
  {X.}~\bibnamefont {Lu}}, \bibinfo {author} {\bibfnamefont {L.-S.}\
  \bibnamefont {Ma}}, \bibinfo {author} {\bibfnamefont {H.}~\bibnamefont
  {M{\"u}ller}}, \bibinfo {author} {\bibfnamefont {N.~R.}\ \bibnamefont
  {Newbury}}, \bibinfo {author} {\bibfnamefont {C.~W.}\ \bibnamefont {Oates}},
  \bibinfo {author} {\bibfnamefont {A.}~\bibnamefont {Peters}}, \bibinfo
  {author} {\bibfnamefont {N.}~\bibnamefont {Poli}}, \bibinfo {author}
  {\bibfnamefont {E.~M.}\ \bibnamefont {Rasel}}, \bibinfo {author}
  {\bibfnamefont {G.}~\bibnamefont {Rosi}}, \bibinfo {author} {\bibfnamefont
  {A.}~\bibnamefont {Roura}}, \bibinfo {author} {\bibfnamefont
  {C.}~\bibnamefont {Salomon}}, \bibinfo {author} {\bibfnamefont
  {S.}~\bibnamefont {Schiller}}, \bibinfo {author} {\bibfnamefont
  {W.}~\bibnamefont {Schleich}}, \bibinfo {author} {\bibfnamefont
  {D.}~\bibnamefont {Schlippert}}, \bibinfo {author} {\bibfnamefont
  {F.}~\bibnamefont {Schreck}}, \bibinfo {author} {\bibfnamefont
  {C.}~\bibnamefont {Schubert}}, \bibinfo {author} {\bibfnamefont
  {F.}~\bibnamefont {Sorrentino}}, \bibinfo {author} {\bibfnamefont
  {U.}~\bibnamefont {Sterr}}, \bibinfo {author} {\bibfnamefont {J.~W.}\
  \bibnamefont {Thomsen}}, \bibinfo {author} {\bibfnamefont {G.}~\bibnamefont
  {Vallone}}, \bibinfo {author} {\bibfnamefont {F.}~\bibnamefont {Vetrano}},
  \bibinfo {author} {\bibfnamefont {P.}~\bibnamefont {Villoresi}}, \bibinfo
  {author} {\bibfnamefont {W.}~\bibnamefont {von Klitzing}}, \bibinfo {author}
  {\bibfnamefont {D.}~\bibnamefont {Wilkowski}}, \bibinfo {author}
  {\bibfnamefont {P.}~\bibnamefont {Wolf}}, \bibinfo {author} {\bibfnamefont
  {J.}~\bibnamefont {Ye}}, \bibinfo {author} {\bibfnamefont {N.}~\bibnamefont
  {Yu}}, \ and\ \bibinfo {author} {\bibfnamefont {M.}~\bibnamefont {Zhan}},\
  }\href {\doibase 10.1140/epjd/e2019-100324-6} {\bibfield  {journal} {\bibinfo
   {journal} {The European Physical Journal D}\ }\textbf {\bibinfo {volume}
  {73}},\ \bibinfo {pages} {228} (\bibinfo {year} {2019})}\BibitemShut
  {NoStop}%
\bibitem [{\citenamefont {Hu}\ \emph {et~al.}(2017)\citenamefont {Hu},
  \citenamefont {Poli}, \citenamefont {Salvi},\ and\ \citenamefont
  {Tino}}]{Hu_2017}%
  \BibitemOpen
  \bibfield  {author} {\bibinfo {author} {\bibfnamefont {L.}~\bibnamefont
  {Hu}}, \bibinfo {author} {\bibfnamefont {N.}~\bibnamefont {Poli}}, \bibinfo
  {author} {\bibfnamefont {L.}~\bibnamefont {Salvi}}, \ and\ \bibinfo {author}
  {\bibfnamefont {G.~M.}\ \bibnamefont {Tino}},\ }\href {\doibase
  10.1103/PhysRevLett.119.263601} {\bibfield  {journal} {\bibinfo  {journal}
  {Phys. Rev. Lett.}\ }\textbf {\bibinfo {volume} {119}},\ \bibinfo {pages}
  {263601} (\bibinfo {year} {2017})}\BibitemShut {NoStop}%
\bibitem [{\citenamefont {del Aguila}\ \emph {et~al.}(2018)\citenamefont {del
  Aguila}, \citenamefont {Mazzoni}, \citenamefont {Hu}, \citenamefont {Salvi},
  \citenamefont {Tino},\ and\ \citenamefont {Poli}}]{Aguila_2018}%
  \BibitemOpen
  \bibfield  {author} {\bibinfo {author} {\bibfnamefont {R.~P.}\ \bibnamefont
  {del Aguila}}, \bibinfo {author} {\bibfnamefont {T.}~\bibnamefont {Mazzoni}},
  \bibinfo {author} {\bibfnamefont {L.}~\bibnamefont {Hu}}, \bibinfo {author}
  {\bibfnamefont {L.}~\bibnamefont {Salvi}}, \bibinfo {author} {\bibfnamefont
  {G.~M.}\ \bibnamefont {Tino}}, \ and\ \bibinfo {author} {\bibfnamefont
  {N.}~\bibnamefont {Poli}},\ }\href {\doibase 10.1088/1367-2630/aab088}
  {\bibfield  {journal} {\bibinfo  {journal} {New Journal of Physics}\ }\textbf
  {\bibinfo {volume} {20}},\ \bibinfo {pages} {043002} (\bibinfo {year}
  {2018})}\BibitemShut {NoStop}%
\bibitem [{\citenamefont {Arvanitaki}\ \emph {et~al.}(2018)\citenamefont
  {Arvanitaki}, \citenamefont {Graham}, \citenamefont {Hogan}, \citenamefont
  {Rajendran},\ and\ \citenamefont {Van~Tilburg}}]{Arvanitaki_2018}%
  \BibitemOpen
  \bibfield  {author} {\bibinfo {author} {\bibfnamefont {A.}~\bibnamefont
  {Arvanitaki}}, \bibinfo {author} {\bibfnamefont {P.~W.}\ \bibnamefont
  {Graham}}, \bibinfo {author} {\bibfnamefont {J.~M.}\ \bibnamefont {Hogan}},
  \bibinfo {author} {\bibfnamefont {S.}~\bibnamefont {Rajendran}}, \ and\
  \bibinfo {author} {\bibfnamefont {K.}~\bibnamefont {Van~Tilburg}},\ }\href
  {\doibase 10.1103/PhysRevD.97.075020} {\bibfield  {journal} {\bibinfo
  {journal} {Phys. Rev. D}\ }\textbf {\bibinfo {volume} {97}},\ \bibinfo
  {pages} {075020} (\bibinfo {year} {2018})}\BibitemShut {NoStop}%
\bibitem [{\citenamefont {Wcis{\l}o}\ \emph {et~al.}(2018)\citenamefont
  {Wcis{\l}o}, \citenamefont {Ablewski}, \citenamefont {Beloy}, \citenamefont
  {Bilicki}, \citenamefont {Bober}, \citenamefont {Brown}, \citenamefont
  {Fasano}, \citenamefont {Ciury{\l}o}, \citenamefont {Hachisu}, \citenamefont
  {Ido} \emph {et~al.}}]{Wcislo_2018}%
  \BibitemOpen
  \bibfield  {author} {\bibinfo {author} {\bibfnamefont {P.}~\bibnamefont
  {Wcis{\l}o}}, \bibinfo {author} {\bibfnamefont {P.}~\bibnamefont {Ablewski}},
  \bibinfo {author} {\bibfnamefont {K.}~\bibnamefont {Beloy}}, \bibinfo
  {author} {\bibfnamefont {S.}~\bibnamefont {Bilicki}}, \bibinfo {author}
  {\bibfnamefont {M.}~\bibnamefont {Bober}}, \bibinfo {author} {\bibfnamefont
  {R.}~\bibnamefont {Brown}}, \bibinfo {author} {\bibfnamefont
  {R.}~\bibnamefont {Fasano}}, \bibinfo {author} {\bibfnamefont
  {R.}~\bibnamefont {Ciury{\l}o}}, \bibinfo {author} {\bibfnamefont
  {H.}~\bibnamefont {Hachisu}}, \bibinfo {author} {\bibfnamefont
  {T.}~\bibnamefont {Ido}},  \emph {et~al.},\ }\href@noop {} {\bibfield
  {journal} {\bibinfo  {journal} {Science Advances}\ }\textbf {\bibinfo
  {volume} {4}},\ \bibinfo {pages} {eaau4869} (\bibinfo {year}
  {2018})}\BibitemShut {NoStop}%
\bibitem [{\citenamefont {Norcia}\ \emph
  {et~al.}(2018{\natexlab{a}})\citenamefont {Norcia}, \citenamefont
  {Lewis-Swan}, \citenamefont {Cline}, \citenamefont {Zhu}, \citenamefont
  {Rey},\ and\ \citenamefont {Thompson}}]{Norcia_SS_2018}%
  \BibitemOpen
  \bibfield  {author} {\bibinfo {author} {\bibfnamefont {M.~A.}\ \bibnamefont
  {Norcia}}, \bibinfo {author} {\bibfnamefont {R.~J.}\ \bibnamefont
  {Lewis-Swan}}, \bibinfo {author} {\bibfnamefont {J.~R.~K.}\ \bibnamefont
  {Cline}}, \bibinfo {author} {\bibfnamefont {B.}~\bibnamefont {Zhu}}, \bibinfo
  {author} {\bibfnamefont {A.~M.}\ \bibnamefont {Rey}}, \ and\ \bibinfo
  {author} {\bibfnamefont {J.~K.}\ \bibnamefont {Thompson}},\ }\href {\doibase
  10.1126/science.aar3102} {\bibfield  {journal} {\bibinfo  {journal}
  {Science}\ }\textbf {\bibinfo {volume} {361}},\ \bibinfo {pages} {259}
  (\bibinfo {year} {2018}{\natexlab{a}})}\BibitemShut {NoStop}%
\bibitem [{\citenamefont {Muniz}\ \emph {et~al.}(2020)\citenamefont {Muniz},
  \citenamefont {Barberena}, \citenamefont {Lewis-Swan}, \citenamefont {Young},
  \citenamefont {Cline}, \citenamefont {Rey},\ and\ \citenamefont
  {Thompson}}]{Muniz2020}%
  \BibitemOpen
  \bibfield  {author} {\bibinfo {author} {\bibfnamefont {J.~A.}\ \bibnamefont
  {Muniz}}, \bibinfo {author} {\bibfnamefont {D.}~\bibnamefont {Barberena}},
  \bibinfo {author} {\bibfnamefont {R.~J.}\ \bibnamefont {Lewis-Swan}},
  \bibinfo {author} {\bibfnamefont {D.~J.}\ \bibnamefont {Young}}, \bibinfo
  {author} {\bibfnamefont {J.~R.~K.}\ \bibnamefont {Cline}}, \bibinfo {author}
  {\bibfnamefont {A.~M.}\ \bibnamefont {Rey}}, \ and\ \bibinfo {author}
  {\bibfnamefont {J.~K.}\ \bibnamefont {Thompson}},\ }\href {\doibase
  10.1038/s41586-020-2224-x} {\bibfield  {journal} {\bibinfo  {journal}
  {Nature}\ }\textbf {\bibinfo {volume} {580}},\ \bibinfo {pages} {602}
  (\bibinfo {year} {2020})}\BibitemShut {NoStop}%
\bibitem [{\citenamefont {Kolkowitz}\ \emph {et~al.}(2016)\citenamefont
  {Kolkowitz}, \citenamefont {Bromley}, \citenamefont {Bothwell}, \citenamefont
  {Wall}, \citenamefont {Marti}, \citenamefont {Koller}, \citenamefont {Zhang},
  \citenamefont {Rey},\ and\ \citenamefont {Ye}}]{Kolkowitz_SOC_2016}%
  \BibitemOpen
  \bibfield  {author} {\bibinfo {author} {\bibfnamefont {S.}~\bibnamefont
  {Kolkowitz}}, \bibinfo {author} {\bibfnamefont {S.~L.}\ \bibnamefont
  {Bromley}}, \bibinfo {author} {\bibfnamefont {T.}~\bibnamefont {Bothwell}},
  \bibinfo {author} {\bibfnamefont {M.~L.}\ \bibnamefont {Wall}}, \bibinfo
  {author} {\bibfnamefont {G.~E.}\ \bibnamefont {Marti}}, \bibinfo {author}
  {\bibfnamefont {A.~P.}\ \bibnamefont {Koller}}, \bibinfo {author}
  {\bibfnamefont {X.}~\bibnamefont {Zhang}}, \bibinfo {author} {\bibfnamefont
  {A.~M.}\ \bibnamefont {Rey}}, \ and\ \bibinfo {author} {\bibfnamefont
  {J.}~\bibnamefont {Ye}},\ }\href {https://doi.org/10.1038/nature20811}
  {\bibfield  {journal} {\bibinfo  {journal} {Nature}\ }\textbf {\bibinfo
  {volume} {542}},\ \bibinfo {pages} {66 EP } (\bibinfo {year}
  {2016})}\BibitemShut {NoStop}%
\bibitem [{\citenamefont {Bromley}\ \emph {et~al.}(2018)\citenamefont
  {Bromley}, \citenamefont {Kolkowitz}, \citenamefont {Bothwell}, \citenamefont
  {Kedar}, \citenamefont {Safavi-Naini}, \citenamefont {Wall}, \citenamefont
  {Salomon}, \citenamefont {Rey},\ and\ \citenamefont {Ye}}]{Bromley_2018}%
  \BibitemOpen
  \bibfield  {author} {\bibinfo {author} {\bibfnamefont {S.~L.}\ \bibnamefont
  {Bromley}}, \bibinfo {author} {\bibfnamefont {S.}~\bibnamefont {Kolkowitz}},
  \bibinfo {author} {\bibfnamefont {T.}~\bibnamefont {Bothwell}}, \bibinfo
  {author} {\bibfnamefont {D.}~\bibnamefont {Kedar}}, \bibinfo {author}
  {\bibfnamefont {A.}~\bibnamefont {Safavi-Naini}}, \bibinfo {author}
  {\bibfnamefont {M.~L.}\ \bibnamefont {Wall}}, \bibinfo {author}
  {\bibfnamefont {C.}~\bibnamefont {Salomon}}, \bibinfo {author} {\bibfnamefont
  {A.~M.}\ \bibnamefont {Rey}}, \ and\ \bibinfo {author} {\bibfnamefont
  {J.}~\bibnamefont {Ye}},\ }\href {\doibase 10.1038/s41567-017-0029-0}
  {\bibfield  {journal} {\bibinfo  {journal} {Nature Physics}\ }\textbf
  {\bibinfo {volume} {14}},\ \bibinfo {pages} {399} (\bibinfo {year}
  {2018})}\BibitemShut {NoStop}%
\bibitem [{\citenamefont {Goban}\ \emph {et~al.}(2018)\citenamefont {Goban},
  \citenamefont {Hutson}, \citenamefont {Marti}, \citenamefont {Campbell},
  \citenamefont {Perlin}, \citenamefont {Julienne}, \citenamefont {D'Incao},
  \citenamefont {Rey},\ and\ \citenamefont {Ye}}]{Goban_2018}%
  \BibitemOpen
  \bibfield  {author} {\bibinfo {author} {\bibfnamefont {A.}~\bibnamefont
  {Goban}}, \bibinfo {author} {\bibfnamefont {R.~B.}\ \bibnamefont {Hutson}},
  \bibinfo {author} {\bibfnamefont {G.~E.}\ \bibnamefont {Marti}}, \bibinfo
  {author} {\bibfnamefont {S.~L.}\ \bibnamefont {Campbell}}, \bibinfo {author}
  {\bibfnamefont {M.~A.}\ \bibnamefont {Perlin}}, \bibinfo {author}
  {\bibfnamefont {P.~S.}\ \bibnamefont {Julienne}}, \bibinfo {author}
  {\bibfnamefont {J.~P.}\ \bibnamefont {D'Incao}}, \bibinfo {author}
  {\bibfnamefont {A.~M.}\ \bibnamefont {Rey}}, \ and\ \bibinfo {author}
  {\bibfnamefont {J.}~\bibnamefont {Ye}},\ }\href {\doibase
  10.1038/s41586-018-0661-6} {\bibfield  {journal} {\bibinfo  {journal}
  {Nature}\ }\textbf {\bibinfo {volume} {563}},\ \bibinfo {pages} {369}
  (\bibinfo {year} {2018})}\BibitemShut {NoStop}%
\bibitem [{\citenamefont {Senaratne}\ \emph {et~al.}(2018)\citenamefont
  {Senaratne}, \citenamefont {Rajagopal}, \citenamefont {Shimasaki},
  \citenamefont {Dotti}, \citenamefont {Fujiwara}, \citenamefont {Singh},
  \citenamefont {Geiger},\ and\ \citenamefont {Weld}}]{Senaratne_2018}%
  \BibitemOpen
  \bibfield  {author} {\bibinfo {author} {\bibfnamefont {R.}~\bibnamefont
  {Senaratne}}, \bibinfo {author} {\bibfnamefont {S.~V.}\ \bibnamefont
  {Rajagopal}}, \bibinfo {author} {\bibfnamefont {T.}~\bibnamefont
  {Shimasaki}}, \bibinfo {author} {\bibfnamefont {P.~E.}\ \bibnamefont
  {Dotti}}, \bibinfo {author} {\bibfnamefont {K.~M.}\ \bibnamefont {Fujiwara}},
  \bibinfo {author} {\bibfnamefont {K.}~\bibnamefont {Singh}}, \bibinfo
  {author} {\bibfnamefont {Z.~A.}\ \bibnamefont {Geiger}}, \ and\ \bibinfo
  {author} {\bibfnamefont {D.~M.}\ \bibnamefont {Weld}},\ }\href {\doibase
  10.1038/s41467-018-04556-3} {\bibfield  {journal} {\bibinfo  {journal}
  {Nature Communications}\ }\textbf {\bibinfo {volume} {9}},\ \bibinfo {pages}
  {2065} (\bibinfo {year} {2018})}\BibitemShut {NoStop}%
\bibitem [{\citenamefont {Norcia}\ \emph {et~al.}(2016)\citenamefont {Norcia},
  \citenamefont {Winchester}, \citenamefont {Cline},\ and\ \citenamefont
  {Thompson}}]{Norcia_SR_2016}%
  \BibitemOpen
  \bibfield  {author} {\bibinfo {author} {\bibfnamefont {M.~A.}\ \bibnamefont
  {Norcia}}, \bibinfo {author} {\bibfnamefont {M.~N.}\ \bibnamefont
  {Winchester}}, \bibinfo {author} {\bibfnamefont {J.~R.~K.}\ \bibnamefont
  {Cline}}, \ and\ \bibinfo {author} {\bibfnamefont {J.~K.}\ \bibnamefont
  {Thompson}},\ }\href@noop {} {\bibfield  {journal} {\bibinfo  {journal}
  {Science Advances}\ }\textbf {\bibinfo {volume} {2}} (\bibinfo {year}
  {2016})}\BibitemShut {NoStop}%
\bibitem [{\citenamefont {Norcia}\ \emph
  {et~al.}(2018{\natexlab{b}})\citenamefont {Norcia}, \citenamefont {Cline},
  \citenamefont {Muniz}, \citenamefont {Robinson}, \citenamefont {Hutson},
  \citenamefont {Goban}, \citenamefont {Marti}, \citenamefont {Ye},\ and\
  \citenamefont {Thompson}}]{Norcia_SRFreq_2018}%
  \BibitemOpen
  \bibfield  {author} {\bibinfo {author} {\bibfnamefont {M.~A.}\ \bibnamefont
  {Norcia}}, \bibinfo {author} {\bibfnamefont {J.~R.~K.}\ \bibnamefont
  {Cline}}, \bibinfo {author} {\bibfnamefont {J.~A.}\ \bibnamefont {Muniz}},
  \bibinfo {author} {\bibfnamefont {J.~M.}\ \bibnamefont {Robinson}}, \bibinfo
  {author} {\bibfnamefont {R.~B.}\ \bibnamefont {Hutson}}, \bibinfo {author}
  {\bibfnamefont {A.}~\bibnamefont {Goban}}, \bibinfo {author} {\bibfnamefont
  {G.~E.}\ \bibnamefont {Marti}}, \bibinfo {author} {\bibfnamefont
  {J.}~\bibnamefont {Ye}}, \ and\ \bibinfo {author} {\bibfnamefont {J.~K.}\
  \bibnamefont {Thompson}},\ }\href {\doibase 10.1103/PhysRevX.8.021036}
  {\bibfield  {journal} {\bibinfo  {journal} {Phys. Rev. X}\ }\textbf {\bibinfo
  {volume} {8}},\ \bibinfo {pages} {021036} (\bibinfo {year}
  {2018}{\natexlab{b}})}\BibitemShut {NoStop}%
\bibitem [{\citenamefont {Pedrozo-Pe{\~n}fiel}\ \emph
  {et~al.}(2020)\citenamefont {Pedrozo-Pe{\~n}fiel}, \citenamefont {Colombo},
  \citenamefont {Shu}, \citenamefont {Adiyatullin}, \citenamefont {Li},
  \citenamefont {Mendez}, \citenamefont {Braverman}, \citenamefont {Kawasaki},
  \citenamefont {Akamatsu}, \citenamefont {Xiao} \emph
  {et~al.}}]{Pedrozo_2020}%
  \BibitemOpen
  \bibfield  {author} {\bibinfo {author} {\bibfnamefont {E.}~\bibnamefont
  {Pedrozo-Pe{\~n}fiel}}, \bibinfo {author} {\bibfnamefont {S.}~\bibnamefont
  {Colombo}}, \bibinfo {author} {\bibfnamefont {C.}~\bibnamefont {Shu}},
  \bibinfo {author} {\bibfnamefont {A.~F.}\ \bibnamefont {Adiyatullin}},
  \bibinfo {author} {\bibfnamefont {Z.}~\bibnamefont {Li}}, \bibinfo {author}
  {\bibfnamefont {E.}~\bibnamefont {Mendez}}, \bibinfo {author} {\bibfnamefont
  {B.}~\bibnamefont {Braverman}}, \bibinfo {author} {\bibfnamefont
  {A.}~\bibnamefont {Kawasaki}}, \bibinfo {author} {\bibfnamefont
  {D.}~\bibnamefont {Akamatsu}}, \bibinfo {author} {\bibfnamefont
  {Y.}~\bibnamefont {Xiao}},  \emph {et~al.},\ }\href@noop {} {\bibfield
  {journal} {\bibinfo  {journal} {arXiv preprint arXiv:2006.07501}\ } (\bibinfo
  {year} {2020})}\BibitemShut {NoStop}%
\bibitem [{\citenamefont {Yasuda}\ and\ \citenamefont
  {Katori}(2004)}]{Yasuda_2004}%
  \BibitemOpen
  \bibfield  {author} {\bibinfo {author} {\bibfnamefont {M.}~\bibnamefont
  {Yasuda}}\ and\ \bibinfo {author} {\bibfnamefont {H.}~\bibnamefont
  {Katori}},\ }\href {\doibase 10.1103/PhysRevLett.92.153004} {\bibfield
  {journal} {\bibinfo  {journal} {Phys. Rev. Lett.}\ }\textbf {\bibinfo
  {volume} {92}},\ \bibinfo {pages} {153004} (\bibinfo {year}
  {2004})}\BibitemShut {NoStop}%
\bibitem [{\citenamefont {Jensen}\ \emph {et~al.}(2011)\citenamefont {Jensen},
  \citenamefont {Ming}, \citenamefont {Westergaard}, \citenamefont
  {Gunnarsson}, \citenamefont {Madsen}, \citenamefont {Brusch}, \citenamefont
  {Hald},\ and\ \citenamefont {Thomsen}}]{Jensen_2011}%
  \BibitemOpen
  \bibfield  {author} {\bibinfo {author} {\bibfnamefont {B.~B.}\ \bibnamefont
  {Jensen}}, \bibinfo {author} {\bibfnamefont {H.}~\bibnamefont {Ming}},
  \bibinfo {author} {\bibfnamefont {P.~G.}\ \bibnamefont {Westergaard}},
  \bibinfo {author} {\bibfnamefont {K.}~\bibnamefont {Gunnarsson}}, \bibinfo
  {author} {\bibfnamefont {M.~H.}\ \bibnamefont {Madsen}}, \bibinfo {author}
  {\bibfnamefont {A.}~\bibnamefont {Brusch}}, \bibinfo {author} {\bibfnamefont
  {J.}~\bibnamefont {Hald}}, \ and\ \bibinfo {author} {\bibfnamefont {J.~W.}\
  \bibnamefont {Thomsen}},\ }\href {\doibase 10.1103/PhysRevLett.107.113001}
  {\bibfield  {journal} {\bibinfo  {journal} {Phys. Rev. Lett.}\ }\textbf
  {\bibinfo {volume} {107}},\ \bibinfo {pages} {113001} (\bibinfo {year}
  {2011})}\BibitemShut {NoStop}%
\bibitem [{\citenamefont {Rosenband}\ \emph {et~al.}(2007)\citenamefont
  {Rosenband}, \citenamefont {Schmidt}, \citenamefont {Hume}, \citenamefont
  {Itano}, \citenamefont {Fortier}, \citenamefont {Stalnaker}, \citenamefont
  {Kim}, \citenamefont {Diddams}, \citenamefont {Koelemeij}, \citenamefont
  {Bergquist},\ and\ \citenamefont {Wineland}}]{Rosenband_2007}%
  \BibitemOpen
  \bibfield  {author} {\bibinfo {author} {\bibfnamefont {T.}~\bibnamefont
  {Rosenband}}, \bibinfo {author} {\bibfnamefont {P.~O.}\ \bibnamefont
  {Schmidt}}, \bibinfo {author} {\bibfnamefont {D.~B.}\ \bibnamefont {Hume}},
  \bibinfo {author} {\bibfnamefont {W.~M.}\ \bibnamefont {Itano}}, \bibinfo
  {author} {\bibfnamefont {T.~M.}\ \bibnamefont {Fortier}}, \bibinfo {author}
  {\bibfnamefont {J.~E.}\ \bibnamefont {Stalnaker}}, \bibinfo {author}
  {\bibfnamefont {K.}~\bibnamefont {Kim}}, \bibinfo {author} {\bibfnamefont
  {S.~A.}\ \bibnamefont {Diddams}}, \bibinfo {author} {\bibfnamefont
  {J.~C.~J.}\ \bibnamefont {Koelemeij}}, \bibinfo {author} {\bibfnamefont
  {J.~C.}\ \bibnamefont {Bergquist}}, \ and\ \bibinfo {author} {\bibfnamefont
  {D.~J.}\ \bibnamefont {Wineland}},\ }\href {\doibase
  10.1103/PhysRevLett.98.220801} {\bibfield  {journal} {\bibinfo  {journal}
  {Phys. Rev. Lett.}\ }\textbf {\bibinfo {volume} {98}},\ \bibinfo {pages}
  {220801} (\bibinfo {year} {2007})}\BibitemShut {NoStop}%
\bibitem [{\citenamefont {Barton}\ \emph {et~al.}(2000)\citenamefont {Barton},
  \citenamefont {Donald}, \citenamefont {Lucas}, \citenamefont {Stevens},
  \citenamefont {Steane},\ and\ \citenamefont {Stacey}}]{Barton_2000}%
  \BibitemOpen
  \bibfield  {author} {\bibinfo {author} {\bibfnamefont {P.~A.}\ \bibnamefont
  {Barton}}, \bibinfo {author} {\bibfnamefont {C.~J.~S.}\ \bibnamefont
  {Donald}}, \bibinfo {author} {\bibfnamefont {D.~M.}\ \bibnamefont {Lucas}},
  \bibinfo {author} {\bibfnamefont {D.~A.}\ \bibnamefont {Stevens}}, \bibinfo
  {author} {\bibfnamefont {A.~M.}\ \bibnamefont {Steane}}, \ and\ \bibinfo
  {author} {\bibfnamefont {D.~N.}\ \bibnamefont {Stacey}},\ }\href {\doibase
  10.1103/PhysRevA.62.032503} {\bibfield  {journal} {\bibinfo  {journal} {Phys.
  Rev. A}\ }\textbf {\bibinfo {volume} {62}},\ \bibinfo {pages} {032503}
  (\bibinfo {year} {2000})}\BibitemShut {NoStop}%
\bibitem [{\citenamefont {Staanum}\ \emph {et~al.}(2004)\citenamefont
  {Staanum}, \citenamefont {Jensen}, \citenamefont {Martinussen}, \citenamefont
  {Voigt},\ and\ \citenamefont {Drewsen}}]{Staanum_2004}%
  \BibitemOpen
  \bibfield  {author} {\bibinfo {author} {\bibfnamefont {P.}~\bibnamefont
  {Staanum}}, \bibinfo {author} {\bibfnamefont {I.~S.}\ \bibnamefont {Jensen}},
  \bibinfo {author} {\bibfnamefont {R.~G.}\ \bibnamefont {Martinussen}},
  \bibinfo {author} {\bibfnamefont {D.}~\bibnamefont {Voigt}}, \ and\ \bibinfo
  {author} {\bibfnamefont {M.}~\bibnamefont {Drewsen}},\ }\href {\doibase
  10.1103/PhysRevA.69.032503} {\bibfield  {journal} {\bibinfo  {journal} {Phys.
  Rev. A}\ }\textbf {\bibinfo {volume} {69}},\ \bibinfo {pages} {032503}
  (\bibinfo {year} {2004})}\BibitemShut {NoStop}%
\bibitem [{\citenamefont {Kreuter}\ \emph {et~al.}(2004)\citenamefont
  {Kreuter}, \citenamefont {Becher}, \citenamefont {Lancaster}, \citenamefont
  {Mundt}, \citenamefont {Russo}, \citenamefont {H\"affner}, \citenamefont
  {Roos}, \citenamefont {Eschner}, \citenamefont {Schmidt-Kaler},\ and\
  \citenamefont {Blatt}}]{Kreuter_2004}%
  \BibitemOpen
  \bibfield  {author} {\bibinfo {author} {\bibfnamefont {A.}~\bibnamefont
  {Kreuter}}, \bibinfo {author} {\bibfnamefont {C.}~\bibnamefont {Becher}},
  \bibinfo {author} {\bibfnamefont {G.~P.~T.}\ \bibnamefont {Lancaster}},
  \bibinfo {author} {\bibfnamefont {A.~B.}\ \bibnamefont {Mundt}}, \bibinfo
  {author} {\bibfnamefont {C.}~\bibnamefont {Russo}}, \bibinfo {author}
  {\bibfnamefont {H.}~\bibnamefont {H\"affner}}, \bibinfo {author}
  {\bibfnamefont {C.}~\bibnamefont {Roos}}, \bibinfo {author} {\bibfnamefont
  {J.}~\bibnamefont {Eschner}}, \bibinfo {author} {\bibfnamefont
  {F.}~\bibnamefont {Schmidt-Kaler}}, \ and\ \bibinfo {author} {\bibfnamefont
  {R.}~\bibnamefont {Blatt}},\ }\href {\doibase 10.1103/PhysRevLett.92.203002}
  {\bibfield  {journal} {\bibinfo  {journal} {Phys. Rev. Lett.}\ }\textbf
  {\bibinfo {volume} {92}},\ \bibinfo {pages} {203002} (\bibinfo {year}
  {2004})}\BibitemShut {NoStop}%
\bibitem [{\citenamefont {Becker}\ \emph {et~al.}(2001)\citenamefont {Becker},
  \citenamefont {Zanthier}, \citenamefont {Nevsky}, \citenamefont {Schwedes},
  \citenamefont {Skvortsov}, \citenamefont {Walther},\ and\ \citenamefont
  {Peik}}]{Becker_2001}%
  \BibitemOpen
  \bibfield  {author} {\bibinfo {author} {\bibfnamefont {T.}~\bibnamefont
  {Becker}}, \bibinfo {author} {\bibfnamefont {J.~v.}\ \bibnamefont
  {Zanthier}}, \bibinfo {author} {\bibfnamefont {A.~Y.}\ \bibnamefont
  {Nevsky}}, \bibinfo {author} {\bibfnamefont {C.}~\bibnamefont {Schwedes}},
  \bibinfo {author} {\bibfnamefont {M.~N.}\ \bibnamefont {Skvortsov}}, \bibinfo
  {author} {\bibfnamefont {H.}~\bibnamefont {Walther}}, \ and\ \bibinfo
  {author} {\bibfnamefont {E.}~\bibnamefont {Peik}},\ }\href {\doibase
  10.1103/PhysRevA.63.051802} {\bibfield  {journal} {\bibinfo  {journal} {Phys.
  Rev. A}\ }\textbf {\bibinfo {volume} {63}},\ \bibinfo {pages} {051802}
  (\bibinfo {year} {2001})}\BibitemShut {NoStop}%
\bibitem [{\citenamefont {Walhout}\ \emph {et~al.}(1994)\citenamefont
  {Walhout}, \citenamefont {Witte},\ and\ \citenamefont
  {Rolston}}]{Walhout_1994}%
  \BibitemOpen
  \bibfield  {author} {\bibinfo {author} {\bibfnamefont {M.}~\bibnamefont
  {Walhout}}, \bibinfo {author} {\bibfnamefont {A.}~\bibnamefont {Witte}}, \
  and\ \bibinfo {author} {\bibfnamefont {S.~L.}\ \bibnamefont {Rolston}},\
  }\href {\doibase 10.1103/PhysRevLett.72.2843} {\bibfield  {journal} {\bibinfo
   {journal} {Phys. Rev. Lett.}\ }\textbf {\bibinfo {volume} {72}},\ \bibinfo
  {pages} {2843} (\bibinfo {year} {1994})}\BibitemShut {NoStop}%
\bibitem [{\citenamefont {Walhout}\ \emph {et~al.}(1995)\citenamefont
  {Walhout}, \citenamefont {Sterr}, \citenamefont {Witte},\ and\ \citenamefont
  {Rolston}}]{Walhout_1995}%
  \BibitemOpen
  \bibfield  {author} {\bibinfo {author} {\bibfnamefont {M.}~\bibnamefont
  {Walhout}}, \bibinfo {author} {\bibfnamefont {U.}~\bibnamefont {Sterr}},
  \bibinfo {author} {\bibfnamefont {A.}~\bibnamefont {Witte}}, \ and\ \bibinfo
  {author} {\bibfnamefont {S.~L.}\ \bibnamefont {Rolston}},\ }\href {\doibase
  10.1364/OL.20.001192} {\bibfield  {journal} {\bibinfo  {journal} {Opt.
  Lett.}\ }\textbf {\bibinfo {volume} {20}},\ \bibinfo {pages} {1192} (\bibinfo
  {year} {1995})}\BibitemShut {NoStop}%
\bibitem [{\citenamefont {D\"orscher}\ \emph {et~al.}(2018)\citenamefont
  {D\"orscher}, \citenamefont {Schwarz}, \citenamefont {Al-Masoudi},
  \citenamefont {Falke}, \citenamefont {Sterr},\ and\ \citenamefont
  {Lisdat}}]{Dorscher_2018}%
  \BibitemOpen
  \bibfield  {author} {\bibinfo {author} {\bibfnamefont {S.}~\bibnamefont
  {D\"orscher}}, \bibinfo {author} {\bibfnamefont {R.}~\bibnamefont {Schwarz}},
  \bibinfo {author} {\bibfnamefont {A.}~\bibnamefont {Al-Masoudi}}, \bibinfo
  {author} {\bibfnamefont {S.}~\bibnamefont {Falke}}, \bibinfo {author}
  {\bibfnamefont {U.}~\bibnamefont {Sterr}}, \ and\ \bibinfo {author}
  {\bibfnamefont {C.}~\bibnamefont {Lisdat}},\ }\href {\doibase
  10.1103/PhysRevA.97.063419} {\bibfield  {journal} {\bibinfo  {journal} {Phys.
  Rev. A}\ }\textbf {\bibinfo {volume} {97}},\ \bibinfo {pages} {063419}
  (\bibinfo {year} {2018})}\BibitemShut {NoStop}%
\bibitem [{\citenamefont {Hutson}\ \emph {et~al.}(2019)\citenamefont {Hutson},
  \citenamefont {Goban}, \citenamefont {Marti}, \citenamefont {Sonderhouse},
  \citenamefont {Sanner},\ and\ \citenamefont {Ye}}]{Hutson_2019}%
  \BibitemOpen
  \bibfield  {author} {\bibinfo {author} {\bibfnamefont {R.~B.}\ \bibnamefont
  {Hutson}}, \bibinfo {author} {\bibfnamefont {A.}~\bibnamefont {Goban}},
  \bibinfo {author} {\bibfnamefont {G.~E.}\ \bibnamefont {Marti}}, \bibinfo
  {author} {\bibfnamefont {L.}~\bibnamefont {Sonderhouse}}, \bibinfo {author}
  {\bibfnamefont {C.}~\bibnamefont {Sanner}}, \ and\ \bibinfo {author}
  {\bibfnamefont {J.}~\bibnamefont {Ye}},\ }\href {\doibase
  10.1103/PhysRevLett.123.123401} {\bibfield  {journal} {\bibinfo  {journal}
  {Phys. Rev. Lett.}\ }\textbf {\bibinfo {volume} {123}},\ \bibinfo {pages}
  {123401} (\bibinfo {year} {2019})}\BibitemShut {NoStop}%
\bibitem [{\citenamefont {Boyd}\ \emph {et~al.}(2007)\citenamefont {Boyd},
  \citenamefont {Zelevinsky}, \citenamefont {Ludlow}, \citenamefont {Blatt},
  \citenamefont {Zanon-Willette}, \citenamefont {Foreman},\ and\ \citenamefont
  {Ye}}]{Boyd_2007}%
  \BibitemOpen
  \bibfield  {author} {\bibinfo {author} {\bibfnamefont {M.~M.}\ \bibnamefont
  {Boyd}}, \bibinfo {author} {\bibfnamefont {T.}~\bibnamefont {Zelevinsky}},
  \bibinfo {author} {\bibfnamefont {A.~D.}\ \bibnamefont {Ludlow}}, \bibinfo
  {author} {\bibfnamefont {S.}~\bibnamefont {Blatt}}, \bibinfo {author}
  {\bibfnamefont {T.}~\bibnamefont {Zanon-Willette}}, \bibinfo {author}
  {\bibfnamefont {S.~M.}\ \bibnamefont {Foreman}}, \ and\ \bibinfo {author}
  {\bibfnamefont {J.}~\bibnamefont {Ye}},\ }\href {\doibase
  10.1103/PhysRevA.76.022510} {\bibfield  {journal} {\bibinfo  {journal} {Phys.
  Rev. A}\ }\textbf {\bibinfo {volume} {76}},\ \bibinfo {pages} {022510}
  (\bibinfo {year} {2007})}\BibitemShut {NoStop}%
\bibitem [{\citenamefont {Porsev}\ and\ \citenamefont
  {Derevianko}(2004)}]{Porsev_2004}%
  \BibitemOpen
  \bibfield  {author} {\bibinfo {author} {\bibfnamefont {S.~G.}\ \bibnamefont
  {Porsev}}\ and\ \bibinfo {author} {\bibfnamefont {A.}~\bibnamefont
  {Derevianko}},\ }\href {\doibase 10.1103/PhysRevA.69.042506} {\bibfield
  {journal} {\bibinfo  {journal} {Phys. Rev. A}\ }\textbf {\bibinfo {volume}
  {69}},\ \bibinfo {pages} {042506} (\bibinfo {year} {2004})}\BibitemShut
  {NoStop}%
\bibitem [{\citenamefont {Santra}\ \emph {et~al.}(2004)\citenamefont {Santra},
  \citenamefont {Christ},\ and\ \citenamefont {Greene}}]{Santra_2004}%
  \BibitemOpen
  \bibfield  {author} {\bibinfo {author} {\bibfnamefont {R.}~\bibnamefont
  {Santra}}, \bibinfo {author} {\bibfnamefont {K.~V.}\ \bibnamefont {Christ}},
  \ and\ \bibinfo {author} {\bibfnamefont {C.~H.}\ \bibnamefont {Greene}},\
  }\href {\doibase 10.1103/PhysRevA.69.042510} {\bibfield  {journal} {\bibinfo
  {journal} {Phys. Rev. A}\ }\textbf {\bibinfo {volume} {69}},\ \bibinfo
  {pages} {042510} (\bibinfo {year} {2004})}\BibitemShut {NoStop}%
\bibitem [{\citenamefont {Norcia}\ \emph {et~al.}(2019)\citenamefont {Norcia},
  \citenamefont {Young}, \citenamefont {Eckner}, \citenamefont {Oelker},
  \citenamefont {Ye},\ and\ \citenamefont {Kaufman}}]{Norcia_2019}%
  \BibitemOpen
  \bibfield  {author} {\bibinfo {author} {\bibfnamefont {M.~A.}\ \bibnamefont
  {Norcia}}, \bibinfo {author} {\bibfnamefont {A.~W.}\ \bibnamefont {Young}},
  \bibinfo {author} {\bibfnamefont {W.~J.}\ \bibnamefont {Eckner}}, \bibinfo
  {author} {\bibfnamefont {E.}~\bibnamefont {Oelker}}, \bibinfo {author}
  {\bibfnamefont {J.}~\bibnamefont {Ye}}, \ and\ \bibinfo {author}
  {\bibfnamefont {A.~M.}\ \bibnamefont {Kaufman}},\ }\href {\doibase
  10.1126/science.aay0644} {\bibfield  {journal} {\bibinfo  {journal}
  {Science}\ }\textbf {\bibinfo {volume} {366}},\ \bibinfo {pages} {93}
  (\bibinfo {year} {2019})}\BibitemShut {NoStop}%
\bibitem [{\citenamefont {Madjarov}\ \emph {et~al.}(2019)\citenamefont
  {Madjarov}, \citenamefont {Cooper}, \citenamefont {Shaw}, \citenamefont
  {Covey}, \citenamefont {Schkolnik}, \citenamefont {Yoon}, \citenamefont
  {Williams},\ and\ \citenamefont {Endres}}]{Madjarov_2019}%
  \BibitemOpen
  \bibfield  {author} {\bibinfo {author} {\bibfnamefont {I.~S.}\ \bibnamefont
  {Madjarov}}, \bibinfo {author} {\bibfnamefont {A.}~\bibnamefont {Cooper}},
  \bibinfo {author} {\bibfnamefont {A.~L.}\ \bibnamefont {Shaw}}, \bibinfo
  {author} {\bibfnamefont {J.~P.}\ \bibnamefont {Covey}}, \bibinfo {author}
  {\bibfnamefont {V.}~\bibnamefont {Schkolnik}}, \bibinfo {author}
  {\bibfnamefont {T.~H.}\ \bibnamefont {Yoon}}, \bibinfo {author}
  {\bibfnamefont {J.~R.}\ \bibnamefont {Williams}}, \ and\ \bibinfo {author}
  {\bibfnamefont {M.}~\bibnamefont {Endres}},\ }\href {\doibase
  10.1103/PhysRevX.9.041052} {\bibfield  {journal} {\bibinfo  {journal} {Phys.
  Rev. X}\ }\textbf {\bibinfo {volume} {9}},\ \bibinfo {pages} {041052}
  (\bibinfo {year} {2019})}\BibitemShut {NoStop}%
\bibitem [{\citenamefont {Matei}\ \emph {et~al.}(2017)\citenamefont {Matei},
  \citenamefont {Legero}, \citenamefont {H\"afner}, \citenamefont {Grebing},
  \citenamefont {Weyrich}, \citenamefont {Zhang}, \citenamefont {Sonderhouse},
  \citenamefont {Robinson}, \citenamefont {Ye}, \citenamefont {Riehle},\ and\
  \citenamefont {Sterr}}]{Matei_2017}%
  \BibitemOpen
  \bibfield  {author} {\bibinfo {author} {\bibfnamefont {D.~G.}\ \bibnamefont
  {Matei}}, \bibinfo {author} {\bibfnamefont {T.}~\bibnamefont {Legero}},
  \bibinfo {author} {\bibfnamefont {S.}~\bibnamefont {H\"afner}}, \bibinfo
  {author} {\bibfnamefont {C.}~\bibnamefont {Grebing}}, \bibinfo {author}
  {\bibfnamefont {R.}~\bibnamefont {Weyrich}}, \bibinfo {author} {\bibfnamefont
  {W.}~\bibnamefont {Zhang}}, \bibinfo {author} {\bibfnamefont
  {L.}~\bibnamefont {Sonderhouse}}, \bibinfo {author} {\bibfnamefont {J.~M.}\
  \bibnamefont {Robinson}}, \bibinfo {author} {\bibfnamefont {J.}~\bibnamefont
  {Ye}}, \bibinfo {author} {\bibfnamefont {F.}~\bibnamefont {Riehle}}, \ and\
  \bibinfo {author} {\bibfnamefont {U.}~\bibnamefont {Sterr}},\ }\href
  {\doibase 10.1103/PhysRevLett.118.263202} {\bibfield  {journal} {\bibinfo
  {journal} {Phys. Rev. Lett.}\ }\textbf {\bibinfo {volume} {118}},\ \bibinfo
  {pages} {263202} (\bibinfo {year} {2017})}\BibitemShut {NoStop}%
\bibitem [{\citenamefont {Zhang}\ \emph {et~al.}(2017)\citenamefont {Zhang},
  \citenamefont {Robinson}, \citenamefont {Sonderhouse}, \citenamefont
  {Oelker}, \citenamefont {Benko}, \citenamefont {Hall}, \citenamefont
  {Legero}, \citenamefont {Matei}, \citenamefont {Riehle}, \citenamefont
  {Sterr},\ and\ \citenamefont {Ye}}]{Zhang_2017}%
  \BibitemOpen
  \bibfield  {author} {\bibinfo {author} {\bibfnamefont {W.}~\bibnamefont
  {Zhang}}, \bibinfo {author} {\bibfnamefont {J.~M.}\ \bibnamefont {Robinson}},
  \bibinfo {author} {\bibfnamefont {L.}~\bibnamefont {Sonderhouse}}, \bibinfo
  {author} {\bibfnamefont {E.}~\bibnamefont {Oelker}}, \bibinfo {author}
  {\bibfnamefont {C.}~\bibnamefont {Benko}}, \bibinfo {author} {\bibfnamefont
  {J.~L.}\ \bibnamefont {Hall}}, \bibinfo {author} {\bibfnamefont
  {T.}~\bibnamefont {Legero}}, \bibinfo {author} {\bibfnamefont {D.~G.}\
  \bibnamefont {Matei}}, \bibinfo {author} {\bibfnamefont {F.}~\bibnamefont
  {Riehle}}, \bibinfo {author} {\bibfnamefont {U.}~\bibnamefont {Sterr}}, \
  and\ \bibinfo {author} {\bibfnamefont {J.}~\bibnamefont {Ye}},\ }\href
  {\doibase 10.1103/PhysRevLett.119.243601} {\bibfield  {journal} {\bibinfo
  {journal} {Phys. Rev. Lett.}\ }\textbf {\bibinfo {volume} {119}},\ \bibinfo
  {pages} {243601} (\bibinfo {year} {2017})}\BibitemShut {NoStop}%
\bibitem [{\citenamefont {Robinson}\ \emph {et~al.}(2019)\citenamefont
  {Robinson}, \citenamefont {Oelker}, \citenamefont {Milner}, \citenamefont
  {Zhang}, \citenamefont {Legero}, \citenamefont {Matei}, \citenamefont
  {Riehle}, \citenamefont {Sterr},\ and\ \citenamefont {Ye}}]{Robinson_2019}%
  \BibitemOpen
  \bibfield  {author} {\bibinfo {author} {\bibfnamefont {J.~M.}\ \bibnamefont
  {Robinson}}, \bibinfo {author} {\bibfnamefont {E.}~\bibnamefont {Oelker}},
  \bibinfo {author} {\bibfnamefont {W.~R.}\ \bibnamefont {Milner}}, \bibinfo
  {author} {\bibfnamefont {W.}~\bibnamefont {Zhang}}, \bibinfo {author}
  {\bibfnamefont {T.}~\bibnamefont {Legero}}, \bibinfo {author} {\bibfnamefont
  {D.~G.}\ \bibnamefont {Matei}}, \bibinfo {author} {\bibfnamefont
  {F.}~\bibnamefont {Riehle}}, \bibinfo {author} {\bibfnamefont
  {U.}~\bibnamefont {Sterr}}, \ and\ \bibinfo {author} {\bibfnamefont
  {J.}~\bibnamefont {Ye}},\ }\href {\doibase 10.1364/OPTICA.6.000240}
  {\bibfield  {journal} {\bibinfo  {journal} {Optica}\ }\textbf {\bibinfo
  {volume} {6}},\ \bibinfo {pages} {240} (\bibinfo {year} {2019})}\BibitemShut
  {NoStop}%
\bibitem [{\citenamefont {Tanji-Suzuki}\ \emph {et~al.}(2011)\citenamefont
  {Tanji-Suzuki}, \citenamefont {Leroux}, \citenamefont {Schleier-Smith},
  \citenamefont {Cetina}, \citenamefont {Grier}, \citenamefont {Simon},\ and\
  \citenamefont {Vuleti\'{c}}}]{TanjiSuzuki_2011}%
  \BibitemOpen
  \bibfield  {author} {\bibinfo {author} {\bibfnamefont {H.}~\bibnamefont
  {Tanji-Suzuki}}, \bibinfo {author} {\bibfnamefont {I.~D.}\ \bibnamefont
  {Leroux}}, \bibinfo {author} {\bibfnamefont {M.~H.}\ \bibnamefont
  {Schleier-Smith}}, \bibinfo {author} {\bibfnamefont {M.}~\bibnamefont
  {Cetina}}, \bibinfo {author} {\bibfnamefont {A.~T.}\ \bibnamefont {Grier}},
  \bibinfo {author} {\bibfnamefont {J.}~\bibnamefont {Simon}}, \ and\ \bibinfo
  {author} {\bibfnamefont {V.}~\bibnamefont {Vuleti\'{c}}},\ }in\ \href
  {\doibase 10.1016/B978-0-12-385508-4.00004-8} {\emph {\bibinfo {booktitle}
  {Advances in Atomic, Molecular, and Optical Physics}}},\ \bibinfo {series}
  {Advances In Atomic, Molecular, and Optical Physics}, Vol.~\bibinfo {volume}
  {60},\ \bibinfo {editor} {edited by\ \bibinfo {editor} {\bibfnamefont
  {P.~B.}\ \bibnamefont {E.~Arimondo}}\ and\ \bibinfo {editor} {\bibfnamefont
  {C.}~\bibnamefont {Lin}}}\ (\bibinfo  {publisher} {Academic Press},\ \bibinfo
  {year} {2011})\ pp.\ \bibinfo {pages} {201 -- 237}\BibitemShut {NoStop}%
\bibitem [{\citenamefont {Jaynes}\ and\ \citenamefont
  {Cummings}(1963)}]{Jaynes_1963}%
  \BibitemOpen
  \bibfield  {author} {\bibinfo {author} {\bibfnamefont {E.}~\bibnamefont
  {Jaynes}}\ and\ \bibinfo {author} {\bibfnamefont {F.}~\bibnamefont
  {Cummings}},\ }\href@noop {} {\bibfield  {journal} {\bibinfo  {journal}
  {Proceedings of the IEEE}\ }\textbf {\bibinfo {volume} {51}},\ \bibinfo
  {pages} {89} (\bibinfo {year} {1963})}\BibitemShut {NoStop}%
\bibitem [{\citenamefont {Kimble}(1998)}]{Kimble_1998}%
  \BibitemOpen
  \bibfield  {author} {\bibinfo {author} {\bibfnamefont {H.~J.}\ \bibnamefont
  {Kimble}},\ }\href {\doibase 10.1238/Physica.Topical.076a00127} {\bibfield
  {journal} {\bibinfo  {journal} {Physica Scripta}\ }\textbf {\bibinfo {volume}
  {1998}},\ \bibinfo {pages} {127} (\bibinfo {year} {1998})}\BibitemShut
  {NoStop}%
\bibitem [{\citenamefont {Ye}\ \emph {et~al.}(2008)\citenamefont {Ye},
  \citenamefont {Kimble},\ and\ \citenamefont {Katori}}]{Ye_2008}%
  \BibitemOpen
  \bibfield  {author} {\bibinfo {author} {\bibfnamefont {J.}~\bibnamefont
  {Ye}}, \bibinfo {author} {\bibfnamefont {H.~J.}\ \bibnamefont {Kimble}}, \
  and\ \bibinfo {author} {\bibfnamefont {H.}~\bibnamefont {Katori}},\ }\href
  {\doibase 10.1126/science.1148259} {\bibfield  {journal} {\bibinfo  {journal}
  {Science}\ }\textbf {\bibinfo {volume} {320}},\ \bibinfo {pages} {1734}
  (\bibinfo {year} {2008})}\BibitemShut {NoStop}%
\bibitem [{\citenamefont {Blatt}\ \emph {et~al.}(2009)\citenamefont {Blatt},
  \citenamefont {Thomsen}, \citenamefont {Campbell}, \citenamefont {Ludlow},
  \citenamefont {Swallows}, \citenamefont {Martin}, \citenamefont {Boyd},\ and\
  \citenamefont {Ye}}]{Blatt_2009}%
  \BibitemOpen
  \bibfield  {author} {\bibinfo {author} {\bibfnamefont {S.}~\bibnamefont
  {Blatt}}, \bibinfo {author} {\bibfnamefont {J.~W.}\ \bibnamefont {Thomsen}},
  \bibinfo {author} {\bibfnamefont {G.~K.}\ \bibnamefont {Campbell}}, \bibinfo
  {author} {\bibfnamefont {A.~D.}\ \bibnamefont {Ludlow}}, \bibinfo {author}
  {\bibfnamefont {M.~D.}\ \bibnamefont {Swallows}}, \bibinfo {author}
  {\bibfnamefont {M.~J.}\ \bibnamefont {Martin}}, \bibinfo {author}
  {\bibfnamefont {M.~M.}\ \bibnamefont {Boyd}}, \ and\ \bibinfo {author}
  {\bibfnamefont {J.}~\bibnamefont {Ye}},\ }\href {\doibase
  10.1103/PhysRevA.80.052703} {\bibfield  {journal} {\bibinfo  {journal} {Phys.
  Rev. A}\ }\textbf {\bibinfo {volume} {80}},\ \bibinfo {pages} {052703}
  (\bibinfo {year} {2009})}\BibitemShut {NoStop}%
\bibitem [{\citenamefont {Chen}\ \emph {et~al.}(2014)\citenamefont {Chen},
  \citenamefont {Bohnet}, \citenamefont {Weiner}, \citenamefont {Cox},\ and\
  \citenamefont {Thompson}}]{Chen_2014}%
  \BibitemOpen
  \bibfield  {author} {\bibinfo {author} {\bibfnamefont {Z.}~\bibnamefont
  {Chen}}, \bibinfo {author} {\bibfnamefont {J.~G.}\ \bibnamefont {Bohnet}},
  \bibinfo {author} {\bibfnamefont {J.~M.}\ \bibnamefont {Weiner}}, \bibinfo
  {author} {\bibfnamefont {K.~C.}\ \bibnamefont {Cox}}, \ and\ \bibinfo
  {author} {\bibfnamefont {J.~K.}\ \bibnamefont {Thompson}},\ }\href {\doibase
  10.1103/PhysRevA.89.043837} {\bibfield  {journal} {\bibinfo  {journal} {Phys.
  Rev. A}\ }\textbf {\bibinfo {volume} {89}},\ \bibinfo {pages} {043837}
  (\bibinfo {year} {2014})}\BibitemShut {NoStop}%
\bibitem [{\citenamefont {Braginsky}\ and\ \citenamefont
  {Khalili}(1996)}]{Braginsky_1996}%
  \BibitemOpen
  \bibfield  {author} {\bibinfo {author} {\bibfnamefont {V.~B.}\ \bibnamefont
  {Braginsky}}\ and\ \bibinfo {author} {\bibfnamefont {F.~Y.}\ \bibnamefont
  {Khalili}},\ }\href {\doibase 10.1103/RevModPhys.68.1} {\bibfield  {journal}
  {\bibinfo  {journal} {Rev. Mod. Phys.}\ }\textbf {\bibinfo {volume} {68}},\
  \bibinfo {pages} {1} (\bibinfo {year} {1996})}\BibitemShut {NoStop}%
\bibitem [{\citenamefont {Appel}\ \emph {et~al.}(2009)\citenamefont {Appel},
  \citenamefont {Windpassinger}, \citenamefont {Oblak}, \citenamefont {Hoff},
  \citenamefont {Kj{\ae}rgaard},\ and\ \citenamefont {Polzik}}]{Appel_2009}%
  \BibitemOpen
  \bibfield  {author} {\bibinfo {author} {\bibfnamefont {J.}~\bibnamefont
  {Appel}}, \bibinfo {author} {\bibfnamefont {P.~J.}\ \bibnamefont
  {Windpassinger}}, \bibinfo {author} {\bibfnamefont {D.}~\bibnamefont
  {Oblak}}, \bibinfo {author} {\bibfnamefont {U.~B.}\ \bibnamefont {Hoff}},
  \bibinfo {author} {\bibfnamefont {N.}~\bibnamefont {Kj{\ae}rgaard}}, \ and\
  \bibinfo {author} {\bibfnamefont {E.~S.}\ \bibnamefont {Polzik}},\ }\href
  {\doibase 10.1073/pnas.0901550106} {\bibfield  {journal} {\bibinfo  {journal}
  {Proceedings of the National Academy of Sciences}\ }\textbf {\bibinfo
  {volume} {106}},\ \bibinfo {pages} {10960} (\bibinfo {year}
  {2009})}\BibitemShut {NoStop}%
\bibitem [{\citenamefont {Chen}\ \emph {et~al.}(2011)\citenamefont {Chen},
  \citenamefont {Bohnet}, \citenamefont {Sankar}, \citenamefont {Dai},\ and\
  \citenamefont {Thompson}}]{Chen_2010}%
  \BibitemOpen
  \bibfield  {author} {\bibinfo {author} {\bibfnamefont {Z.}~\bibnamefont
  {Chen}}, \bibinfo {author} {\bibfnamefont {J.~G.}\ \bibnamefont {Bohnet}},
  \bibinfo {author} {\bibfnamefont {S.~R.}\ \bibnamefont {Sankar}}, \bibinfo
  {author} {\bibfnamefont {J.}~\bibnamefont {Dai}}, \ and\ \bibinfo {author}
  {\bibfnamefont {J.~K.}\ \bibnamefont {Thompson}},\ }\href {\doibase
  10.1103/PhysRevLett.106.133601} {\bibfield  {journal} {\bibinfo  {journal}
  {Phys. Rev. Lett.}\ }\textbf {\bibinfo {volume} {106}},\ \bibinfo {pages}
  {133601} (\bibinfo {year} {2011})}\BibitemShut {NoStop}%
\bibitem [{\citenamefont {Schleier-Smith}\ \emph {et~al.}(2010)\citenamefont
  {Schleier-Smith}, \citenamefont {Leroux},\ and\ \citenamefont
  {Vuleti\ifmmode~\acute{c}\else \'{c}\fi{}}}]{Schleier-Smith_2010}%
  \BibitemOpen
  \bibfield  {author} {\bibinfo {author} {\bibfnamefont {M.~H.}\ \bibnamefont
  {Schleier-Smith}}, \bibinfo {author} {\bibfnamefont {I.~D.}\ \bibnamefont
  {Leroux}}, \ and\ \bibinfo {author} {\bibfnamefont {V.}~\bibnamefont
  {Vuleti\ifmmode~\acute{c}\else \'{c}\fi{}}},\ }\href {\doibase
  10.1103/PhysRevLett.104.073604} {\bibfield  {journal} {\bibinfo  {journal}
  {Phys. Rev. Lett.}\ }\textbf {\bibinfo {volume} {104}},\ \bibinfo {pages}
  {073604} (\bibinfo {year} {2010})}\BibitemShut {NoStop}%
\bibitem [{\citenamefont {Bohnet}\ \emph {et~al.}(2014)\citenamefont {Bohnet},
  \citenamefont {Cox}, \citenamefont {Norcia}, \citenamefont {Weiner},
  \citenamefont {Chen},\ and\ \citenamefont {Thompson}}]{Bohnet2014}%
  \BibitemOpen
  \bibfield  {author} {\bibinfo {author} {\bibfnamefont {J.~G.}\ \bibnamefont
  {Bohnet}}, \bibinfo {author} {\bibfnamefont {K.~C.}\ \bibnamefont {Cox}},
  \bibinfo {author} {\bibfnamefont {M.~A.}\ \bibnamefont {Norcia}}, \bibinfo
  {author} {\bibfnamefont {J.~M.}\ \bibnamefont {Weiner}}, \bibinfo {author}
  {\bibfnamefont {Z.}~\bibnamefont {Chen}}, \ and\ \bibinfo {author}
  {\bibfnamefont {J.~K.}\ \bibnamefont {Thompson}},\ }\href
  {https://doi.org/10.1038/nphoton.2014.151} {\bibfield  {journal} {\bibinfo
  {journal} {Nature Photonics}\ }\textbf {\bibinfo {volume} {8}},\ \bibinfo
  {pages} {731 EP } (\bibinfo {year} {2014})}\BibitemShut {NoStop}%
\bibitem [{\citenamefont {Hosten}\ \emph {et~al.}(2016)\citenamefont {Hosten},
  \citenamefont {Engelsen}, \citenamefont {Krishnakumar},\ and\ \citenamefont
  {Kasevich}}]{Hosten2016}%
  \BibitemOpen
  \bibfield  {author} {\bibinfo {author} {\bibfnamefont {O.}~\bibnamefont
  {Hosten}}, \bibinfo {author} {\bibfnamefont {N.~J.}\ \bibnamefont
  {Engelsen}}, \bibinfo {author} {\bibfnamefont {R.}~\bibnamefont
  {Krishnakumar}}, \ and\ \bibinfo {author} {\bibfnamefont {M.~A.}\
  \bibnamefont {Kasevich}},\ }\href {https://doi.org/10.1038/nature16176}
  {\bibfield  {journal} {\bibinfo  {journal} {Nature}\ }\textbf {\bibinfo
  {volume} {529}},\ \bibinfo {pages} {505 EP } (\bibinfo {year}
  {2016})}\BibitemShut {NoStop}%
\bibitem [{\citenamefont {Cox}\ \emph {et~al.}(2016)\citenamefont {Cox},
  \citenamefont {Greve}, \citenamefont {Weiner},\ and\ \citenamefont
  {Thompson}}]{Cox2016}%
  \BibitemOpen
  \bibfield  {author} {\bibinfo {author} {\bibfnamefont {K.~C.}\ \bibnamefont
  {Cox}}, \bibinfo {author} {\bibfnamefont {G.~P.}\ \bibnamefont {Greve}},
  \bibinfo {author} {\bibfnamefont {J.~M.}\ \bibnamefont {Weiner}}, \ and\
  \bibinfo {author} {\bibfnamefont {J.~K.}\ \bibnamefont {Thompson}},\ }\href
  {\doibase 10.1103/PhysRevLett.116.093602} {\bibfield  {journal} {\bibinfo
  {journal} {Phys. Rev. Lett.}\ }\textbf {\bibinfo {volume} {116}},\ \bibinfo
  {pages} {093602} (\bibinfo {year} {2016})}\BibitemShut {NoStop}%
\bibitem [{\citenamefont {Martin}\ \emph {et~al.}(2011)\citenamefont {Martin},
  \citenamefont {Meiser}, \citenamefont {Thomsen}, \citenamefont {Ye},\ and\
  \citenamefont {Holland}}]{Martin_2011}%
  \BibitemOpen
  \bibfield  {author} {\bibinfo {author} {\bibfnamefont {M.~J.}\ \bibnamefont
  {Martin}}, \bibinfo {author} {\bibfnamefont {D.}~\bibnamefont {Meiser}},
  \bibinfo {author} {\bibfnamefont {J.~W.}\ \bibnamefont {Thomsen}}, \bibinfo
  {author} {\bibfnamefont {J.}~\bibnamefont {Ye}}, \ and\ \bibinfo {author}
  {\bibfnamefont {M.~J.}\ \bibnamefont {Holland}},\ }\href {\doibase
  10.1103/PhysRevA.84.063813} {\bibfield  {journal} {\bibinfo  {journal} {Phys.
  Rev. A}\ }\textbf {\bibinfo {volume} {84}},\ \bibinfo {pages} {063813}
  (\bibinfo {year} {2011})}\BibitemShut {NoStop}%
\bibitem [{\citenamefont {Christensen}\ \emph {et~al.}(2015)\citenamefont
  {Christensen}, \citenamefont {Henriksen}, \citenamefont {Sch\"affer},
  \citenamefont {Westergaard}, \citenamefont {Tieri}, \citenamefont {Ye},
  \citenamefont {Holland},\ and\ \citenamefont {Thomsen}}]{Christensen_2015}%
  \BibitemOpen
  \bibfield  {author} {\bibinfo {author} {\bibfnamefont {B.~T.~R.}\
  \bibnamefont {Christensen}}, \bibinfo {author} {\bibfnamefont {M.~R.}\
  \bibnamefont {Henriksen}}, \bibinfo {author} {\bibfnamefont {S.~A.}\
  \bibnamefont {Sch\"affer}}, \bibinfo {author} {\bibfnamefont {P.~G.}\
  \bibnamefont {Westergaard}}, \bibinfo {author} {\bibfnamefont
  {D.}~\bibnamefont {Tieri}}, \bibinfo {author} {\bibfnamefont
  {J.}~\bibnamefont {Ye}}, \bibinfo {author} {\bibfnamefont {M.~J.}\
  \bibnamefont {Holland}}, \ and\ \bibinfo {author} {\bibfnamefont {J.~W.}\
  \bibnamefont {Thomsen}},\ }\href {\doibase 10.1103/PhysRevA.92.053820}
  {\bibfield  {journal} {\bibinfo  {journal} {Phys. Rev. A}\ }\textbf {\bibinfo
  {volume} {92}},\ \bibinfo {pages} {053820} (\bibinfo {year}
  {2015})}\BibitemShut {NoStop}%
\bibitem [{SI()}]{SI}%
  \BibitemOpen
  \href@noop {} {\ }\bibinfo {note} {See supplementary information for
  details.}\BibitemShut {Stop}%
\bibitem [{\citenamefont {Yariv}(1991)}]{Yariv1991}%
  \BibitemOpen
  \bibfield  {author} {\bibinfo {author} {\bibfnamefont {A.}~\bibnamefont
  {Yariv}},\ }\href@noop {} {\emph {\bibinfo {title} {Optical electronics}}}\
  (\bibinfo  {publisher} {Saunders College Publ.},\ \bibinfo {year}
  {1991})\BibitemShut {NoStop}%
\bibitem [{\citenamefont {Nicholson}\ \emph {et~al.}(2015)\citenamefont
  {Nicholson}, \citenamefont {Campbell}, \citenamefont {Hutson}, \citenamefont
  {Marti}, \citenamefont {Bloom}, \citenamefont {McNally}, \citenamefont
  {Zhang}, \citenamefont {Barrett}, \citenamefont {Safronova}, \citenamefont
  {Strouse}, \citenamefont {Tew},\ and\ \citenamefont {Ye}}]{Nicholson_2015}%
  \BibitemOpen
  \bibfield  {author} {\bibinfo {author} {\bibfnamefont {T.~L.}\ \bibnamefont
  {Nicholson}}, \bibinfo {author} {\bibfnamefont {S.~L.}\ \bibnamefont
  {Campbell}}, \bibinfo {author} {\bibfnamefont {R.~B.}\ \bibnamefont
  {Hutson}}, \bibinfo {author} {\bibfnamefont {G.~E.}\ \bibnamefont {Marti}},
  \bibinfo {author} {\bibfnamefont {B.~J.}\ \bibnamefont {Bloom}}, \bibinfo
  {author} {\bibfnamefont {R.~L.}\ \bibnamefont {McNally}}, \bibinfo {author}
  {\bibfnamefont {W.}~\bibnamefont {Zhang}}, \bibinfo {author} {\bibfnamefont
  {M.~D.}\ \bibnamefont {Barrett}}, \bibinfo {author} {\bibfnamefont {M.~S.}\
  \bibnamefont {Safronova}}, \bibinfo {author} {\bibfnamefont {G.~F.}\
  \bibnamefont {Strouse}}, \bibinfo {author} {\bibfnamefont {W.~L.}\
  \bibnamefont {Tew}}, \ and\ \bibinfo {author} {\bibfnamefont
  {J.}~\bibnamefont {Ye}},\ }\href {https://doi.org/10.1038/ncomms7896}
  {\bibfield  {journal} {\bibinfo  {journal} {Nature Communications}\ }\textbf
  {\bibinfo {volume} {6}},\ \bibinfo {pages} {6896 EP } (\bibinfo {year}
  {2015})},\ \bibinfo {note} {article}\BibitemShut {NoStop}%
\bibitem [{\citenamefont {Drozdowski}\ \emph {et~al.}(1997)\citenamefont
  {Drozdowski}, \citenamefont {Ignaciuk}, \citenamefont {Kwela},\ and\
  \citenamefont {Heldt}}]{Drozdowski_1997}%
  \BibitemOpen
  \bibfield  {author} {\bibinfo {author} {\bibfnamefont {R.}~\bibnamefont
  {Drozdowski}}, \bibinfo {author} {\bibfnamefont {M.}~\bibnamefont
  {Ignaciuk}}, \bibinfo {author} {\bibfnamefont {J.}~\bibnamefont {Kwela}}, \
  and\ \bibinfo {author} {\bibfnamefont {J.}~\bibnamefont {Heldt}},\ }\href
  {\doibase 10.1007/s004600050300} {\bibfield  {journal} {\bibinfo  {journal}
  {Zeitschrift f{\"u}r Physik D Atoms, Molecules and Clusters}\ }\textbf
  {\bibinfo {volume} {41}},\ \bibinfo {pages} {125} (\bibinfo {year}
  {1997})}\BibitemShut {NoStop}%
\bibitem [{\citenamefont {Zelevinsky}\ \emph {et~al.}(2006)\citenamefont
  {Zelevinsky}, \citenamefont {Boyd}, \citenamefont {Ludlow}, \citenamefont
  {Ido}, \citenamefont {Ye}, \citenamefont {Ciury\l{}o}, \citenamefont
  {Naidon},\ and\ \citenamefont {Julienne}}]{PhysRevLett.96.203201}%
  \BibitemOpen
  \bibfield  {author} {\bibinfo {author} {\bibfnamefont {T.}~\bibnamefont
  {Zelevinsky}}, \bibinfo {author} {\bibfnamefont {M.~M.}\ \bibnamefont
  {Boyd}}, \bibinfo {author} {\bibfnamefont {A.~D.}\ \bibnamefont {Ludlow}},
  \bibinfo {author} {\bibfnamefont {T.}~\bibnamefont {Ido}}, \bibinfo {author}
  {\bibfnamefont {J.}~\bibnamefont {Ye}}, \bibinfo {author} {\bibfnamefont
  {R.}~\bibnamefont {Ciury\l{}o}}, \bibinfo {author} {\bibfnamefont
  {P.}~\bibnamefont {Naidon}}, \ and\ \bibinfo {author} {\bibfnamefont {P.~S.}\
  \bibnamefont {Julienne}},\ }\href {\doibase 10.1103/PhysRevLett.96.203201}
  {\bibfield  {journal} {\bibinfo  {journal} {Phys. Rev. Lett.}\ }\textbf
  {\bibinfo {volume} {96}},\ \bibinfo {pages} {203201} (\bibinfo {year}
  {2006})}\BibitemShut {NoStop}%
\bibitem [{\citenamefont {Lodewyck}\ \emph {et~al.}(2009)\citenamefont
  {Lodewyck}, \citenamefont {Westergaard},\ and\ \citenamefont
  {Lemonde}}]{Lodewyck_2009}%
  \BibitemOpen
  \bibfield  {author} {\bibinfo {author} {\bibfnamefont {J.}~\bibnamefont
  {Lodewyck}}, \bibinfo {author} {\bibfnamefont {P.~G.}\ \bibnamefont
  {Westergaard}}, \ and\ \bibinfo {author} {\bibfnamefont {P.}~\bibnamefont
  {Lemonde}},\ }\href {\doibase 10.1103/PhysRevA.79.061401} {\bibfield
  {journal} {\bibinfo  {journal} {Phys. Rev. A}\ }\textbf {\bibinfo {volume}
  {79}},\ \bibinfo {pages} {061401} (\bibinfo {year} {2009})}\BibitemShut
  {NoStop}%
\bibitem [{\citenamefont {B\'eguin}\ \emph {et~al.}(2014)\citenamefont
  {B\'eguin}, \citenamefont {Bookjans}, \citenamefont {Christensen},
  \citenamefont {S\o{}rensen}, \citenamefont {M\"uller}, \citenamefont
  {Polzik},\ and\ \citenamefont {Appel}}]{Beguin_2014}%
  \BibitemOpen
  \bibfield  {author} {\bibinfo {author} {\bibfnamefont {J.-B.}\ \bibnamefont
  {B\'eguin}}, \bibinfo {author} {\bibfnamefont {E.~M.}\ \bibnamefont
  {Bookjans}}, \bibinfo {author} {\bibfnamefont {S.~L.}\ \bibnamefont
  {Christensen}}, \bibinfo {author} {\bibfnamefont {H.~L.}\ \bibnamefont
  {S\o{}rensen}}, \bibinfo {author} {\bibfnamefont {J.~H.}\ \bibnamefont
  {M\"uller}}, \bibinfo {author} {\bibfnamefont {E.~S.}\ \bibnamefont
  {Polzik}}, \ and\ \bibinfo {author} {\bibfnamefont {J.}~\bibnamefont
  {Appel}},\ }\href {\doibase 10.1103/PhysRevLett.113.263603} {\bibfield
  {journal} {\bibinfo  {journal} {Phys. Rev. Lett.}\ }\textbf {\bibinfo
  {volume} {113}},\ \bibinfo {pages} {263603} (\bibinfo {year}
  {2014})}\BibitemShut {NoStop}%
\bibitem [{\citenamefont {Norcia}\ and\ \citenamefont
  {Thompson}(2016)}]{Norcia_2016}%
  \BibitemOpen
  \bibfield  {author} {\bibinfo {author} {\bibfnamefont {M.~A.}\ \bibnamefont
  {Norcia}}\ and\ \bibinfo {author} {\bibfnamefont {J.~K.}\ \bibnamefont
  {Thompson}},\ }\href {\doibase 10.1103/PhysRevA.93.023804} {\bibfield
  {journal} {\bibinfo  {journal} {Phys. Rev. A}\ }\textbf {\bibinfo {volume}
  {93}},\ \bibinfo {pages} {023804} (\bibinfo {year} {2016})}\BibitemShut
  {NoStop}%
\bibitem [{\citenamefont {Vallet}\ \emph {et~al.}(2017)\citenamefont {Vallet},
  \citenamefont {Bookjans}, \citenamefont {Eismann}, \citenamefont {Bilicki},
  \citenamefont {Targat},\ and\ \citenamefont {Lodewyck}}]{Vallet_2017}%
  \BibitemOpen
  \bibfield  {author} {\bibinfo {author} {\bibfnamefont {G.}~\bibnamefont
  {Vallet}}, \bibinfo {author} {\bibfnamefont {E.}~\bibnamefont {Bookjans}},
  \bibinfo {author} {\bibfnamefont {U.}~\bibnamefont {Eismann}}, \bibinfo
  {author} {\bibfnamefont {S.}~\bibnamefont {Bilicki}}, \bibinfo {author}
  {\bibfnamefont {R.~L.}\ \bibnamefont {Targat}}, \ and\ \bibinfo {author}
  {\bibfnamefont {J.}~\bibnamefont {Lodewyck}},\ }\href {\doibase
  10.1088/1367-2630/aa7c84} {\bibfield  {journal} {\bibinfo  {journal} {New
  Journal of Physics}\ }\textbf {\bibinfo {volume} {19}},\ \bibinfo {pages}
  {083002} (\bibinfo {year} {2017})}\BibitemShut {NoStop}%
\bibitem [{\citenamefont {Taichenachev}\ \emph {et~al.}(2006)\citenamefont
  {Taichenachev}, \citenamefont {Yudin}, \citenamefont {Oates}, \citenamefont
  {Hoyt}, \citenamefont {Barber},\ and\ \citenamefont
  {Hollberg}}]{Taichenachev_2006}%
  \BibitemOpen
  \bibfield  {author} {\bibinfo {author} {\bibfnamefont {A.~V.}\ \bibnamefont
  {Taichenachev}}, \bibinfo {author} {\bibfnamefont {V.~I.}\ \bibnamefont
  {Yudin}}, \bibinfo {author} {\bibfnamefont {C.~W.}\ \bibnamefont {Oates}},
  \bibinfo {author} {\bibfnamefont {C.~W.}\ \bibnamefont {Hoyt}}, \bibinfo
  {author} {\bibfnamefont {Z.~W.}\ \bibnamefont {Barber}}, \ and\ \bibinfo
  {author} {\bibfnamefont {L.}~\bibnamefont {Hollberg}},\ }\href {\doibase
  10.1103/PhysRevLett.96.083001} {\bibfield  {journal} {\bibinfo  {journal}
  {Phys. Rev. Lett.}\ }\textbf {\bibinfo {volume} {96}},\ \bibinfo {pages}
  {083001} (\bibinfo {year} {2006})}\BibitemShut {NoStop}%
\bibitem [{\citenamefont {Origlia}\ \emph {et~al.}(2018)\citenamefont
  {Origlia}, \citenamefont {Pramod}, \citenamefont {Schiller}, \citenamefont
  {Singh}, \citenamefont {Bongs}, \citenamefont {Schwarz}, \citenamefont
  {Al-Masoudi}, \citenamefont {D\"orscher}, \citenamefont {Herbers},
  \citenamefont {H\"afner}, \citenamefont {Sterr},\ and\ \citenamefont
  {Lisdat}}]{Origlia_2018}%
  \BibitemOpen
  \bibfield  {author} {\bibinfo {author} {\bibfnamefont {S.}~\bibnamefont
  {Origlia}}, \bibinfo {author} {\bibfnamefont {M.~S.}\ \bibnamefont {Pramod}},
  \bibinfo {author} {\bibfnamefont {S.}~\bibnamefont {Schiller}}, \bibinfo
  {author} {\bibfnamefont {Y.}~\bibnamefont {Singh}}, \bibinfo {author}
  {\bibfnamefont {K.}~\bibnamefont {Bongs}}, \bibinfo {author} {\bibfnamefont
  {R.}~\bibnamefont {Schwarz}}, \bibinfo {author} {\bibfnamefont
  {A.}~\bibnamefont {Al-Masoudi}}, \bibinfo {author} {\bibfnamefont
  {S.}~\bibnamefont {D\"orscher}}, \bibinfo {author} {\bibfnamefont
  {S.}~\bibnamefont {Herbers}}, \bibinfo {author} {\bibfnamefont
  {S.}~\bibnamefont {H\"afner}}, \bibinfo {author} {\bibfnamefont
  {U.}~\bibnamefont {Sterr}}, \ and\ \bibinfo {author} {\bibfnamefont
  {C.}~\bibnamefont {Lisdat}},\ }\href {\doibase 10.1103/PhysRevA.98.053443}
  {\bibfield  {journal} {\bibinfo  {journal} {Phys. Rev. A}\ }\textbf {\bibinfo
  {volume} {98}},\ \bibinfo {pages} {053443} (\bibinfo {year}
  {2018})}\BibitemShut {NoStop}%
\bibitem [{\citenamefont {Peik}\ and\ \citenamefont {Tamm}(2003)}]{Peik_2003}%
  \BibitemOpen
  \bibfield  {author} {\bibinfo {author} {\bibfnamefont {E.}~\bibnamefont
  {Peik}}\ and\ \bibinfo {author} {\bibfnamefont {C.}~\bibnamefont {Tamm}},\
  }\href {\doibase 10.1209/epl/i2003-00210-x} {\bibfield  {journal} {\bibinfo
  {journal} {Europhysics Letters ({EPL})}\ }\textbf {\bibinfo {volume} {61}},\
  \bibinfo {pages} {181} (\bibinfo {year} {2003})}\BibitemShut {NoStop}%
\bibitem [{\citenamefont {von~der Wense}\ \emph {et~al.}(2016)\citenamefont
  {von~der Wense}, \citenamefont {Seiferle}, \citenamefont {Laatiaoui},
  \citenamefont {Neumayr}, \citenamefont {Maier}, \citenamefont {Wirth},
  \citenamefont {Mokry}, \citenamefont {Runke}, \citenamefont {Eberhardt},
  \citenamefont {D{\"u}llmann}, \citenamefont {Trautmann},\ and\ \citenamefont
  {Thirolf}}]{vonderWense_2016}%
  \BibitemOpen
  \bibfield  {author} {\bibinfo {author} {\bibfnamefont {L.}~\bibnamefont
  {von~der Wense}}, \bibinfo {author} {\bibfnamefont {B.}~\bibnamefont
  {Seiferle}}, \bibinfo {author} {\bibfnamefont {M.}~\bibnamefont {Laatiaoui}},
  \bibinfo {author} {\bibfnamefont {J.~B.}\ \bibnamefont {Neumayr}}, \bibinfo
  {author} {\bibfnamefont {H.-J.}\ \bibnamefont {Maier}}, \bibinfo {author}
  {\bibfnamefont {H.-F.}\ \bibnamefont {Wirth}}, \bibinfo {author}
  {\bibfnamefont {C.}~\bibnamefont {Mokry}}, \bibinfo {author} {\bibfnamefont
  {J.}~\bibnamefont {Runke}}, \bibinfo {author} {\bibfnamefont
  {K.}~\bibnamefont {Eberhardt}}, \bibinfo {author} {\bibfnamefont {C.~E.}\
  \bibnamefont {D{\"u}llmann}}, \bibinfo {author} {\bibfnamefont {N.~G.}\
  \bibnamefont {Trautmann}}, \ and\ \bibinfo {author} {\bibfnamefont {P.~G.}\
  \bibnamefont {Thirolf}},\ }\href {https://doi.org/10.1038/nature17669}
  {\bibfield  {journal} {\bibinfo  {journal} {Nature}\ }\textbf {\bibinfo
  {volume} {533}},\ \bibinfo {pages} {47 EP } (\bibinfo {year}
  {2016})}\BibitemShut {NoStop}%
\bibitem [{\citenamefont {Seiferle}\ \emph {et~al.}(2019)\citenamefont
  {Seiferle}, \citenamefont {von~der Wense}, \citenamefont {Bilous},
  \citenamefont {Amersdorffer}, \citenamefont {Lemell}, \citenamefont
  {Libisch}, \citenamefont {Stellmer}, \citenamefont {Schumm}, \citenamefont
  {D{\"u}llmann}, \citenamefont {P{\'a}lffy},\ and\ \citenamefont
  {Thirolf}}]{Seiferle_2019}%
  \BibitemOpen
  \bibfield  {author} {\bibinfo {author} {\bibfnamefont {B.}~\bibnamefont
  {Seiferle}}, \bibinfo {author} {\bibfnamefont {L.}~\bibnamefont {von~der
  Wense}}, \bibinfo {author} {\bibfnamefont {P.~V.}\ \bibnamefont {Bilous}},
  \bibinfo {author} {\bibfnamefont {I.}~\bibnamefont {Amersdorffer}}, \bibinfo
  {author} {\bibfnamefont {C.}~\bibnamefont {Lemell}}, \bibinfo {author}
  {\bibfnamefont {F.}~\bibnamefont {Libisch}}, \bibinfo {author} {\bibfnamefont
  {S.}~\bibnamefont {Stellmer}}, \bibinfo {author} {\bibfnamefont
  {T.}~\bibnamefont {Schumm}}, \bibinfo {author} {\bibfnamefont {C.~E.}\
  \bibnamefont {D{\"u}llmann}}, \bibinfo {author} {\bibfnamefont
  {A.}~\bibnamefont {P{\'a}lffy}}, \ and\ \bibinfo {author} {\bibfnamefont
  {P.~G.}\ \bibnamefont {Thirolf}},\ }\href {\doibase
  10.1038/s41586-019-1533-4} {\bibfield  {journal} {\bibinfo  {journal}
  {Nature}\ }\textbf {\bibinfo {volume} {573}},\ \bibinfo {pages} {243}
  (\bibinfo {year} {2019})}\BibitemShut {NoStop}%
\bibitem [{\citenamefont {Gardiner}\ and\ \citenamefont
  {Collett}(1985)}]{Gardiner_1985}%
  \BibitemOpen
  \bibfield  {author} {\bibinfo {author} {\bibfnamefont {C.~W.}\ \bibnamefont
  {Gardiner}}\ and\ \bibinfo {author} {\bibfnamefont {M.~J.}\ \bibnamefont
  {Collett}},\ }\href {\doibase 10.1103/PhysRevA.31.3761} {\bibfield  {journal}
  {\bibinfo  {journal} {Phys. Rev. A}\ }\textbf {\bibinfo {volume} {31}},\
  \bibinfo {pages} {3761} (\bibinfo {year} {1985})}\BibitemShut {NoStop}%
\bibitem [{\citenamefont {Barberena}\ \emph {et~al.}(2019)\citenamefont
  {Barberena}, \citenamefont {Lewis-Swan}, \citenamefont {Thompson},\ and\
  \citenamefont {Rey}}]{Barberena_2018}%
  \BibitemOpen
  \bibfield  {author} {\bibinfo {author} {\bibfnamefont {D.}~\bibnamefont
  {Barberena}}, \bibinfo {author} {\bibfnamefont {R.~J.}\ \bibnamefont
  {Lewis-Swan}}, \bibinfo {author} {\bibfnamefont {J.~K.}\ \bibnamefont
  {Thompson}}, \ and\ \bibinfo {author} {\bibfnamefont {A.~M.}\ \bibnamefont
  {Rey}},\ }\href {\doibase 10.1103/PhysRevA.99.053411} {\bibfield  {journal}
  {\bibinfo  {journal} {Phys. Rev. A}\ }\textbf {\bibinfo {volume} {99}},\
  \bibinfo {pages} {053411} (\bibinfo {year} {2019})}\BibitemShut {NoStop}%
\bibitem [{\citenamefont {Muniz}\ \emph {et~al.}(2018)\citenamefont {Muniz},
  \citenamefont {Norcia}, \citenamefont {Cline},\ and\ \citenamefont
  {Thompson}}]{Muniz_2018}%
  \BibitemOpen
  \bibfield  {author} {\bibinfo {author} {\bibfnamefont {J.~A.}\ \bibnamefont
  {Muniz}}, \bibinfo {author} {\bibfnamefont {M.~A.}\ \bibnamefont {Norcia}},
  \bibinfo {author} {\bibfnamefont {J.~R.}\ \bibnamefont {Cline}}, \ and\
  \bibinfo {author} {\bibfnamefont {J.~K.}\ \bibnamefont {Thompson}},\
  }\href@noop {} {\bibfield  {journal} {\bibinfo  {journal} {arXiv preprint
  arXiv:1806.00838}\ } (\bibinfo {year} {2018})}\BibitemShut {NoStop}%
\bibitem [{\citenamefont {Sansonetti}\ and\ \citenamefont
  {Nave}(2010)}]{Sansonetti_2010}%
  \BibitemOpen
  \bibfield  {author} {\bibinfo {author} {\bibfnamefont {J.~E.}\ \bibnamefont
  {Sansonetti}}\ and\ \bibinfo {author} {\bibfnamefont {G.}~\bibnamefont
  {Nave}},\ }\href {\doibase 10.1063/1.3449176} {\bibfield  {journal} {\bibinfo
   {journal} {Journal of Physical and Chemical Reference Data}\ }\textbf
  {\bibinfo {volume} {39}},\ \bibinfo {pages} {033103} (\bibinfo {year}
  {2010})}\BibitemShut {NoStop}%
\bibitem [{\citenamefont {Scully}\ and\ \citenamefont
  {Zubairy}(1997)}]{scully1997quantum}%
  \BibitemOpen
  \bibfield  {author} {\bibinfo {author} {\bibfnamefont {M.}~\bibnamefont
  {Scully}}\ and\ \bibinfo {author} {\bibfnamefont {M.}~\bibnamefont
  {Zubairy}},\ }\href {https://books.google.com/books?id=20ISsQCKKmQC} {\emph
  {\bibinfo {title} {Quantum Optics}}}\ (\bibinfo  {publisher} {Cambridge
  University Press},\ \bibinfo {year} {1997})\BibitemShut {NoStop}%
\bibitem [{\citenamefont {Weisskopf}\ and\ \citenamefont
  {Wigner}(1930)}]{Weisskopf1930}%
  \BibitemOpen
  \bibfield  {author} {\bibinfo {author} {\bibfnamefont {V.}~\bibnamefont
  {Weisskopf}}\ and\ \bibinfo {author} {\bibfnamefont {E.}~\bibnamefont
  {Wigner}},\ }\href {\doibase 10.1007/BF01336768} {\bibfield  {journal}
  {\bibinfo  {journal} {Zeitschrift f{\"u}r Physik}\ }\textbf {\bibinfo
  {volume} {63}},\ \bibinfo {pages} {54} (\bibinfo {year} {1930})}\BibitemShut
  {NoStop}%
\end{thebibliography}%

\newpage
\clearpage

\begin{center}
  \textbf{\large Supplementary information}\\[.2cm]
  \vspace{0.2cm}
\end{center}

\setcounter{figure}{0}
\setcounter{equation}{0}

\renewcommand{\theequation}{S\arabic{equation}}
\renewcommand{\thefigure}{S\arabic{figure}}

\title{Supplementary Material: Determining the natural linewidth of the $^{87}$Sr millihertz clock transition with 30 $\mu$Hz resolution}


\maketitle

This document contains several additional details about our experiment. We derive the atomic-like phase shift, both on the $^3$P$_0$ and $^3$P$_1$ transitions, as well as the scaling of the excitation fraction for a given measurement window. We then describe the experimental setup as well as the different frequency components that are used to evaluate the phase shifts on both transitions. In order to present a precise measurement of the clock transition natural linewidth we present a detailed analysis of the systematic effects that need to be taken into account in this measurement, as well as a detailed description  of the low power measurement presented in the main text. We offer a detailed description of how we extract $(g_0/g_1)^2$ from our dispersive measurements. Finally, we include a discussion of any unknown $N$-dependent effects on our measurements.

\section{Light-matter interactions \label{sec:sec1}}
In this section we are going to explore the light-matter interactions and the phase shift acquired by a probe detuned from the atomic transition.

\subsection{Optical Bloch equations \label{sec:obes}}

Following the conventions adopted in Ref.~\cite{Chen_2014}, we consider $N$ two-level atoms, described by the usual Pauli operators $\hat{\sigma}^i_j$, with $i= x,y ,z$ and $j =1,...,N$, equally coupled to a single cavity mode with annihilation operator $\hat{c}$. In this case the Jaynes-Cummings Hamiltonian \cite{Jaynes_1963,Kimble_1998} in the atomic frame is:
\begin{equation}
    H = \hbar\delta_c \hat{c}^\dagger\hat{c}+\hbar g (\hat{c}\hat{J}^++\hat{c}^\dagger\hat{J}^-),
    \label{eq:rotwaveNH}
\end{equation}
where $\hat{J}^{\pm} = \sum_{i=1}^N \hat{\sigma}^\pm_i$ are the collective raising and lowering operators for the atoms, $\delta_c = \omega_c-\omega_a$ is the cavity detuning from the atomic transition ($\omega_a$), and $2g$ is the single photon Rabi frequency. We can further define $\hat{J}^z = \frac{1}{2} \sum_{i=1}^N \hat{\sigma}^z_j$, in order to have a closed angular momentum algebra.
 
If we add a cavity drive $c_i(t)$ at frequency $\omega_p$, where $c_i$ has units of $\sqrt{\textrm{photons/s}}$, we can use the input-output formalism \cite{Chen_2014,Gardiner_1985} to write the Heisenberg-Langevin equations of motion for the cavity and atomic mean operators ($O = \langle \hat{O}\rangle$), on the atomic frame, as follows:
\begin{eqnarray}
    &\dot{c} &= -\left(i\delta_c +\frac{\kappa}{2}\right)c-igJ^-+\sqrt{\kappa_m}c_i(t);\nonumber\\
    &\dot{J}^- &= i2gJ^zc-\gamma_\perp J^-;\nonumber\\
    &\dot{J}^z &= -ig\left(cJ^+-c^*J^-\right) - \gamma\left(\frac{N}{2}+J^z\right),
    \label{eq:fulleq}
\end{eqnarray}
where we have included the spontaneous emission rate $\gamma$, a transverse dephasing term $\gamma_\perp$ and cavity losses characterized by its linewidth $\kappa$. The single mirror transmission is characterized by $\sqrt{\kappa_m}$ ($\kappa=2\kappa_m$). In the rotating frame, the incident cavity field is $c_i(t) = c_{i0}e^{-i\delta_p t}$, with $\delta_p = \omega_p - \omega_a$ the drive detuning from the optical transition.

\subsection{Steady-state solution}
In the presence of a driving field with detuning $\delta_p$ ($c_i(t) = c_{i0}e^{-i\delta_p t}$), the steady state solution is characterized by observables of the form $J^- = \tilde{J}^-e^{-i\delta_p t}$ and $c = \tilde{c}e^{-i\delta_p t}$. Working in the weak probe approximation, such that all the atoms remain in the ground state, the transmitted field ($\tilde{c}_t = \sqrt{\kappa_m}\tilde{c}$) satisfies $\tilde{c}_t = T(\delta_p) c_{i0}$ for a transfer function $T(\delta_p)$ given by:
\begin{equation}
    T(\delta_p) = \frac{1}{1-i\left(\frac{\delta_p-\delta_c}{\kappa/2}\right)+\frac{NC\gamma/2}{\gamma_\perp-i\delta_p}},
    \label{eq:ssc}
\end{equation}
and for the collective atomic coherence $\tilde{J}^-$ is:
\begin{equation}
    \tilde{J}^- = \frac{igN \tilde{c}}{(i\delta_p-\gamma_\perp)}.
    \label{eq:ssJm}
\end{equation}
Here we defined the cooperativity parameter as $C = (2g)^2/(\gamma\kappa)$

The phase $\delta\varphi_t(\delta_p)$ acquired by the transmitted field $(\tilde{c}_t = |T(\delta_p)|e^{i\delta\varphi_t(\delta_p)}c_{i0})$ satisfies
\begin{equation}
    \tan(\delta\varphi_t(\delta_p)) = \frac{-2\left(\delta_p g^2 N + (\delta_c-\delta_p)(\delta_p^2+\gamma_\perp^2)\right)}{\gamma_\perp 2 g^2 N+\kappa(\gamma_\perp^2+\delta_p^2)}
    \label{eq:phitss_1p}
\end{equation}

Under the following hierarchy, realized in our system at the clock transition \cite{Norcia_SS_2018}:
\begin{eqnarray}
    &\sqrt{N}g &\sim \delta_p\ll \kappa ;\nonumber\\
    &\gamma_\perp &\ll \delta_p ; \nonumber\\
    &|\delta_c - \delta_p| &\sim \delta_p,\nonumber\\  
\end{eqnarray}
we can approximate 
\begin{equation}
    \tan(\delta\varphi_t(\delta_p)) \approx -\frac{2g^2 N}{\kappa\delta_p} - \frac{2(\delta_c-\delta_p)}{\kappa},
    \label{eq:phitapprox_1p}
\end{equation}
which can be rewritten as 
\begin{equation}
    \tan(\delta\varphi_t(\delta_p)) \approx -\frac{NC\gamma}{2\delta_p}  -\frac{2(\delta_c-\delta_p)}{\kappa}
    \label{eq:phitapprox_1p}
\end{equation}

The first therm on the r.h.s. of Eq.~\ref{eq:phitapprox_1p} is the atomic like phase shift, while the other term is a cavity-like phase shift, independent of the atoms. For our measurements, we are interested in the first term as it encodes the collective interactions ($Ng^2$), while the last term can be measured independently and subtracted, by for example measuring the phase shift of an identical tone one free spectral range away. For our system, the atomic-like phase shift ($NC\gamma/(2\delta_p)$) is typically 30~mrad, while the on-resonance cavity-like phase shift ($2\delta_p/\kappa$) is typically 15~mrad, for $\delta_p/(2\pi) = 1$~kHz. 

As described in the main text, we take the difference between the pair-wise phases on two consecutive cavity modes. A detailed description is given in the next section. The total phase shift, defined as $\Delta\varphi_0$ in the main text, becomes
\begin{eqnarray}
    &\Delta\varphi_{0} = - \frac{NC\gamma}{\delta_p}\left(1 - 4\frac{\delta_c^2}{\kappa^2}+4\frac{\delta_p^2}{\kappa^2}+\frac{(NC\gamma)^2}{12\delta_p^2} - \frac{NC\gamma}{\kappa} \right).
    \label{eq:phase_error_sausage}
\end{eqnarray}
Higher order terms in Eq.~\ref{eq:phase_error_sausage} show higher order corrections on the phase shift, that will be consider as corrections (see systematic section later).

Finally we want to note that at short times, there is an initial transient ringing, of duration $NC\gamma$, associated with the homogeneous solution of the optical Bloch equations, that in our system lasts about 2~ms.

\subsection{Excitation fraction}

If we break the weak probe power approximation, and allow the inversion to change during the probing, using Eq.~\ref{eq:fulleq} and Eq.~\ref{eq:ssJm} we find that the steady state excitation fraction is 
\begin{equation}
    N_e/N \approx \frac{\gamma_\perp C}{\delta_p^2+\gamma_\perp^2}|c_{i0}|^2,
\end{equation}
showing the characteristic $1/\delta_p^2$ dependence of these dispersive measurements.

In the absence of dephasing ($\gamma_\perp = \gamma/2$), the steady state is reached on a $1/\gamma$ timescale. Therefore, for a measurement window $T_m$, the fraction of atoms in the excited state at the end of the measurement will be 
\begin{equation}
    \left(N_e/N\right)_{T_m}  \approx \gamma T_m \frac{\gamma C}{2\delta_p^2}|c_{i0}|^2.
    \label{eq:NeoverNss}
\end{equation}

We will adjust the power, i.e. $|c_{i0}|^2$, such that $ \left(N_e/N\right)_{T_m}\ll 1$ over the measurement window. Ideally, this constrain, and the final quantum efficiency, will limit the resolution of the atom counting as a non destructive process. 

Furthermore, this suggests that even if the inversion is changing because of a different process rather than the excitation caused by the tones, the atomic-like phase shift $\Delta\varphi_0$ can be used to dynamically track the atomic inversion $J_z(t)$ \cite{Barberena_2018}.

\section{Experiment details}\label{sec:sec2}

\begin{figure*}[!htb]
\includegraphics[width=7in]{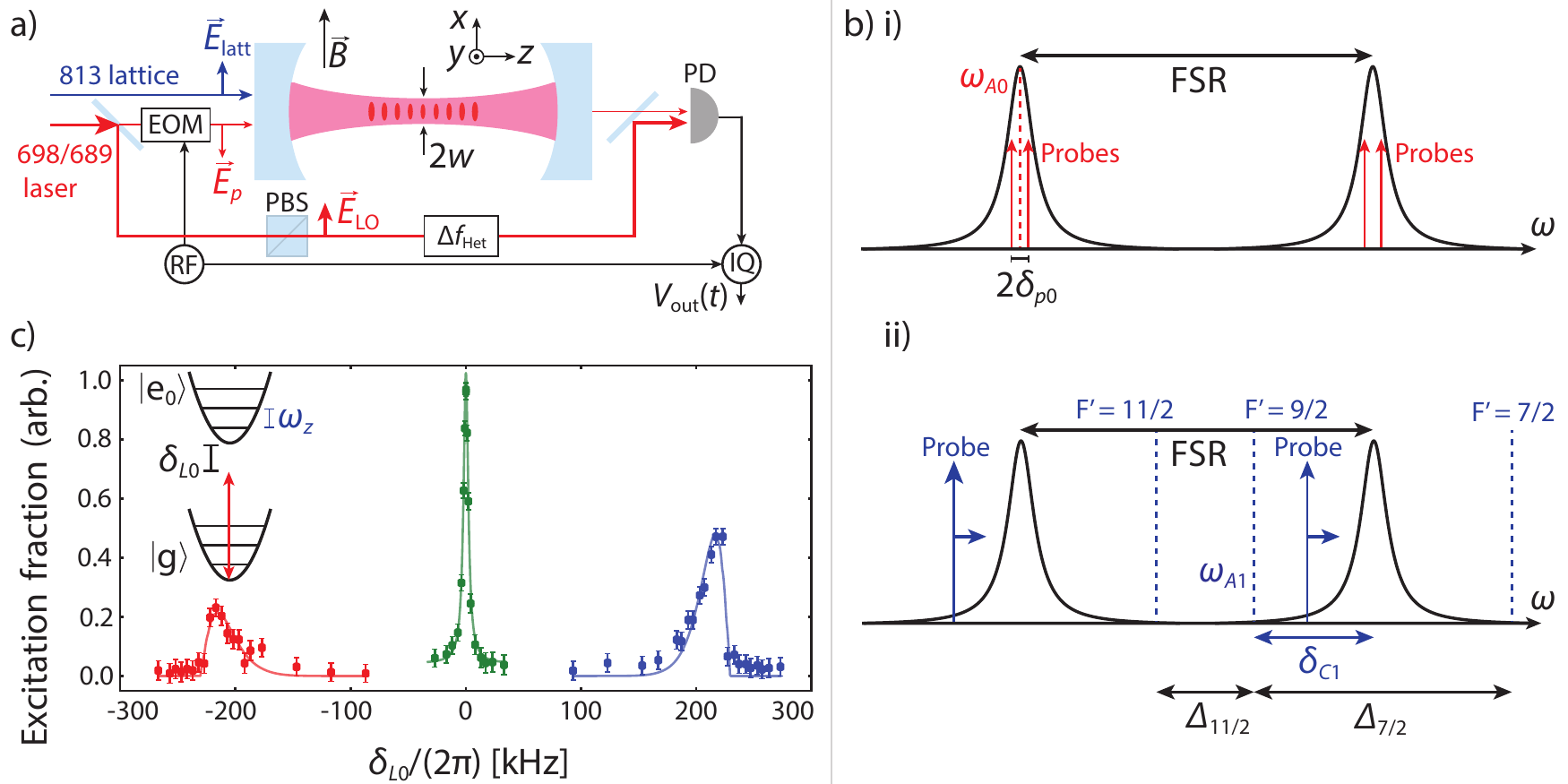}
\caption{(a) Experimental set up. An 813~nm optical lattice confines atoms at 14(1)~$\mu$K. A 689/698~nm laser addresses the atomic transitions between the ground $^1$S$_0$ state and the $^3$P$_1$ and $^3$P$_0$ states, respectively. The different frequency tones to generate the probes, as described in the main text, are generated by an in-fiber EOM and RF function generators operating at the different necessary frequencies. Most of the light constitutes an optical local oscillator. The local oscillator is frequency shifted by $\Delta f_{\textrm{Het}}$, polarization filtered along $\hat{x}$ with a polarization beam splitter, and is beated with the transmitted probes into a photo-diode. The photo-current is properly demodulated by each RF frequency and recorded as a different voltage $V_{\textrm{out}}(t)$. (b) The probe tones used to address the clock transition are shown in (i), while the tones and hyperfine levels for the excited $^3$P$_1$ manifold are shown in (ii). The detuning $\delta_{C1}$ is defined with respect to the $F=9/2$ manifold. Frequencies not to scale. (c) Vibrational spectroscopy on the clock transition. We scan the detuning $\delta_{L0}$ of a strong 698~nm probe and record the excitation fraction. We follow the procedure in Ref.~\cite{Blatt_2009} to fit the occupation number, temperature and trap frequency.}
\label{fig:SI1}
\end{figure*}

A detailed scheme of the experimental setup is given in Fig.~\ref{fig:SI1}(a) and it has been already detailed in Ref.~\cite{Norcia_SR_2016,Norcia_2016,Norcia_SRFreq_2018}. Atoms are loaded into a 813~nm near magic wavelength intracavity optical lattice, following cooling and trapping on the 7.5~kHz transition ($^1$S$_0\to \ ^3$P$_1$) at 689~nm \cite{Muniz_2018}. The lattice is near magical for the $^1$S$_0\to \ ^3$P$_0$ millihertz clock transition at 698~nm, with a detuning of maximally $\sim2$~GHz from the magic wavelength (half free spectral range). As in the main text, from here on all the 0(1) subindices refer to quantities defined on the $^1$S$_0\to \ ^3$P$_0 ~(^3$P$_1)$ transition. 

For both the clock and the 7.5~kHz transitions, a small fraction of the light is coupled into the same in-fiber electro-optical  phase modulator and sent to the optical resonator. Which light enters the modulator at a given time is controlled by acousto-optic modulators prior to the phase modulator. Both probes are polarized along the quantization direction $\hat{z}$, established by a static magnetic field $\vec{\textit{B}}$. The different probe tones are driven by different RF sources, indicated by a generic RF generator in Fig.~\ref{fig:SI1}(a). The different cavity modes probed for each transition are shown in Fig.~\ref{fig:SI1}(b), which also indicates the relative frequency difference of the hyperfine states of the $^3$P$_1$ excited state. Further details for the individual transitions are provided below.

\subsection{Probing $^3$P$_0$}
The clock transition $^3$P$_0$ is probed with two tones nominally detuned by $\pm \delta_{p0} \pm 1$ kHz from atomic resonance (see Fig.~\ref{fig:SI1}(b)(i). An identical pair of probe tones offset by one free spectral range ($\Delta_{\textrm{FSR},0}$) probes the phase shifts induced by the empty cavity resonance. Since this second pair of probe tones is far from resonance, we can use much more power in these probe tones to reduce their photon shot noise contributions to the final signal to noise.

The probe tones are typically applied for 20~ms to 40~ms. We remove the first interval of width $\textrm{T}_\textrm{H} \approx 5$~ms, where the initial transient is large. Using two probe tones near the atomic transition provides first order insensitivity to laser frequency uncertainty and noise relative to the atomic transition frequency.  Using the full four-tone probe technique provides cavity and laser frequency noise rejection, as well as automatic rejection of the empty cavity phase shift.

A large fraction of the laser light is picked off prior to the phase modulator and frequency shifted by $\Delta f_{LO}$ to provide a local oscillator $\vec{\textit{E}}_{LO}$, linearly polarized along $\hat{x}$. The LO is frequency shifted by $\Delta f_{\textrm{LO}} = 20$~kHz, polarization filtered by a polarization beam splitter that transmits light polarized along $\hat{x}$, and combined with the transmitted tones onto a fast photo-detector (PD), forming a heterodyne beat-note [Fig.~\ref{fig:SI1}(a)] with photon shot-noise limited sensitivity. 

After amplification, both pairs of probe tones are separately IQ demodulated to a base band of 20~kHz using the same RF sources used to drive the phase modulator. The demodulated IQ voltage signals $V_{\mathrm{IQ}}(t)$ are digitally sampled into the computer and then  fitted to extract the difference in the phases for a single pair. For example, the near-resonant pair of tones that probe the clock transition are demodulated to $20\pm1$~kHz, for $\delta_{p0}/(2\pi)=1$~kHz. 

After computing the four individual phases $\delta\varphi_0(\pm\delta_p)$ and $\delta\varphi_0(\Delta_{\textrm{FSR},0}\pm\delta_p)$, we compute the appropriate pair-wise differences to arrive at an estimate of the atomic phase shift $\Delta\varphi_{0} = \left(\delta\varphi_0(\delta_p)-\delta\varphi_0(-\delta_p)\right)-\left(\delta\varphi_0(\Delta_{\textrm{FSR},0}+\delta_p)-\delta\varphi_0(\Delta_{\textrm{FSR},0}-\delta_p) \right)$ that for the small angles here is related to the atomic linewidth $\gamma_0$ by $\Delta\varphi_0= -NC_0\gamma_0/\delta_{p0}$.

\subsection{Probing $^3$P$_1$}
The 7.5~kHz transition $^1$S$_0$ to $^3$P$_1$ is probed using a total of two tones separated by one free spectral range (see Fig.~\ref{fig:SI1}(b)(ii). Both tones are linearly swept in frequency at the same time, and as described above, the transmitted probe light is heterodyne detected, amplified, the individual probe tones are IQ demodulated, and digitally sampled into the computer. The IQ data is fitted to extract the resonance frequency $\delta\omega_1$ of each cavity mode up to a common offset, and an estimate of the differential frequency shift between the two cavity modes $\Delta \omega_1$ is then computed.

\subsection{Axial Sideband Spectroscopy}
We determine the mean occupation number, trap frequency and temperature via axial sideband spectroscopy using a cavity probe near resonance with the clock transition, following the approach of Ref.~\cite{Blatt_2009}. The fraction of atoms excited by this probe as we scan its frequency is shown in Fig.~\ref{fig:SI1}(c). We consistently measure the axial trap frequency to be $\omega_z/(2\pi) = 230(1)$~kHz, the temperature to be $T = 14(1)~\mu$K and the mean occupation number $\overline{n}_z$ = 0.9(1). The Lamb-Dicke parameter computed for the $^3$P$_0$ transition is $\eta_0 = 0.1425(6)$, and $\eta_1 = 0.1443(6)$ for the $^3$P$_1$ transition. Remarkably, the measurement is dispersive ($\delta_{p0}\gg NC_0\gamma_0$), but addresses the carrier transition ($\delta_{p0}\ll \omega_z$), which is achievable for this set of ultra-narrow optical transitions.

\subsection{Cavity Geometry Determination}\label{sec:CavityGeometry}
\begin{table*}[htb!]
\caption{\label{tab:CavityAtomParams} Summary of Cavity and Atomic Parameters.}
\begin{ruledtabular}
\begin{tabular}{llll}
\textrm{Description}&
\textrm{Symbol}&
\textrm{Value}
&\textrm{Unit}\\
\colrule
Probe wavelength 0 - $^3$P$_0$ probe \cite{Sansonetti_2010} & $\lambda_0$ & 698.4457 & nm\\
Probe wavelength 1 - $^3$P$_1$ probe \cite{Sansonetti_2010} &$\lambda_1$ & 689.4485 & nm\\
Trap wavelength & $\lambda_{trap}$ & 813.4257(2) & nm \\
Cavity FWHM 0 for probe polarized along $\hat{x}$ & $\kappa_0/2\pi$ & 140.9(3) & kHz\\
Cavity FWHM 1 for probe polarized along $\hat{x}$ & $\kappa_1/2\pi$ & 153.0(4) & kHz\\
Mode waist 0 & $\textrm{w}_0$ & 73.85(7) & $\mu$m\\
Mode waist 1 &$\textrm{w}_1$ & 73.37(7) & $\mu$m\\
Lattice waist & $\textrm{w}_{trap}$ & 79.7(1) & $\mu$m\\
Rayleigh Range 0,1 & $z_R$ & 2.453(5) & cm\\
Free Spectral Range 0 & $\Delta_{\textrm{FSR},0}/2\pi$ & 3.71461(3) & GHz\\
Free Spectral Range 1 & $\Delta_{\textrm{FSR},1}/2\pi$ & 3.71459(2) & GHz\\
Cavity Length 0 & $\textrm{L}_0$ & 4.03532(3) & cm\\
Cavity Length 1 & $\textrm{L}_1$ & 4.03534(2) & cm\\
Axial trap frequency on axis & $\omega_z/2\pi$ & 230(1) & kHz\\
Radial trap frequency & $\omega_r/2\pi$ & 528(2) & Hz\\
Axial temperature & $T_z$ & 14(1) & $\mu$K\\
Radial temperature & $T_r$ & 12(2) & $\mu$K \\
RMS thermal radius & $\sigma_r$ & 14(1) & $\mu$m \\
RMS longitudinal cloud radius & $\sigma_{long}$ & 0.30(5) & mm\\
Axial vibrational quanta & $\bar{n}_z$ &0.9(1) & \\
Axial Lamb-Dicke parameter 0 & $\eta_0$ & 0.1425(6) & \\
Axial Lamb-Dicke parameter 1 & $\eta_1$ & 0.1443(6) & \\
Cavity detuning 0 & $\delta_{C0}/2\pi$ & 0(10)& kHz\\
Cavity detuning 1 & $\delta_{C1}/2\pi$ & 277.5(8) & MHz\\
Birefringent cavity mode full splitting 0 & $\delta_{b0}/2\pi$ & 23(3) & kHz\\
Birefringent cavity mode full splitting 1 & $\delta_{b1}/2\pi$ & 24(3) & kHz\\
Birefringent cavity mode polar angle on Poincar\'e sphere (Jones vector) & $\theta_b$ & 30(2) & \text{deg}\\
Birefringent cavity mode azimuthal angle on Poincar\'e sphere (Jones vector) & $\varphi_b$ & $\pm 14(4)$ & \text{deg}\\
$^3$P$_1$ linewidth \cite{Nicholson_2015} & $\gamma_1/2\pi$ & 7.48(1) & kHz\\
$^3$P$_1$ $F'= 11/2$ detuning from $9/2$ \cite{Sansonetti_2010} & $\Delta_{11/2}/2\pi $ & -1463.15(6) & MHz\\ 
$^3$P$_1$ $F'= 7/2$ detuning from $9/2$ \cite{Sansonetti_2010} & $\Delta_{7/2}/2\pi $ & 1130.26(6)& MHz\\ 
\end{tabular}
\end{ruledtabular}
\end{table*}

In order to set the detuning of the cavity to the $F=9/2$ $^3$P$_1$ manifold ($\delta_{C1}$), while keeping the cavity on resonance with the clock transition, we heat up our ceramic cavity spacer with a set of lights by about 10~K from room temperature. At these settings, we measure a free spectral range (FSR) of $\Delta_{\textrm{FSR},0} = 2\pi\times3.71461(3)$~GHz for the clock transition and $\Delta_{\textrm{FSR},1} = 2\pi\times3.71459(2)$~GHz for the 7.5~kHz transition. The cavity waist at the clock transition $^3$P$_0$ wavelength is determined to be $\textrm{w}_0=73.85(7)~\mu$m. For the broader transition $^3$P$_1$ transition's wavelength, the waist is $\textrm{w}_1=73.37(7)~\mu$m. These values are predicted by the Gaussian beam propagation theory using the known wavelengths, mirror radius of curvature, and cavity free spectal range. The waist sizes have been independently verified to agree at the 0.1\% level by measuring the spacing between the TEM$_{00}$ mode and the TEM$_{1,0/0,1}$ modes relative to the measured free spectral range of the cavity \cite{Yariv1991}.  

Atoms are loaded at the cavity center (in between the two mirrors), as confirmed by taking fluorescence images of the loaded atoms. The Rayleigh length of the modes ($\sim 2.453(5)~$cm) is typically much longer than the longitudinal extent of the cold atomic cloud ($\sigma_{long} = 0.30(5)~$mm). 
Finally, we performed cavity ring-down measurements to determine the cavity linewidth. For these measurements we probed the cavity on resonance with light polarized along $\hat{x}$ and after quickly turning off the probe light with the EOM, we observed the photo-current on a fast DC coupled photodiode directly positioned after the cavity. We determine a linewidth of $\kappa_0/(2\pi) = 140.9(3)$~kHz at the clock transition and $\kappa_1/(2\pi) = 153.0(4)$~kHz at the 689~nm transition, after taking statistics over several trials. The cavity and atomic parameters can be found in Table ~\ref{tab:CavityAtomParams}.

\subsection{Computing $(g_0/g_1)^2$ from measured quantities}
We measure the ratio $(g_0/g_1)^2$ by interleaved measurements of the  atomic induced phase shift between two probes near resonant with the clock transition, $\Delta\varphi_0$, and cavity frequency shift $\Delta\omega_1$ on the 7.5~kHz transition, from which we calculate the associated phase shift $\Delta\varphi_1 = \Delta\omega_1/(\kappa_1/2)$. An ideal measurement assumes that all the atoms are homogeneously coupled to the cavity mode, they do not move, they are optically pumped to the $m_F=\pm 9/2$ states, all the atoms remain in the ground state during the probing, both probes are $\pi$-polarized and that the cavity mode is aligned to the clock transition, while the 689 $F=9/2\rightarrow F^\prime = 9/2$ transition is detuned by $\delta_{C1}$. The validity of these approximations will be taken into consideration when analyzing the systematic corrections. 

The atomic contribution to the phase shift on the clock transition between tones at $\pm\delta_{p0}$ with respect to the atomic transition is
\begin{equation}
\Delta \varphi_{0}  = \frac{4 N c_C^2 g_0^2 }{\kappa_0 \delta_{p0}},\label{eq:deltaphi_full}     
\end{equation} 
where $c_C$ is the Clebsch-Gordan coefficient for $\pi$-polarized light probing the stretched states ($c_C = \sqrt{\frac{9}{11}}$), $2g_0$ is the single photon Rabi frequency for the clock transition, and $\kappa_0$ is the cavity linewidth at 698~nm.

The difference between the cavity frequency shifts for the two 689~nm modes, as shown in Fig.~\ref{fig:SI1}(b)(ii) is
\begin{widetext}
\begin{align}
    \begin{split}
        \Delta \omega_1= N g_1^2 \left\lbrace \left( \frac{c_{N1}^2}{\delta_{C1}} + \frac{c_{N2}^2}{\delta_{C1} - \Delta_{11/2}} \right) - \left( \frac{c_{N1}^2}{\delta_{C1}-\Delta_{\textrm{FSR},1}} + \frac{c_{N2}^2}{\delta_{C1} -\Delta_{11/2} - \Delta_{\textrm{FSR},1}}\right)\right\rbrace,
    \label{eq:deltaomega_full}
    \end{split}
\end{align}
\end{widetext}
where $c_{N1}$ and $c_{N2}$ are the Clebsch-Gordan coefficient for $\pi$-polarized light probing the stretched states on the $F=9/2\rightarrow F^\prime = 9/2$ and $F=9/2\rightarrow F^\prime = 11/2$ transitions, respectively ($c_{N1} = \sqrt{\frac{9}{11}},c_{N2} = \sqrt{\frac{2}{11}}$), $\Delta_{11/2}/(2\pi) = -1463.15(6)~$MHz is detuning of $F^\prime = 11/2$ with respect to the $F^\prime = 9/2$ manifold \cite{Sansonetti_2010}, and $\Delta_{\textrm{FSR},1}/(2\pi)$ is the cavity free spectral range at 689~nm. Note that the $F^\prime=7/2$ manifold, detuned by $\Delta_{7/2}/(2\pi) = 1130.260(6)$MHz from the $F^\prime=9/2$ transition, does not contribute to the expression in Eq.~\ref{eq:deltaomega_full} as we are assuming atoms are optically pumped to $m_F = \pm 9/2$ and the probe is $\pi$-polarized. 

To finally reveal the ratio between $g_0$ and $g_1$, we need to precisely know all the numerical factors in Eq.~\ref{eq:deltaphi_full} and Eq.~\ref{eq:deltaomega_full}, as well as characterize all systematic corrections that must be applied to account for deviations of the experiment from the idealized situation above.  

\section{Systematic effects}\label{sec:sec3}

This measurement approach for determining natural lifetimes of long-lived states is new, and it is important to think broadly about potential systematic corrections that must be applied, as well as the uncertainties on these corrections. In this section, we will discuss nearly 20 different systematic corrections. Most of these are small enough to be ignored, but we include them for completeness and for the sake of future applications of the technique, in which details of the experimental system might make these effects larger.

We roughly break up the systematics discussion into three categories: those that affect individually the cavity phase shift measurement $\Delta\varphi_1$ on the 7.5~kHz transition, those that affect the phase shift measurement $\Delta\varphi_0$ on the clock transition, and those that affect the measured ratio $(\Delta\varphi_0/\Delta\varphi_1)$. We define the correction factors $F_C$ as the ratio of the ideal quantity $Q^i$ and the actually measured quantity $Q^m$ such that the ideal quantity can be recovered from the measured quantity as $Q^i= F_C Q^m $. 

\subsection{Corrections on the phase shift $\Delta\varphi_1 $}

In this section we discuss effects that affect the measured cavity frequency shift $\Delta\varphi_1 $ for the two consecutive TEM$_{00}$ modes on the 7.5~kHz transition at 689~nm. The magnitude of the correction factors $F_C$ are shown in the following table.

\begin{table}[htb!]
\caption{\label{tab:correction689} Correction factors $F_C$ for $\Delta\varphi_1 $ measurement.}
\begin{ruledtabular}
\begin{tabular}{lcc}
\textrm{Effect}&
\textrm{$1-F_C$}&
\textrm{Error on $F_C$}\\
\colrule
Polarization uncertainty  & $2\times10^{-4}$ & $6\times10^{-5}$ \\
Differential lattice shift  & $-2.3\times10^{-3}$ & $8\times10^{-4}$ \\
Saturation during probe &  $-2\times10^{-4}$ & $2\times10^{-4}$ \\
Probe optical pumping and losses & $-4\times10^{-4}$ & $4\times10^{-4}$\\
Zeeman shift & $2\times10^{-8}$ & $1 \times 10^{-9}$ \\
Higher order corrections & $-2.5\times10^{-4}$ & $1\times10^{-4}$ \\
\end{tabular}
\end{ruledtabular}
\end{table}

\subsubsection{Polarization uncertainty in $^3$P$_1$ probe}
The polarization uncertainty effect refers to the fact that the probe light's polarization might not have been perfectly $\pi$-polarized. To optimize the probe polarization's orientation relative to the magnetic field, we performed a measurement of $\Delta\varphi_1$ at a (variable) value of the transverse magnetic field $\textit{B}_t$ first, and within 4~ms we measure it again at another magnetic field $\textit{B}^{ref}_t$ that we believe to be close to the value that cancels the transverse components. Magnetic fields along $\hat{x},\hat{y},\hat{z}$ are generated by three respective sets of Helmholtz coils driven by a stabilized current source, which allow us to rapidly perform small changes in the $y$ and $z$ components of $\textit{B}_t$ in order to perform this measurement (see Fig.~\ref{fig:SI1}(a)). In this way, we have the ability to compute the ratio in each experimental repetition, which gives us further insensitivity with respect to other quantities that fluctuate shot-to-shot, like atom number.

The ratio of the two measurements, $\Delta\varphi_1(\textit{B}_t)/\Delta\varphi_1(\textit{B}^{ref}_t)$, is maximized when the $y$ and $z$ components are nulled, as our model shows. An example of this measurement is shown in Fig.~\ref{fig:SIfig2}(c). If the reference field $\textrm{B}^{ref}_t$ was not properly chosen, we can change it accordingly and evaluate the ratio again, until we consistently find the right value for $\textit{B}^{ref}_t$, where we would like to operate the experiment. Typically, we observed day to day shifts of 3~mG as we repeat this procedure before any of the measurements to establish the linewidth ratio. The associated correction factor $F_C$ takes into account the effect of a small magnetic field fluctuations of magnitude 3~mG on typical data sets as shown in Fig.~\ref{fig:SIfig2}(c). Based on our model, we find that we can realize a probe with 98\% pure $\pi$ polarization. The fitted quadratic dependence of the phase shift magnitude along with this 3~mG uncertainty is used to estimate the correction factor for this effect.

\subsubsection{Differential lattice shift in $^3$P$_1$ probe}
The differential lattice shift is due to the fact that the lattice is not quite magic for the $^3$P$_1$ states. However, the expected differential AC Stark shifts, around $0.7(2)~$MHz for our trap depth, are very small compared to the cavity detuning from the atomic transition $\delta_{C1}$. Experimentally, we determine $\Delta\varphi_1 $ to change by less than 1\% for trap depths changing by 50\%. The error is estimated based on a combination of trap depth uncertainty, resonance frequency uncertainty and cavity detuning uncertainty. 

\subsubsection{Saturation and optical pumping due to the $^3$P$_1$ probe}

While probing the cavity phase shift $\Delta\varphi_1$, the probe itself can excite atoms, especially if the probe power is large. To characterize this effect, we measure the change in $\Delta\varphi_1$ as a function of the probe power. In each experimental repetition, we perform three consecutive measurements as depicted in Fig.~\ref{fig:SIfig2}(b): first at a (variable) power \textit{P}$_1^H$, then at a (low) reference power \textit{P}$_1^L$, then again at a (variable) power \textit{P}$_1^H$. This allow us to be insensitive to, for instance, shot to shot variations in atom number, while allowing us to characterize the effect of a high power probe on each measurement.

In each experimental shot, we obtain three cavity phase shifts denoted $\Delta\varphi^1_1$, $\Delta\varphi^2_1$, and $\Delta\varphi^3_1$, respectively. We model the effect of any possible probe power related effect, in the low power limit, as a modification in the measured phase shift as $\Delta\varphi_1(\textit{P}) = \Delta\varphi_1(\textit{P}=0) (1-2(\textit{P}/\textit{P}_0))$, where $\textit{P}$ is the probe optical power, $\textit{P}_0$ is some parameter that works as an effective saturation power in this model, and $\Delta\varphi_1(\textit{P}=0)$ is the zero-power phase shift that is the interest of our measurement. In order to characterize this behaviour, we varied the power \textit{P}$_1^H$ and measured the effect on $\Delta\varphi_1$ using the ratio $s_S = ((\Delta\varphi^1_1+\Delta\varphi^3_1)/2-\Delta\varphi^2_1)/(\Delta\varphi^2_1)$; for low excitation fractions, this is linear in the input power \textit{P}$_1^H$ and scales as $s_S = (\textit{P}_0-2\textit{P}_1^H)/(\textit{P}_0-2\textit{P}_1^L)$. For a given probe power \textit{P}, the correction factor would be $F_C = 1/(1-2(\textit{P}/\textit{P}_0))$.

The measurement is shown in Fig.~\ref{fig:SIfig2}(b), where we fit a linear function (red line) to the input probe power \textit{P}$_1^H$. This allows us to establish that the excitation fraction is not significant ($<0.0001$) for powers below 1~nW. The value quoted for the associated correction factor, takes into account the maximum power used for the data presented in Fig.~3(b) in the main text, which was around 1~nW and was decreased together with the clock transition probe power to extract the ratio $\Delta\varphi_0/\Delta\varphi_1$. The uncertainty is taken to cover the full range of power used in the $\Delta\varphi_0/\Delta\varphi_1$ measurement presented in the main text. We assign a value $F_C = 1.0002(2)$ for the correction factor.

Furthermore, the probe itself can also either cause atom loss or optically pump atoms to different magnetic sub-levels, modifying $\Delta\varphi_1$ and eventually $\Delta\varphi_0$ on consecutive measurement sequences. To characterize this effect, we again use three consecutive measurements to obtain $\Delta\varphi^1_1$, $\Delta\varphi^2_1$, and $\Delta\varphi^3_1$: first at a (low) reference power \textit{P}$_1^L$, then at a (variable) power \textit{P}$_1^H$, then again at a (low) reference power \textit{P}$_1^L$. This scheme allow us to, in a single shot, characterize the change that occurs after applying a relative high power probe. 

The low power probes fall into the low power region in the previous analysis, such that for simplicity we will consider they do not have a significant effect. However, we will consider the second high power probe has a more permanent effect. For example, we consider the case that during, and after, the second pulse with power \textit{P}$_1^H$ the measured phase shift is modified by an effective value $(1-(\textit{P}_1^H/\textit{P}_{OP})$ from the zero power value. Again, this $\textit{P}_{OP}$ tries to capture any effect such as redistribution in the ground state hyperfine state manifold consequence of a higher probe power. 

To characterize this effect, we consider the parameter $s_{OP} = 1-(\Delta\varphi^3_1)/(\Delta\varphi^1_1)$, that measures the differential phase shift after applying a higher probe power in between the two pulses. In particular, our model predicts a behaviour of the form $s_{OP} = (\textit{P}_1^H/\textit{P}_{OP})$. We show the result of these measurements in Fig.~\ref{fig:SIfig2}(c), where we varied the optical power of the second probe $\textit{P}_1^H$. For the final optical power used in the phase shifts ratio measurement presented in the text, we use maximum probe powers in the 689~nm transition on the order of 1~nW. From this characterization, we assign a correction factor $F_C = 1.0004(4)$, to cover the full range of variation for the used powers.

\subsubsection{Zeeman shifts in $^3$P$_1$ probe}
The Zeeman shift on the different ground and excited states changes the effective atom-cavity detuning by a few hundred ~kHz, which is much smaller than the cavity detuning $\delta_{C1}$. The typical magnetic field that we use is 95~mG, as calibrated using the splitting between the peaks in the superradiant pulses \cite{Norcia_SRFreq_2018} and corroborated by the splitting measured in Fig.~2(a) in the main text, for example. Taking this effect into account we expect a very small correction to the ratio of phase shifts.

\subsubsection{Higher order corrections on $\Delta\varphi_1$ measurement}

Higher order corrections on the cavity frequency shift manifest in $\delta\omega_1$ as $\delta\omega_1 = Ng^2_1/\delta_{C1}(1-2N g^2_1/\delta^2_{C1})$ \cite{Chen_2014}. We note that this correction is $N$-dependent. For typical experimental parameters we have $2N g_1^2/\delta^2_{C1}\lesssim  1\times10^{-3}$. We calculate the correction factor $F_C$ based on an independent atom number calibration using florescence imaging, and its error is estimated assuming extreme 50\% fluctuations in typical $N$.

\begin{figure*}[!htb]
\includegraphics[width=7in]{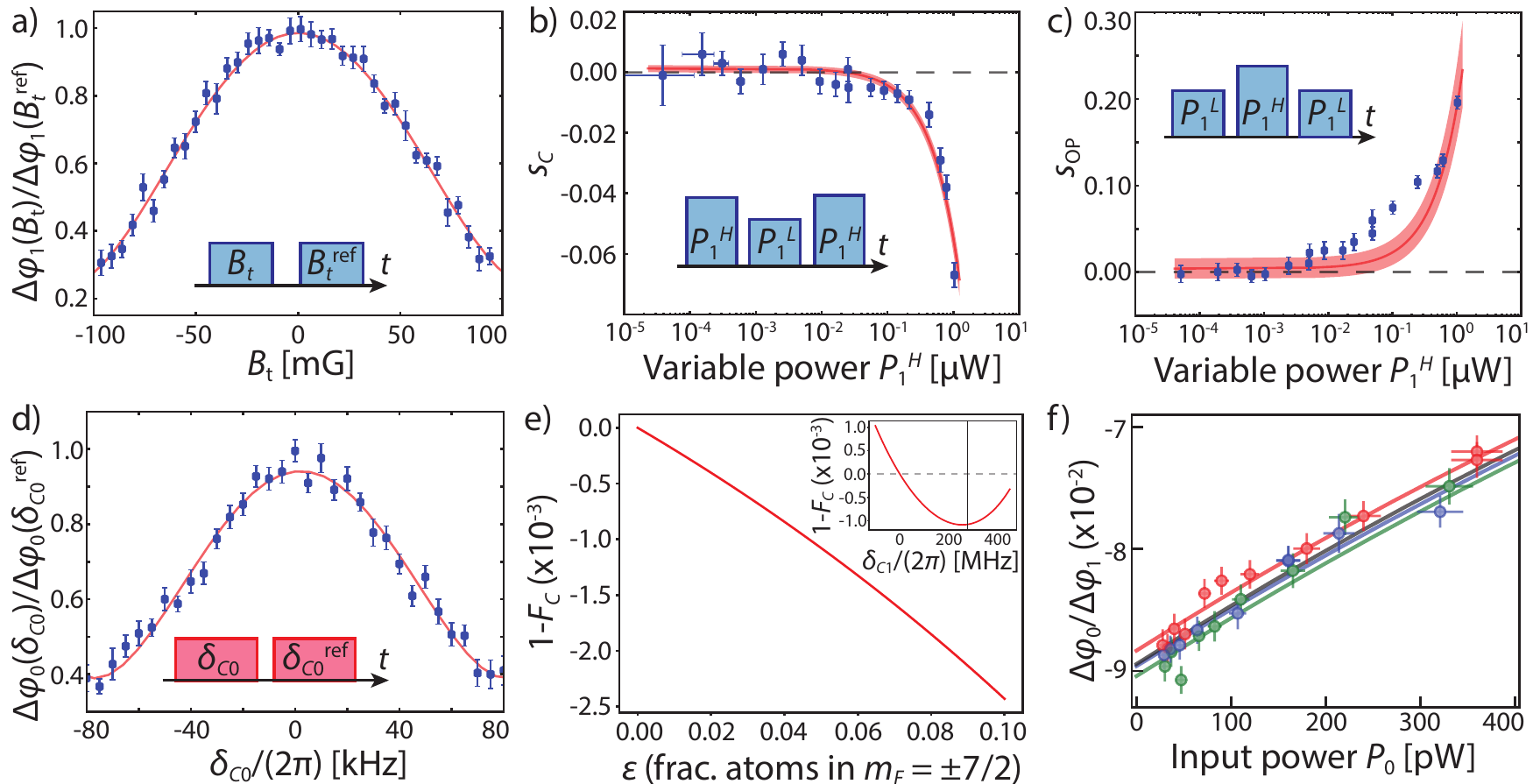}
\caption{(a) Polarization alignment using cavity phase shift measurements while changing one of the transverse magnetic field components. In a single shot, we compute the ratio between two measurements at a variable transverse field, $\textrm{B}_t$, and a reference transverse field $\textrm{B}^{ref}_t$. Typical day to day fluctuations are around 3~mG. (b) Saturation effects on the 689~nm transition probe. We use interleave $\Delta\varphi_1 $ measurements at different powers to determine the relative reduction on $\Delta\varphi_1$ as the probe power increases. (c) Power induced reduction on the 689~nm transition phase shift. We use interleave $\Delta\varphi_1$ measurements at different powers to determine the relative reduction on $\Delta\varphi_1 $ as the probe power increases. (d) Influence of the cavity detuning $\delta_{C0}$ on the atom-like phase shift $\Delta\varphi_0$ on the clock transition. We tune the cavity at two different cavity detunings, one fixed $\delta_{C0,ref}$, and one variable $\delta_{C0}$, and in a single show we compute the ratio $\Delta\varphi_0(\delta_{C0})/\Delta\varphi_0(\delta_{C0,ref})$ that maximizes when $\delta_{C0} = \delta_{C0,ref} = 0$. (e) Optical pumping correction factor, $F_C$, where we take the fractional population on the $\pm 7/2$ states to be $\varepsilon$, on the $\pm 5/2$ states is $\varepsilon^2$, on the $\pm 3/2$ states is $\varepsilon^3$, and on the $\pm 1/2$ states is $\varepsilon^4$. Inset shows the dependence of the correction factor for $\varepsilon = 0.05$ versus the cavity detuning $\delta_{C1}$ to the $^3$P$_1$, $F^{\prime} = 9/2$ manifold. (f) Zoom in for low powers for Fig.~3(c) in the main text. Results and fits shown for estimator $E_3$ as described in this text.}
\label{fig:SIfig2}
\end{figure*}

\subsection{Corrections on the phase shift $\Delta\varphi_{0} $}

In this section we discuss effects that affect the measured atomic-like phase shift $\Delta\varphi_0$ on the millihertz transition at 698~nm. The magnitude of the correction factors $F_C$ are shown in the following table.

\begin{table}[htb!]
\caption{\label{tab:correction698} Correction factors $F_C$ for $\Delta\varphi_0$ measurement.}
\begin{ruledtabular}
\begin{tabular}{lcc}
\textrm{Effect}&
\textrm{$1-F_C$}&
\textrm{Error on $F_C$}\\
\colrule
Polarization uncertainty  & $-3\times10^{-4}$ & $3\times10^{-4}$ \\
Atomic resonance uncertainty  & $6\times10^{-3}$ & $6\times10^{-3}$ \\
Cavity resonance drift &  $-8\times10^{-3}$ & $6\times10^{-3}$ \\
Zeeman shift & $2.3\times10^{-3}$ & $5\times10^{-4}$ \\
Higher order corrections & $4\times 10^{-4}$ & $2\times 10^{-4}$ \\
\end{tabular}
\end{ruledtabular}
\end{table}

\subsubsection{Polarization uncertainty in $^3$P$_0$ probe}

The polarization uncertainty error refers to the purity of the probe polarization. Based on the measurements for $\Delta\omega_1$ presented before, we model in a very similar way what the effect would have been for the atomic phase measurement $\Delta\varphi_0$ with a typical 3~mG uncertainty on the transverse magnetic fields. 

\subsubsection{Atomic resonance uncertainty in $^3$P$_0$ probe}

The clock transition is addressed with light from a state-of-the-art laser, used in the $^{87}$Sr optical lattice clock experiments at JILA \cite{Oelker_2019,Campbell_2017,Robinson_2019}. To determine the atomic resonance, we perform Rabi spectroscopy, measuring the excitation fraction versus the probe light's frequency. The probe frequency is changed by changing the in-fiber EOM driving frequency. For sufficiently low power, we are able to determine the central frequency with less than 10~Hz uncertainty, but the full-width at half maximum of the spectroscopic feature is typically between 50~Hz and 100~Hz, similar to the data shown in Fig.~2(b) in the main text. The frequency might be shifted from the natural $^{87}$Sr frequency because of different atomic frequency shifts, i.e. DC Stark shifts, Doppler shifts, collective shifts, lattice detuning from the magic wavelength to name a few. However, we have already fully characterized clock transitions in Ref.~\cite{Norcia_SRFreq_2018} to be well below 100~Hz.

Because we are using two symmetric tones to address the atomic transition, the associated correction factor $F_C$ to the measured phase shift scales as $(1-\left(\delta_{L0}/\delta_{p0}\right)^2)$ with $\delta_{L0}$ the detuning from the tones central frequency to the atomic transition, as defined in the main text. Note that this effects increases the absolute value of the measured phase shift $\Delta\varphi_0$, as can be seen in Fig.~2(b) in the main text. The correction factor is calculated by taking an rms average on the variation of $\Delta\varphi_0$ when $|\delta_{L0}/(2\pi)|<100~$Hz. Its error is computed to cover the full range.  

\subsubsection{Cavity resonance uncertainty in $^3$P$_0$ probe}

By probing and subtracting the phase shifts for two consecutive TEM$_{00}$ modes, one on resonance with the atomic clock transition, we guarantee that any instantaneous cavity length fluctuation will be instantaneously removed from our measurement. However, if the initial cavity detuning from the clock transition, $\delta_{C0}$, is non-zero, the phase shift will be modified by a factor  $(1+\left(\delta_{C0}/(\kappa_0/2)\right)^2)$, as noted in Eq.~\ref{eq:phase_error_sausage}. Typically, we can align the initial cavity length and minimize cavity drifts such that $\left|\delta_{C0}\right|/(2\pi)\leq 10$~kHz during each of the measurement in Fig.~3(b) in the main text. We had verified analytically and experimentally what would be the effect of a cavity resonance drift, with good agreement. For example, Fig.~\ref{fig:SIfig2}(d) shows the relative change in $\Delta\varphi_0$ as $\delta_{C0}$ is intentionally changed. For this measurement, we are able to change the cavity detuning by changing the drive voltage on the PZTs after the atoms are already loaded in the lattice, as shown in the inset. We first measure $\Delta\varphi_0$ for a variable detuning $\delta_{C0}$ and then change the cavity length to a reference detuning $\delta^{ref}_{C0}$, which allow us to remove unwanted effects, such as atom number drifts, from our measurements as we did when we analyze the impact of transverse components of the magnetic field on the phase measurements. Based on these results, and a precise determination of the cavity FSR, we estimate a correction of less than 1\% if we average over cavity detunings below a maximum 10~kHz drift. In fact the drifts in the zero-power value for $\Delta\varphi_0/\Delta\varphi_1$ reported in Fig.~3(b) in the main text are consistent with cavity frequency misalignment within our 10~kHz uncertainty.

\subsubsection{Zeeman shift in $^3$P$_0$ probe}
The small magnetic field present to define the quantization axis generates a Zeeman splitting between the two ground states, of typical magnitude 100~Hz, smaller than the probes splitting $2\delta_{p0}/(2\pi) = 2$~kHz. We can accurately calibrate the magnetic field by observing the splitting between superradiant pulses, as in Ref.~\cite{Norcia_SRFreq_2018}. By using two symmetric tones to address the clock transition, the phase shift will be only second order sensitive to the Zeeman splitting. We calculate this value and assign and uncertainty based on a 5\% uncertainty on the determination of the magnetic field along the quantization axis. 

\subsubsection{Higher order corrections on the $\Delta\varphi_0$ measurement}
Higher order corrections on the four-tones phase shift method are derived in Eq.~\ref{eq:phase_error_sausage}. The second correction factor ($(2\delta_c/\kappa)^2$) is the cavity resonance drift considered above. The other higher order terms, remnant from the small angle approximation, contribute at the level of $10^{-4}$ for typical atom numbers, obtained through an independent calibration of our fluorescence imaging. Its error is estimated assuming extreme 50\% fluctuations in typical $N$. 

\subsection{Corrections on the ratio $\left(\Delta \varphi_{0}/\Delta\varphi_1\right)$}

In this section we discuss effects that modify both the measured atomic-like phase shift $\Delta\varphi_0$ on the millihertz transition at 698~nm and the cavity phase shift $\Delta\varphi_1$ on the 7.5~kHz transition at 689~nm. The magnitude of the correction factors $F_C$ are shown in the following table.

\begin{table}[htb!]
\caption{\label{tab:correctionratio} This table contains the correction factors $F_C$ to correct the ratio $\left(\Delta \varphi_{0}/\Delta\varphi_1\right) $}
\begin{ruledtabular}
\begin{tabular}{lcc}
\textrm{Effect}&
\textrm{$1-F_C$}&
\textrm{Error on $F_C$}\\
\colrule
Offsets in $\Delta\phi_{0,1}$ & $0$ & $1\times 10^{-2}$ \\
Axial inhomogeneous probe coupling & 0 & $2\times10^{-3}$\\
Differential radial average  & $7\times10^{-5}$ & $1\times10^{-5}$ \\
Finite axial confinement/ &  $-6.2\times10^{-2}$ & $4\times10^{-3}$ \\
Resolved carrier correction &  &  \\
Ground state $m_F$ distribution  & $-1\times10^{-3}$ & $1\times10^{-3}$ \\
Lifetime in the lattice & $-2\times10^{-3}$ & $2\times10^{-3}$\\
Cavity birefringence & $-1.2\times 10^{-2}$ & $5 \times 10^{-3}$ \\
\end{tabular}
\end{ruledtabular}
\end{table}

\subsubsection{Offsets in $\Delta\phi_{0,1}$}

The effect of non cancelled offsets in our measurements is to alter the measured values of $\Delta\phi_{0}$ and $\Delta\phi_{1}$. In particular, because the desired phase shifts are collective, while the offset are not, it can cause an $N$-dependent correction to the ratio $\Delta\phi_{0}/\Delta\phi_{1}$.

Assuming single atom phase shifts $\Delta\phi^a_{0,1}$ and offsets $\Delta\phi^{off}_{0,1}$ on each measurement, we can express the desired ratio as 
\begin{equation}\label{eq:offsets}
    \frac{\Delta\phi_0}{\Delta\phi_1} = \frac{N \Delta\phi^a_{0} + \Delta\phi^{off}_{0}}{N \Delta\phi^a_{1} + \Delta\phi^{off}_{1}}.
\end{equation}

For the three different sets shown in Fig. 3(b) in the main text we measured low power sets with no atoms in the cavity, and verified that $\Delta\phi^{off}_{0,1} = 0$ within error bars. To be specific, we typically measure $\Delta\phi^{off}_{0} = 0(0.25)$~mrad and $\Delta\phi^{off}_{1} = 0(4)$~mrad, while the low power phase shifts for $N = 80\times10^3$ atoms are approximately $\Delta\phi_{0} = 40$~mrad and $\Delta\phi_{1} = 400$~mrad. Therefore, the offsets do not alter the measured ratios at the 1\% level, limited by the uncertainty in our determinations of the offsets.

We consider in this case the correction factor to be $F_C = (\Delta\phi^a_{0}/\Delta\phi^a_{1})/(\Delta\phi_0/\Delta\phi_1)$ that for small offsets ($\Delta\phi^{off}_{0,1} \ll N \Delta\phi^a_{0,1}$) is approximately $F_C = 1+\left((\Delta\phi^{off}_{1}/\Delta\phi^{a}_{1})-((\Delta\phi^{off}_{0}/\Delta\phi^{a}_{0})\right)/N$. For the values just quoted and summing in quadrature the errors for each phase shift, we have $F_C = 1.00(1)$, which represents the largest single uncertainty contribution to the final $F_C$. It is worth noticing that the uncertainty in the phase shifts measurements can potentially be improved by, for example, increasing the probe detunings and their power, and improving the final quantum efficiency of the detection system.

\subsubsection{Axial inhomogeneous probe coupling}

The optical lattice at $\lambda_{\mathrm{trap}} = 813$~nm, the probe at $\lambda_1=689$~nm and the probe at $\lambda_0=698$~nm all form standing waves in the cavity that are all incommensurate with each other. Focusing on just the two probes, the couplings vary approximately as $g_{0/1}^2=g_{m,0/1}^2 \cos^2\left(2\pi z/\lambda_{0,1} + \psi_{0/1}\right)$ where $z$ is the location along the cavity axis, $z=0$ is at the center of the cavity, and the spatial phase of the standing waves are $\psi_{0/1}= 0$~or~$\pi/2$, depending the relative parity of the modes. The maximum coupling at an antinode is $g_{0/1,m}^2$.

As one moves along the cavity axis, the standing wave of the two probes continuously transform every 13~$\mu$m from being aligned (having antinode aligned to antinode) to anti-aligned (having anitnodes aligned to nodes.) As a result, the probes do not interact with exactly the same set of atoms. However the atoms are loaded into lattice sites spanning approximately 0.6~mm along the cavity axis (rms diameter) so that one expects the reduction in the coupling due to spatial averaging to be nearly identical and thus cancel in the ratio of the measured couplings. Assuming atoms are only located every $\lambda_{\mathrm{trap}}/2$, and are spread uniformly along $0.6$~mm, the ratio of averaged couplings is modified by  $<2\times10^{-3}$. If a more reasonable Gaussian envelope with standard deviation $0.3$~mm (rms radius) is used to describe the loading of the lattice sites, the ratio of averaged couplings is changed by many orders of magnitude less. Here, we conservatively apply a correction $F_C =1$ with an error on $F_C$ of $2\times10^{-3}$.

\subsubsection{Radial inhomogeneous probe coupling}
The measured phase shifts are also modified when averaging over the radial positions of the atoms due to the finite difference in the probe mode waist sizes $\textrm{w}_0$ and $\textrm{w}_1$ characterizing the $1/e^2$ in intensity radius of the Gaussian TEM$_{00}$ probe modes (see Table~\ref{tab:CavityAtomParams}). The ratio of waists scales as $\textrm{w}_1/ \textrm{w}_0=\lambda_1/\lambda_0\approx 1 - 1.29\times10^{-2}.$  For an atom at a distance $r$ away from the cavity axis, the ratio of the couplings $g_0^2/g_1^2$ is modified by the factor $f$ compared to its on-axis value

\begin{equation}
   f= e^{-2\left(\frac{r}{\bar{\textrm{w}}}\right)^2\left(\frac{ \textrm{w}_1^2-\textrm{w}_0^2}{\bar{\textrm{w}}^2}\right)},
\end{equation}
\noindent where $\bar{\textrm{w}}=\sqrt{\textrm{w}_0 \textrm{w}_1}$ is the geometric mean of the waists. For scale, at the rms thermal radius of the atomic cloud  $\sigma_r = 14~\mu$m, the correction factor is $f=1.0007$. After averaging over the atomic radial distribution, the averaged coupling  $g_0^2$ and $g_1^2$ are both reduced by about 4\% but the ratio $g_0^2/g_1^2$ is changed by less than $10^{-4}$.  We expect that the rms thermal radius is common to both measurements because we do interleaved nondestructive probes and because we interpolate to zero probe power so that any potential mechanical forces on the atoms is also interpolated to zero.

\subsubsection{Finite Axial Confinement/Resolved Carrier Correction}
The largest systematic correction that must be applied arises from the finite localization of the atoms along the axial direction. The atoms are trapped in the Lamb-Dicke regime along the cavity axis with spacing dictated by the lattice wavelength (813~nm). The probe tones almost exclusively interact with the well-resolved motional carrier transition since $\delta_{p0}\ll\omega_z$. Lastly, because of the finite localization of the atomic wave-function (i.e.~ finite Lamb-Dicke parameter) the effective strength of the carrier transition (i.e.~the effective $g_0^2$ is reduced by an estimated 6.2(4)\% for which we apply a correction.

In order to evaluate the apparent modification to $g_0^2$ from this effect, we estimate the  probability distribution $P(n)$ of finding an atom in the $n^{th}$ axial vibrational level. The estimate is made using sideband spectroscopy measurements as shown in Fig.~\ref{fig:SI1}(c), following Ref.~\cite{Blatt_2009}. Based on this probability distribution, we calculate the average correction to $g_0^2$. We model the light-matter coupling as $g_0^2(\phi,\hat{z}) =g_{0,m}^2 \cos^2(k_p \hat{z} + \phi)$, where $k_p$ is the probe wave-vector, $\hat{z}$ is the harmonic oscillator position operator, $g_{0,m}$ is the value of $g_0$ at a probe anti-node, and $\phi$ is a uniformly distributed phase between 0 and $2\pi$ that accounts for the inhomogeneous coupling of the trap atoms to the probe. This is justified as the probe and the axial atomic distribution are incommensurate and the beating length is much shorter than the cloud extent. 

Furthermore, the radial spreading of the cloud means that each atom will have a slightly different axial trap frequency. To leading order, an atom at distance $r$ from the center will have an axial frequency $\omega_z(r) = \omega_{z,0}(1-(r/\textrm{w}_\textrm{trap})^2)$, where $\omega_{z,0}$ is the maximum axial frequency ($\omega_{z,0}/2\pi = 230(1)$~kHz) and $\textrm{w}_\textrm{trap}$ is the trap waist ($\textrm{w}_\textrm{trap} = 79.7~\mu$m). As both directions are decoupled, we have that the average axial frequency over the atomic ensemble is $\langle \omega_z \rangle = \omega_{z,0}(1-(\langle r^2 \rangle /\textrm{w}_\textrm{trap}^2)$. For a Gaussian radial density distribution profile, we have $\langle r^2 \rangle= 2\sigma^2_r= 2 k_B T_r/ (m\omega_r^2)$, with $T_r$ the radial temperature determined from the motional sideband fit, $k_B$ is the Boltzmann constant, $\omega_r$ the radial trap frequency, and $\sigma_r$ the rms thermal radius of the cloud along its radial direction.   

We calculate the average value $\left< g^2_0 \right>$ over this distribution for this effective axial trap frequency $\langle \omega_z \rangle$, that is the value that enters in our measurement result for $\Delta\varphi_0$, as  
\begin{equation}
    \left< g^2_0 \right> = \sum_{n=0}^{n=Nz} \frac{1}{2\pi}\int_0^{2\pi} P(n) \bra{n}  g_{0,m}^2 \cos^2(k_p \hat{z} + \phi) \ket{n} d\phi, \label{eq:g_axial}
\end{equation}
where $N_z$ is the maximum harmonic level on the trap ($N_z\sim17$) \cite{Blatt_2009} and $\ket{n}$ are the eigenstates of the unperturbed harmonic potential along the $z$-direction. We follow a similar procedure for the 689~nm probe, but taking into account that we are probing every transition, i.e. we sum over all possible initial and final states correcting the relative detuning between each harmonic oscillator state. Based on the reconstructed probability distribution $P(n)$, we obtain a correction factor $F_C = 1.062(4)$, dominated by the error on the fitted temperature on the axial and radial coordinates. This is the biggest correction we apply to the measured ratio $\left(\Delta \varphi_{0}/\Delta\varphi_1\right)$.   

We emphasize that for the $^3$P$_1$ probe, where the probe detuning is much bigger than the trap frequency ($\delta_{C1}\gg \omega_z$), the vibrational degrees of freedom do not play a significant role. However, the average over the phase $\varphi$ in Eq.~\ref{eq:g_axial} gives a 1/2 reduction on $g^2_1$, that is also present in the $g^2_0$ term, making this a common mode effect whose impact is highly suppressed. Imperfect cancellation of this factor is taken into account on the previous section \textit{Axial inhomogeneous probe coupling}.

\subsubsection{Ground state $m_F$ distribution}

The initial distribution among the different magnetic $m_F$ sub-levels in the ground states is extremely important. For example, if there are atoms in any other $m_F$ state other than $\pm9/2$, both $\Delta\varphi_0$ and $\Delta\varphi_1$ (or the measured $\Delta\omega_1$) will be affected. Measuring the frequency splitting between the superradiant pulses on the clock transition \cite{Norcia_SRFreq_2018}, confirmed that the initial optical pumping efficiency is at least 95\% to the $\pm 9/2$ states. In order to estimate the correction to the dispersive phase shift ratio, we assume a conservative bound of 5\% of the atoms in the wrong state. We model the measured ratio $\left(\Delta \varphi_{0}/\Delta\varphi_1\right)$ when 5\% of the atoms are allowed to be in any of the other $m_F$ levels as a function of the detuning $\delta_{C1}$. Because the position of the different hyperfine levels on the $^3$P$_1$ state relative to the cavity modes (hyperfine splitting is comparable to cavity free spectral range - see fig.~\ref{fig:SI1}(b)(ii)), and the fact that each transition has a different set of Clebsch-Gordan coefficients, the correction factor is highly sensitive to the cavity detuning $\delta_{C1}$. A detail explanation follows below. 

We consider a realization of the atomic distribution among the ground hyperfine state levels $P_G$ that contains the list of probabilities of finding an atom in each ground state $m_F$. Ideally, $P_G = \left\lbrace 1/2,0,...,0,1/2\right\rbrace$, for the set $m_F=\left\lbrace-9/2,-7/2,...,7/2,9/2\right\rbrace$. For the phase shift on $^1$S$_0 \rightarrow ^3$P$_0$ transition at 698~nm, $\Delta\varphi_0$, the new phase shift for an arbitrary distribution $P_G$ on the $m_F$ manifold is
\begin{equation}
    \Delta\varphi_{0,{\left\lbrace m_F\right\rbrace}} = \sum_{m_F = -9/2}^{m_F = 9/2} P_G(m_F) \frac{4 N (c^0_\pi(m_F))^2 g_0^2 }{\kappa_0 \delta_{p0}}, \label{eq:phi0_groundmf}
\end{equation}
where $c^0_\pi(m_F)$ is the Clebsch-Gordan coefficient for $\pi$-polarized light probing the $m_F$ hyperfine ground state on the $^1$S$_0 \rightarrow ^3$P$_0$ transition, populated with probability $P_G(m_F)$.

For the phase shift at the $^1$S$_0 \rightarrow ^3$P$_1$ transition at 689~nm, $\Delta\varphi_1$, the equivalent modification is

\begin{widetext}
    \begin{equation}
        \begin{split}
            \Delta\varphi_{1,{\left\lbrace m_F\right\rbrace}} = &\sum_{k = \left\lbrace 0,1 \right\rbrace} \sum_{m_F = -9/2}^{m_F = 9/2} P_G(m_F) \frac{2 \pi N g_1^2}{\kappa_1} (-1)^{k} \\
            &\times \left( \frac{(c^0_{\pi, 9/2}(m_F))^2}{\delta_{C1}-k\times\Delta_{\textrm{FSR},1}} + \frac{(c^0_{\pi, 11/2/2}(m_F))^2}{\delta_{C1} - \Delta_{11/2}-k\times\Delta_{\textrm{FSR},1}}+\frac{(c^0_{\pi, 7/2}(m_F))^2}{\delta_{C1} - \Delta_{7/2}-k\times\Delta_{\textrm{FSR},1}} \right),
        \end{split}
    \end{equation}
\end{widetext} 
where $c^0_{\pi, 9/2}(m_F)$, $c^0_{\pi, 9/2}(m_F)$ and $c^0_{\pi, 9/2}(m_F)$ are the Clebsch-Gordan coefficients for the $\pi$-polarized transition on the $F = 9/2,11/2,7/2$ hyperfine manifolds for each $m_F$ state, populated with probability $P_G(m_F)$. Note that the sum subtracts the shifts on the two cavity modes ($k$ index), as shown in Fig.~\ref{fig:SI1}(b)(ii), and the signs on $\Delta_{9/2}$ and $\Delta_{7/2}$ are taken to be consistent with the cavity detuning definition ($\delta_{C1}$).

Corrections on the measured ratio are shown in Fig.~\ref{fig:SIfig2}(e) for the case where the fractional population on the $\pm 7/2$ states is $\varepsilon$, on the $\pm 5/2$ states is $\varepsilon^2$,  on the $\pm 3/2$ states is $\varepsilon^3$, and on the $\pm 1/2$ states is $\varepsilon^4$. We determine the value of $F_C$ as the one for $\varepsilon = 0.05$, and its error the one associated to its spread in order to cover up to $\varepsilon = 0.1$, giving $F_C = 1.001(1)$. We point out again the dependence on the cavity detuning to the $^3$P$_1$ manifold, $\delta_{C1}$, on the correction factor $F_C$ on Fig.~\ref{fig:SIfig2}(e) inset for $\varepsilon=0.05$. For the value we choose to operate ($\delta_{C1}/(2\pi) = 277.5(8)$~MHz) we are near the maximum correction factor, but we gain in insensitivity with respect to the cavity detuning.

\subsubsection{Finite lifetime on the optical lattice}

Any of our measurement sequences that involve a few consecutive measurements per experimental trial are susceptible to atom loss from the trap. In particular, the lifetime in the lattice is $\tau_{\textrm{latt}}\sim 500$~ms (limited by parametric heating), while typical measurements on the clock transition last $T_m \sim 20$~ms typically. By combining 5 of these measurements, as in Fig.~3(b) in the main text, we can use the different outcomes and partially cancel the effect of the trap lifetime, by retaining a correction $1+\alpha (T_m/\tau_{\textrm{latt}})^2$, where $\alpha$ can vary from 0 to 1 according to the way we combine the measurement outcomes (see next sections). The magnitude and error on the correction contemplates a uniform spread of $\alpha$.

\subsubsection{Cavity birefringence}
In an ideal atom-cavity system, light polarized along the atoms' quantization axis will only interact with $\pi$ transitions. However, the presence of cavity birefringence featuring normal modes misaligned with this axis leads to a coupling between $\pi$-polarized light and atomic transitions normally driven by circularly polarized light that is quadratic in the birefringent energy splitting. This effect introduces corrections to both phase shift measurements which do not cancel in their ratio, leading to a systematic on the ratio measurement. Calculating these corrections requires modifying the cavity transfer function shown in equation \ref{eq:ssc}.

In the presence of cavity birefringence, a single longitudinal mode splits into two resonances characterized by polarization eigenmodes $\hat{c}_\pm$, such that
\begin{equation}
    \hat{H}_\text{cav} = (\omega_c - \frac{\delta_b}{2}) \hat{c}_-^\dagger \hat{c}_- + (\omega_c + \frac{\delta_b}{2}) \hat{c}_+^\dagger \hat{c}_+
\end{equation}
for birefringent splitting $\delta_b$. Since the probe beam polarization and quantization axis are aligned to a common vertical direction ($\hat{x}$ as in Fig.~\ref{fig:SI1}(a)), it makes sense to express these eigenmodes in this basis as well. This is accomplished using two parameters $\theta_b, \varphi_b$:
\begin{equation}
\begin{split}
    \hat{c}_- &= \big[\cos(\theta_b/2) \big]\, \hat{v} + \big[-\sin(\theta_b/2)e^{-i \varphi_b}\big]\, \hat{h}\\
    \hat{c}_+ &= \big[\sin(\theta_b/2) e^{i \varphi_b}\big]\, \hat{v} + \big[\cos(\theta_b/2)\big]\, \hat{h},
\end{split}
\end{equation}
such that light along $\hat{h}$ ($\hat{y}$ as in Fig.~\ref{fig:SI1}(a)) and $\hat{v}$ polarizations interact with $\sigma$ and $\pi$ transitions respectively. The above expressions are essentially Jones vector representations of the eigenmodes; correspondingly, $\theta_b$ and $\varphi_b$ can be thought of as spherical coordinates for the eigenmodes on the Poincar\'e sphere with poles defined by $h$ and $v$ polarizations.

As long as the atoms occupy stretched states ($m_F = \pm 9/2$), there is only one $\sigma$ transition. Therefore one can unambiguously define collective spin operators along the two transitions, denoted by $J_{\pi/\sigma}^\pm = \sum_{i=1}^N \sigma_{i,\pi/\sigma}^\pm$ and  $J_{\pi/\sigma}^z = \tfrac{1}{2}\sum_{i=1}^N \sigma_{i,\pi/\sigma}^z$ for single particle operators $\sigma_{i,\pi/\sigma}^*$. We go into the rotating frame of the atoms, assuming the two transitions are degenerate in frequency, to construct the following Hamiltonian:
\begin{widetext}
    \begin{equation}
        \hat{H} = \Big[\big(\delta_c - \frac{\delta_b}{2}\cos\theta_b\big) \hat{v}^\dagger \hat{v} + \big(\delta_c + \frac{\delta_b}{2}\cos\theta_b\big) \hat{h}^\dagger \hat{h} + \frac{\delta_b}{2}\sin\theta_b \big( \hat{h}^\dagger \hat{v} e^{i \varphi_b} + \hat{h} \hat{v}^\dagger e^{-i \varphi_b} \big)\Big] 
    + \Big[g_\pi \big( \hat{v} \hat{J}_\pi^+ + \hat{v}^\dagger \hat{J}_\pi^- \big)
    + g_\sigma \big( \hat{h} \hat{J}_\sigma^+ + \hat{h}^\dagger \hat{J}_\sigma^- \big)\Big].
    \end{equation}
\end{widetext}
Analogously to the derivation at the start of this document, one can derive Optical Bloch equations to analyze mean-field behavior ($O = \langle\hat{O}\rangle$). Assuming a vertically polarized cavity drive $v_i(t)$ at detuning $\delta_p$ from atomic resonance, these equations are given by
\begin{widetext}
\begin{equation}
\begin{split}
    \dot{v} &= -i \Big[\big(\delta_c - \frac{\delta_b}{2}\cos \theta_b\big)v + \big(\frac{\delta_b}{2}\sin \theta_b\,e^{-i \varphi_b}\big)h\Big] - i g_\pi J_\pi^- - \frac{\kappa}{2} v + \sqrt{\kappa_m} v_i(t)\\
    \dot{h} &= -i \Big[\big(\delta_c + \frac{\delta_b}{2}\cos \theta_b\big)h + \big(\frac{\delta_b}{2}\sin \theta_b\,e^{i \varphi_b}\big)v\Big] - i g_\sigma J_\sigma^- - \frac{\kappa}{2} h \\
    \dot{J}_\pi^- &= 2i g_\pi v J_\pi^z - \gamma_\pi^\perp J_\pi^- \qquad\qquad\quad\;\;\,\,
    \dot{J}_\sigma^- = 2i g_\sigma h J_\sigma^z - \gamma_\sigma^\perp J_\sigma^- \\
    \dot{J}_\pi^z &= i g_\pi \left(v^\dagger J_\pi^- - v J_\pi^+\right) - \gamma_\pi N_\pi \qquad
    \dot{J}_\sigma^z = i g_\sigma \left( h^\dagger J_\sigma^- - h J_\sigma^+\right) - \gamma_\sigma N_\sigma,
\end{split}
\end{equation}
\end{widetext}
where $N_{\pi/\sigma}$ represents the number of atoms excited along the $\pi/\sigma$ transition. In the weak probe limit, both of these go to $0$ as there are no excited atoms available to decay.

From these equations, one can determine how the input probe $v_i(t) = \tilde{v}_i e^{-i\delta_p t}$ changes in transmission through the atom-cavity system. In general for a birefringent cavity, the transmitted light's polarization may be different from the probe due to different resonance conditions for the two normal polarization modes. In our experiment, we beat the transmitted light with a vertically polarized local oscillator to measure the light in heterodyne, so the signal of interest is the vertical component of any transmitted light. We are therefore interested in the transfer function $T_v(\delta_p)$, defined by
$\tilde{v}_t = T_v(\delta_p) \tilde{v}_i$. It turns out that $T_v$ can be expressed in terms of the following transfer functions, which decouple the horizontal and vertical excitations:
\begin{equation}
    \begin{split}
        T_\pi(\delta_p) &= \frac{1}{1 - i(\tfrac{\delta_p - \delta_c + \delta_b/2\, \cos\theta_b}{\kappa/2}) + \tfrac{NC_\pi\gamma_\pi/2}{\gamma^\perp_\pi - i \delta_p}} \\
        T_\sigma(\delta_p) &= \frac{1}{1 - i(\tfrac{\delta_p - \delta_c - \delta_b/2\, \cos\theta_b}{\kappa/2}) + \tfrac{NC_\sigma\gamma_\sigma/2}{\gamma^\perp_\sigma - i \delta_p}}.
    \end{split}
\end{equation}
Then the full transfer function is given by
\begin{equation}
    \label{eq:tf_biref}
    T_v(\delta_p) = \frac{T_\pi(\delta_p)}{1 + (\tfrac{\delta_b}{\kappa}\sin\theta_b)^2 T_\sigma(\delta_p) T_\pi(\delta_p)}.
\end{equation}
Note that the transfer function does not depend on the azimuthal angle $\varphi_b$; this holds as long as the transmitted light is only measured along $v$. For small birefringent splitting, $F_v$ can be calculated perturbatively by expanding in powers of $(\tfrac{\delta_b}{\kappa} \sin\theta_b)^2$. It follows that leading order corrections to the cavity shifts will be quadratic in $\delta_b$.

Using a simple polarimetry setup consisting of PBSs, waveplates, and photodiodes, we were able to measure $\delta_{b0}/\kappa_0 = +0.16(2)$, $\delta_{b1}/\kappa_1 = +0.16(2)$, and $\theta_b = 30(2)\degree$. This implies $(\tfrac{\delta_b}{\kappa}\sin\theta_b)^2 = 0.006(2)$ along both transitions, justifying a perturbative treatment. From this, we can calculate the modified shifts and derive a correction factor for the shift ratio $\Delta\varphi_0/\Delta\varphi_1$, which turns out to be $F_C = 1.012(3)$. This value accounts for all differential shift measurements, as well as the full hyperfine landscape.

Considering the effect of cavity birefringence opens up new potential sources of uncertainty. First, one might imagine that an imperfect optical pumping scheme might conspire with the cavity's birefringence to produce larger corrections than previously discussed. In fact, numerical calculations show the two effects are largely decoupled and can be treated separately. Second, if the local oscillator is misaligned from vertical polarization by some small angle $\alpha$, all phase shifts will receive a linear correction proportional to $\alpha\,\tfrac{\delta_b}{\kappa}\sin\theta_b$. However, the experiment's differential probe design leads to partial cancellation of these shifts. Assuming $\alpha$ is off by a much as $5\degree$, the additional uncertainty on $F_C$ is at most $0.003$. Finally, if the two birefringent normal modes exhibit slightly different linewidths, the optical Bloch equations change accordingly and lead to further phase shift modifications. Data used to determine $\delta_b$ allows us to constrain any linewidth difference to $\delta\kappa \lesssim 0.05\kappa$, which limits the correction on $F_C$ to $\lesssim 0.001$. Incorporating these additional sources of uncertainty into the birefringence correction factor gives $F_C = 1.012(5)$.

This experiment was performed before the discovery of cavity birefringence in our system. In principle, for future experiments one could mitigate the effect of this systematic by aligning all beam polarizations and the atoms' quantization axis along the birefringent eigenmode axis. If the eigenmodes are linearly polarized, the probe beam will only excite one of the two modes, completely removing any birefringent coupling. Otherwise, any ellipticity the eigenmodes possess will limit the ability to suppress the coupling with a linearly polarized probe beam, which is necessary for this experiment. In our system, the effect of birefringence could be suppressed by approximately $\sim 17$ by such an alignment.

\subsection{Summary - Full systematic correction}

Taking all this effects into account we infer a correction factor $F_C = 1.074(16)$ on the measured ratio $\left(\Delta\varphi_{0}/\Delta\varphi_1\right)$. Its value is determined by multiplying the systematic corrections detailed in Tables \ref{tab:correction689}, \ref{tab:correction698}, and \ref{tab:correctionratio}, while its error is properly summed in quadrature. The error on $F_C$ is dominated mostly by technical issues, such as the uncertainty in the clock atomic frequency, the cavity alignment with the atomic resonance, and alignment of the probe polarization with respect to the cavity eigenmode axis considering birefringence, which can be further improved. Furthermore, its uncertainty is also dominated by technical aspects such as signal to noise in our data and its influence on determining the phase shifts offsets, as well as uncertainty in the atomic transition frequency and cavity alignment on the clock transition.

\section{Extrapolating $\Delta\varphi_0/\Delta\varphi_1$ to zero probe power and interleaved measurements}

In this section, we will discuss the details of the low power measurement presented in Fig.~3(b) in the main text. Absent systematic corrections, it remains to determine the zero-probe-power value for the ratio $\left(\Delta \varphi_{0}/\Delta\varphi_1\right)$, that we will name $\left(\Delta \varphi_{0}/\Delta\varphi_1\right)_{P=0}$.

We measured the ratio of the atomic phase shift to cavity frequency shift while simultaneously decreasing both 698 and 689 probe powers, \textit{P}$_0$ and \textit{P}$_1$ respectively, and taking longer sets to accumulate similar statistics for lower optical power measurements, as expected from the photon-shot noise scaling. The ratio is expected to strongly depend on both powers, although the maximum 689~nm optical power was already low enough to be a significant effect, according to the results shown in panels b and c in Fig.~\ref{fig:SIfig2}. We measure $\Delta\varphi_{0}$ and $\Delta\varphi_1$ in an interleave form, to gain insensitivity with respect to lattice lifetime. We realize five measurements every $T_c =25$~ms, as shown in Fig.~3(b) in the main text, interleaving three $\Delta\varphi_1$ short measurements ($\sim 2~$ms) with two longer $\Delta\varphi_0$ measurements ($\sim 25~$ms). Upon further detailed inspection, the ratio of the average of two $\Delta\varphi_{0}$'s and the average of the three $\Delta\varphi_1$'s measurements will have the same linear sensitivity to atom loss, therefore a ratio of the two quantities will be quadratically sensitive to $T_c/\tau_{\textrm{latt}}$.  

The 698~nm clock transition probe could excite atoms to $\ket{e_0}$, and those atoms will not be counted by the following dispersive 689~nm probe. We assume, in the weak probe power limit, that each clock transition probe excites a fraction $\beta_0$ of atoms into $\ket{e_0}$ every $T_c/2$ interval while they are being probed. Reversely, if there are atoms in the excited state, a fraction $\beta_0$ is transferred to $\ket{g}$. For the 689~nm probe, we assume an excitation fraction $\beta_1$, but also that any atom in the excited state is reset to the ground state before the following $\Delta\varphi_0$ measurement, as the spontaneous emission decay time is only 21~$\mu$s. Furthermore, losses from the lattice are treated as an exponential loss decay with time constant $\tau_{\textrm{latt}}$, which was experimentally verified repeatedly.

We use the measurement outcomes of the different $\Delta\varphi_{0}$ and $\Delta\varphi_1$ measurements to construct different estimators for the zero-power ratio $\left(\Delta \varphi_{0}/\Delta\varphi_1\right)_{P=0}$. Examples of these estimators, to name a few, are 
\begin{eqnarray}
    &E_1 =& \frac{3}{2}\left(\frac{\Delta\varphi^{1}_0+\Delta\varphi^{2}_0}{\Delta\varphi^{1}_1+\Delta\varphi^{2}_1+\Delta\varphi^{3}_1}\right) \\
    &E_2 =& 2\left(\frac{\Delta\varphi^{1}_0+\Delta\varphi^{2}_0}{\Delta\varphi^{1}_1+2\Delta\varphi^{2}_1+\Delta\varphi^{3}_1}\right) \\
    &E_3 =& 4 \left(\frac{\Delta\varphi^{1}_0+\Delta\varphi^{2}_0}{\Delta\varphi^{1}_1+6\Delta\varphi^{1}_1+\Delta\varphi^{3}_1}\right),
\end{eqnarray}
where the super-index orders each of the five measurements, i.e. $\Delta\varphi^{2}_1$ is the second $\Delta\varphi_1$ measurement.

For low optical power $\beta_0$ ($\beta_1$) is proportional to the probe optical power \textit{P}$_0$ (\textit{P}$_1$) on the clock transition (7.5~kHz transition) and satisfies $\beta_0 \ (\beta_1) \ll 1$). In this case we can compute how populating the $\ket{e_0}$ and $\ket{e_1}$ states during the measurement sequence affects the estimators, for example, 
\begin{widetext}
\begin{align}\begin{split}
    &E_1 =  \left(\frac{\Delta\varphi_0}{\Delta\varphi_1}\right)_{P=0} \left(1-\beta_0+2\beta_1 -\frac{5}{6}\left(T_c/\tau_{\textrm{latt}}\right)^2+\mathcal{O}\left(\beta_0,\beta_1,T_c/\tau_{\textrm{latt}}\right)\right)\\
    &E_2 = \left(\frac{\Delta\varphi_0}{\Delta\varphi_1}\right)_{P=0} \left(1-\beta_0+2\beta_1-\frac{1}{2}\left(T_c/\tau_{\textrm{latt}}\right)^2+\mathcal{O}\left(\beta_0,\beta_1,T_c/\tau_{\textrm{latt}}\right)\right) \\
    &E_3 = \left(\frac{\Delta\varphi_0}{\Delta\varphi_1}\right)_{P=0} \left(1-\beta_0+2\beta_1+\mathcal{O}\left(\beta_0,\beta_1,T_c/\tau_{\textrm{latt}}\right)\right),
\end{split}\end{align}
\end{widetext}
where $\mathcal{O}\left(\beta_0,\beta_1,T_c/\tau_{\textrm{latt}}\right)$ refers to higher order terms in combinations of $\beta_0,\beta_1$, and $T_c/\tau_{\textrm{latt}}$, and $\left(\frac{\Delta\varphi_0}{\Delta\varphi_1}\right)_{P=0}$ is the zero-power ratio that we want to determine. 

In Fig.~3(b) in the main text we show the result for the so-called $E_3$ estimator above and show quadratic polynomial fits in the optical power \textit{P}$_0$ for the clock transition ($\beta_0\propto\textit{P}_0$). In Table~\ref{tab:estimators} we show the fitted $\left(\frac{\Delta\varphi_0}{\Delta\varphi_1}\right)_{P=0}$ for different estimators and fit methods, as a way to show a consistent method-independent value. The data is not corrected by any systematic. We also point out that we did not take $\beta_1$ or $P_1$ into consideration for either of these fits, as doing so does not significantly modify the other fitted parameters, because the maximum value that $P_1$ takes on all the Fig.~3(b) measurements is already low enough to cause significant population in $\ket{e_1}.$

\begin{table*}[htb!]
\caption{\label{tab:estimators} This table contains different fits for several estimators for the ratio $\left(\Delta \varphi_{0}/\Delta\varphi_1\right)$}
\begin{ruledtabular}
\begin{tabular}{lccc}
\textrm{Estimator}&
\textrm{Fit method/origin}&
\textrm{$\left(\frac{\Delta\varphi_0}{\Delta\varphi_1}\right)_{P=0}~\times 10^{-2}$}&
\textrm{$\chi_\nu^2$}\\
\colrule
Estimator $E_1$ & Quadratic on \textit{P}$_0$. Mean value for crossing. & -8.92(6) & \\
\colrule
Estimator $E_2$ & Quadratic on \textit{P}$_0$. Mean value for crossing. & -8.92(6) &  \\
\colrule
Estimator $E_3$ & Quadratic on \textit{P}$_0$. Mean value for crossing. & -8.95(6) &  \\
\colrule
Estimator $E_3$ & Linear on \textit{P}$_0$ (\textit{P}$_0\leq400$~pW). Mean value for crossing. & -8.95(6) \\
\colrule
Estimator $E_3$& Quadratic on \textit{P}$_0$. Mean value for crossing. & -8.95(6) &  \\
removing 7~ms data & & & \\
\colrule
Estimator $E_3$  & Quadratic on \textit{P}$_0$.  & -8.86(6) & 0.7 \\
only for red set & & & \\
\colrule
Estimator $E_3$ & Quadratic on \textit{P}$_0$.  & -9.02(7) & 1.2 \\
only for green set & & & \\
\colrule
Estimator $E_3$ & Quadratic on \textit{P}$_0$.  & -8.95(4) & 0.6 \\
only for blue set & & & \\
\colrule
Estimator $E_3$ & Quadratic on \textit{P}$_0$. Using a global fit to the three sets. & -8.95(4) & 1.1\\
\end{tabular}
\end{ruledtabular}
\end{table*}

The results are consistent with a zero power crossing $\left(\frac{\Delta\varphi_0}{\Delta\varphi_1}\right)_{P=0} = -8.95(9)\times 10^{-2}$.
We finally note that for single measurement instances as represented by the red, green and blue data sets, independently of the estimators we compute, the data is spread consistently with a 10~kHz uncertainty on the alignment of the cavity resonance frequency to the clock atomic transition ($\delta_{C0}$), as described previously. A zoom in of the data presented in Fig.~3(b) in the main text is shown in Fig.~\ref{fig:SIfig2}(f). Fits are for the estimator $E_3$ for each set, and the black solid line is a global fit for all the data sets.  

\section{Extracting $\gamma_0/\gamma_1$ from the measured $\Delta\varphi_0/\Delta\varphi_1$}\label{sec:sec4}

In order to extract the ratio between $\left(g_0/g_1\right)^2$, which in conjunction with the known $^3$P$_1$ linewidth \cite{Nicholson_2015} could determine the $^3$P$_0$ natural radiative linewidth, we use Eq.~\ref{eq:deltaphi_full} and Eq.~\ref{eq:deltaomega_full}, the systematic correction $F_C$ and the measured value for $\left(\Delta \varphi_{0}/\Delta\varphi_1\right)_{P=0}$ from the zero-power crossing measurement. Furthermore, we need to determine all the numerical factors that appear in Eq.~\ref{eq:deltaphi_full} and Eq.~\ref{eq:deltaomega_full} with their uncertainty. Most of these factors were already described above and appear on Table~\ref{tab:CavityAtomParams}.

We obtain $\left(\Delta\varphi_{0}/\Delta\varphi_1\right)_{P=0} = -9.61(17)\times 10^{-2}$ after applying the systematic correction factor $F_C$, summing both statistical and systematic errors contributions in quadrature. Using Eq.~\ref{eq:deltaphi_full} and Eq.~\ref{eq:deltaomega_full}, we determine $(g_0/g_1)^2 = 1.83(3)\times10^{-7}$.  

The radiative excited state linewidths are $\gamma_0 \propto d_0^2 (\omega_{A0})^3$ and $\gamma_1 \propto d_1^2 (\omega_{A1})^3$ \cite{scully1997quantum,Weisskopf1930}, where $d_0$ and $d_1$ are the electric dipole moments between the $^1$S$_0$ and the $^3$P$_0$ state, and the $^1$S$_0$ and the $^3$P$_1$ state, respectively. On the other hand, $g_0\propto d_0 \sqrt{\frac{\omega_{A0}}{\textrm{w}^2_0 \textrm{L}_0}}$ and $g_1\propto d_1 \sqrt{\frac{\omega_{A1}}{\textrm{w}^2_1 \textrm{L}_1}}$ for the same pair of transitions \cite{Jaynes_1963,Kimble_1998,TanjiSuzuki_2011}. Here $\textrm{w}_{0,1}$ and $\textrm{L}_{0,1}$ refer to the cavity mode waist ($1/e^2$ radio) and cavity length for the same transitions as before. All the numerical proportionality constants that we are omitting are physical constants, independent of the transitions we use. Finally, we have 
\begin{equation}
    \frac{\gamma_0}{\gamma_1}  = \left(\frac{g_0}{g_1}\right)^2 \left(\frac{\textrm{w}_0}{\textrm{w}_1}\right)^2 \left(\frac{\omega_{A0}}{\omega_{A1}}\right)^2 \frac{\textrm{L}_0}{\textrm{L}_1}
\end{equation}

Using the values measured in this work and the best reported value for $\gamma_1$ to our knowledge \cite{Nicholson_2015}, we report $\gamma_0/(2\pi) = 1.35\pm 0.03$~mHz, that implies a lifetime of $118 \pm 3$~s.  

\section{Constraining $N$ dependent effects on $\Delta\varphi_0/\Delta\varphi_1$ measurements.}\label{sec:sec5}

As discussed on a few of the systematic corrections presented previously, sometimes we can find atom number dependent corrections that do not completely cancel when measuring the ratio $\Delta\varphi_0/\Delta\varphi_1$. For example, when discussing the independent phase shifts offsets or higher order corrections. 

In order to check the influence of these effects, and lacking an underlying model to believe they would impact our measurement, we decided to perform $\Delta\varphi_0/\Delta\varphi_1$ measurements for different atom number $N$. These measurements are presented as an inset in Fig.~3(b) in the main text. 

For that set, our phase shifts measurements present some non-zero phase shifts offsets that were properly measured. The most simplistic model, as introduced in Eq.~\ref{eq:offsets}, serves us as a proxy to further investigate any unknown variations with $N$ and $1/N$ on the ratio measurements. All in all, by considering different variations of these fits, taking into account the $N$ = 0 point and the offsets we measured, we find agreement at the 2\% level with the weighed average of the measured phase shift ratio (the value that we would assign for the ratio if no $N$ dependent effect were present). This uncertainty is dominated by the signal to noise ratio on the current data set.

Therefore, we constrain any unknown $N$ dependent effect on the ratio at the 2\% level, which is at the level of our final uncertainty on the phase shift ratio and linewidth ratio. We consider this experiment as a sanity check, but we do not use this independent constraint to modify our final uncertainty.

\end{document}